\documentclass[preprint,5p,number,sort&compress,times]{elsarticle}
\usepackage{mathtools, nccmath}
\usepackage{amsfonts}

\usepackage{hypernat}
\usepackage[pdfencoding=auto, psdextra,hidelinks]{hyperref}
\usepackage[boxruled,vlined,linesnumbered]{algorithm2e}

\makeatletter
\newcommand{\removelatexerror}{\let\@latex@error\@gobble}
\makeatother

\usepackage{multicol}
\usepackage{listings}

\usepackage[dvipsnames,table]{xcolor}
\usepackage{array}
\newcolumntype{M}[1]{>{\centering\arraybackslash}m{#1}}
\usepackage{afterpage}
\usepackage{placeins}
\usepackage{stackengine}
\usepackage{anyfontsize}

\makeatletter
\AtBeginEnvironment{algorithm}{\let\c@figure\c@nalgo}
\makeatother

\usepackage[nameinlink,capitalize]{cleveref}

\setstackgap{L}{0.75\normalbaselineskip}
\stackMath 
\newcommand\showdim[3]{\stackunder{#1}{\scalebox{0.7}{$\color{Blue} \left(#2\! \times\! #3\right)$}}}
\newcommand\showdimv[2]{\stackunder{#1}{\scalebox{0.7}{$\color{Blue} \left(#2\right)$}}}
\SetKwInput{KwTemp}{Temporary Variables}
\SetKwInput{KwTId}{\texttt{MPI} rank}
\SetKwInput{KwBIdx}{\texttt{blockIdx.x}}
\SetKwInput{KwBIdy}{\texttt{blockIdx.y}}
\SetKwInput{KwTIdx}{\texttt{threadIdx.x}}
\SetKwInput{KwTIdy}{\texttt{threadIdx.y}}

\SetCommentSty{mycommfont}

\definecolor{codegreen}{rgb}{0,0.6,0}
\definecolor{codegray}{rgb}{0.5,0.5,0.5}
\definecolor{codepurple}{rgb}{0.58,0,0.82}
\definecolor{backcolour}{rgb}{0.95,0.95,0.92}
\usepackage[many]{tcolorbox}
\tcbuselibrary{listings}
\tcbuselibrary{breakable}
\tcolorboxenvironment{lstlisting}{
  spartan,boxrule=0pt,
  colframe=Gray!15,
  boxsep=0mm,breakable,
  left=3mm,right=1.5mm,top=0mm,bottom=0mm,
  colback=Gray!15,
}
\lstdefinestyle{mystyle}{
    language=C++,
    backgroundcolor=\color{Gray!15}, 
    commentstyle=\color{Maroon},
    keywordstyle=\color{Blue},
    directivestyle={\color{green}},
    numberstyle=\color{black}\scriptsize\textbf,
    stringstyle=\color{codepurple},
    basicstyle=\ttfamily,
    breakatwhitespace=false,         
    breaklines=true,                 
    captionpos=b,                    
    keepspaces=false,                 
    numbers=left,                    
    numbersep=1pt,                  
    showspaces=false,                
    showstringspaces=false,
    showtabs=false,                  
    tabsize=1,
    otherkeywords={constexpr,__m512d},
    escapeinside={||}
}

\usepackage{amssymb}

\usepackage[utf8]{inputenc}
\usepackage{graphicx}
\usepackage{float}
\usepackage{stfloats}
\usepackage{microtype}
\graphicspath{{images/}}
\usepackage{amsmath}
\usepackage{bm}
\usepackage{wrapfig}
\usepackage{caption}
\usepackage{subcaption}
\usepackage{scalerel}
\usepackage{xcolor}

\newcommand{\del}{\nabla}

\newcommand{\bb}{\boldsymbol{b}}

\newcommand{\bn}{\boldsymbol{n}}

\newcommand{\bu}{\boldsymbol{u}}

\newcommand{\bx}{\boldsymbol{\textbf{x}}}

\newcommand{\bA}{\normalfont\boldsymbol{\textbf{A}}}
\newcommand{\bB}{\boldsymbol{\textbf{B}}}
\newcommand{\bC}{\normalfont\boldsymbol{\textbf{C}}}
\newcommand{\bD}{\boldsymbol{\textbf{D}}}
\newcommand{\bE}{\normalfont\boldsymbol{\textbf{E}}}
\newcommand{\bF}{\boldsymbol{\textbf{F}}}
\newcommand{\bG}{\boldsymbol{\mathcal{G}}}
\newcommand{\bH}{\boldsymbol{\textbf{H}}}
\newcommand{\bI}{\boldsymbol{\textbf{I}}}
\newcommand{\bJ}{\boldsymbol{\textbf{J}}}
\newcommand{\bK}{\boldsymbol{\textbf{K}}}

\newcommand{\bX}{\boldsymbol{\textbf{X}}}
\newcommand{\bY}{\boldsymbol{\textbf{Y}}}
\newcommand{\bZ}{\boldsymbol{\textbf{Z}}}

\newcommand{\bM}{\boldsymbol{\textbf{M}}}
\newcommand{\bN}{\normalfont\boldsymbol{\textbf{N}}}

\newcommand{\bP}{\normalfont\boldsymbol{\textbf{P}}}
\newcommand{\bQ}{\normalfont\boldsymbol{\textbf{Q}}}
\newcommand{\bR}{\normalfont\boldsymbol{\textbf{R}}}

\newcommand{\bT}{\normalfont\boldsymbol{\textbf{T}}}
\newcommand{\bU}{\normalfont\boldsymbol{\textbf{U}}}
\newcommand{\bUt}{\boldsymbol{\mathcal{U}}}
\newcommand{\bTt}{\boldsymbol{\mathcal{T}}}
\newcommand{\bRt}{\boldsymbol{\mathcal{R}}}
\newcommand{\bV}{\normalfont\boldsymbol{\textbf{V}}}
\newcommand{\bW}{\boldsymbol{\textbf{W}}}

\newcommand{\bxi}{\boldsymbol{\xi}}

\newcommand{\dx}{\,d\bx}

\newcommand{\clstcomment}{\color{Maroon}\small}

\newcommand{\cn}{\color{black}}

\newcommand\norm[1]{\lVert#1\rVert}
\newcommand\wh[1]{\hstretch{2}{\hat{\hstretch{.5}{#1}}}}
\def\appendixname{}

\journal{Journal of Parallel and Distributed Computing}
\hypersetup{
  pdftitle={Fast hardware-aware matrix-free algorithm for higher-order finite-element discretized matrix-multivector products},
  pdfauthor={Gourab Panigrahi, Nikhil Kodali, Debashis Panda and Phani Motamarri}
}

\begin{document}

\begin{frontmatter}



\title{\LARGE Fast hardware-aware matrix-free algorithm for higher-order finite-element discretized matrix multivector products on distributed systems}


\author[inst1]{Gourab Panigrahi\fnref{fn1}}
\author[inst1]{Nikhil Kodali\fnref{fn1}}
\author[inst1]{Debashis Panda\fnref{fn2}}
\author[inst1]{Phani Motamarri}
\fntext[fn1]{Gourab Panigrahi and Nikhil Kodali contributed equally to this work.}
\fntext[fn2]{Currently at Department of Chemical Engineering in Imperial College London, U.K.}
\ead{phanim@iisc.ac.in}
\affiliation[inst1]{organization={Department of Computational and Data Sciences, Indian Institute of Science},
            addressline={CV Raman Road}, 
            city={Bengaluru},
            postcode={560012}, 
            state={Karnataka},
            country={India}}


\begin{abstract}
Recent hardware-aware matrix-free algorithms for higher-order finite-element (FE) discretized matrix-vector multiplications reduce floating point operations and data access costs compared to traditional sparse matrix approaches. This work proposes efficient matrix-free algorithms for evaluating FE discretized matrix-multivector products on both multi-node CPU and GPU architectures. We address a critical gap in existing matrix-free implementations, which are well suited only for the action of FE discretized matrices on a single vector. We employ batched evaluation strategies, with the batchsize tailored to underlying hardware architectures, leading to better data locality and enabling further parallelization. On CPUs, we utilize even-odd decomposition, SIMD vectorization, and overlapping computation and communication strategies. On GPUs, we employ strategies to overlap compute and data movement in conjunction with GPU shared memory, constant memory, and kernel fusion to reduce data accesses. Our implementation outperforms the baselines for Helmholtz operator action, achieving up to 1.4x improvement on one CPU node and up to 2.8x on one GPU node, while reaching up to 4.4x and 1.5x improvement on multiple nodes for CPUs ($\sim 3000$ cores) and GPUs ($\sim$ 25 GPUs), respectively. We further benchmark the performance of the proposed implementation for solving a model eigenvalue problem for 1024 smallest eigenvalue-eigenvector pairs by employing the Chebyshev Filtered Subspace Iteration method, achieving up to 1.5x improvement on one CPU node and up to 2.2x on one GPU node while reaching up to 3.0x and 1.4x improvement on multinode CPUs ($\sim 3000$ cores) and GPUs ($\sim$ 25 GPUs), respectively.
\end{abstract}



\begin{keyword}
Matrix-free \sep Finite Element Method \sep Sum factorization \sep Scalable algorithms for heterogeneous architectures
\end{keyword}
\end{frontmatter}
\clearpage
\section{Introduction}
Finite-element (FE) based computational methodologies are routinely employed to numerically solve partial differential equations (PDEs) arising in various domains of science and engineering.  The underlying FE basis functions are usually the compactly supported piecewise-continuous Lagrange polynomials. The numerical solution of a partial differential equation employing the FE basis usually involves constructing an FE discretized operator, which is a sparse matrix due to the compact support of these FE basis functions. Consequently, the PDE reduces to a sparse system of linear equations or sparse matrix eigenvalue problems. These sparse matrix problems are traditionally solved using iterative solvers, which require computing the action of the sparse matrix on trial FE discretized fields for the solution of a linear system of equations or eigenvalue problems. Evaluation of the product of the sparse matrix and the vector (FE discretized field) is usually the computationally demanding step. It is traditionally computed using sparse-matrix vector multiplication algorithms \cite{Kirby2006OptimizingMatrices,Hughes2012TheAnalysis}. However, previous works~\cite{Carey1988Element-by-elementComputations,Hughes1987Large-scaleGradients,Cantwell2011FromElements} note that the evaluation of such sparse matrix-vector products for higher-order finite-elements can be performed more efficiently on multithreaded architectures using FE-cell level dense matrix-vector multiplications followed by the assembly of FE-cell level product vectors. \citet{Motamarri2020DFT-FECalculations,Das2022DFT-FEDiscretization} have recently employed this strategy on multi-node CPU and GPU architectures for evaluating the FE discretized matrix-multivector products involving a large number of vectors ($>$300). They have demonstrated a good throughput performance for the solution of FE discretized large-scale nonlinear eigenvalue problems arising in the field of quantum modeling of materials using density functional theory. However, recent hardware-aware algorithms for evaluating such matrix-vector multiplications suggest that computing on-the-fly matrix-vector products without storing the FE-cell level dense matrices reduces arithmetic complexity, data movement and memory footprint \cite{Ljungkvist2017Matrix-FreeMeshes,Kronbichler2012AApplication,Davydov2020AMultigrid,Fischer2020ScalabilitySolvers}. These algorithms, referred to as matrix-free approaches, exploit the tensor-structured nature of the FE basis functions and recast the 3D integrals involved in the matrix-vector products as a sequence of tensor contractions.
The open-source implementations of the above matrix-free methods currently available to the community~\cite{Arndt2019TheLibrary,Anderson2021MFEM:Library,Brown2021LibCEED:Discretizations,Swirydowicz2019AccelerationMethodsb} are neither optimal nor directly applicable for the action of an FE discretized operator on a large number of vectors. Such situations are often encountered when solving FE discretized eigenvalue problems~\cite{Hughes2012TheAnalysis,Sun2016FiniteProblems} using iterative orthogonal projection approaches or solving linear systems of equations arising from FE discretizations with multiple RHS vectors. These problems arise in real-space quantum modeling of materials \cite{Tsuchida1996AdaptiveCalculations,Das2022DFT-FEDiscretization,Ghosh2019All-electronSolids}, electroelastics~\cite{Martynova2023TheMaterials}, modal analysis~\cite{Fan2014ParallelAnalysis,Fan2015SomeFramework}, and scientific machine learning to train ML models with the solutions of FE discretized PDEs involving multiple forcing vectors \cite{Markidis2021TheSolvers}. Although some preliminary works, such as interpolation of mulitvectors to quadrature points \cite{Beams2020High-OrderGPUs} and evaluation of FE operator action on sparse multivectors \cite{Davydov2020AlgorithmsMulti-Vectors} exist in this regard, no efficient algorithm exists for performing generic FE discretized matrix-multivector multiplication efficiently under the matrix-free paradigm. This work proposes an efficient hardware-aware matrix-free algorithm and implementation strategies to compute such FE discretized matrix-multivector products on multi-node CPU-only and multi-node GPU architectures. 

To this end, as traditionally done in the finite-element literature, we partition the physical domain into non-overlapping subdomains, each assigned to an \texttt{MPI} task, and use the \texttt{MPI} paradigm to communicate the boundary data across multiple nodes. The tensor contractions involved in the matrix-free approach are recast as small dense matrix-matrix multiplications involving the FE shape function matrices. On CPU architectures, to compute these small dense matrix-matrix multiplications, we utilize the SIMD vectorization capabilities of modern CPUs along with optimal implementation strategies that exploit the symmetry of the FE shape function matrices (such as the \emph{even-odd} decomposition \cite{Kopriva2009ImplementingEquations,Solomonoff1992ADifferentiation}) to minimize the computation time and use non-blocking \texttt{MPI} communications to overlap computation and communication, allowing for higher scaling efficiencies. Our proposed implementation utilizes a batched layout for the storage of the multivector, which improves data locality and allows for efficient use of the SIMD capabilities on modern CPUs, and the \emph{even-odd} decomposition strategy reduces the floating point operations required to compute the matrix multivector products by half at the cost of increased data movement. On GPU architectures, the proposed matrix-free implementation efficiently utilizes the GPU shared memory and registers to pipeline data access and computation in conjunction with the proposed batched layout. The small matrix-matrix multiplications arising in the matrix-free approach are performed as a linear combination of columns of FE shape function matrices, which are stored in constant memory, to overlap computation with data movement from device memory. Furthermore, constant memory is utilized to broadcast accesses of the FE shape function matrices and reduces shared memory usage and bank conflicts. The proposed implementation also utilizes the concept of kernel fusion to minimize data access by combining various implementation steps in a single kernel. This has the added benefit of reducing the memory footprint further. We also employ CUDA-Aware MPI to optimize communications and a mixed precision strategy to communicate data on the shared subdomain boundary to reduce the amount of data that needs to be communicated.

In \cref{sec:sec2}, we provide a concise account of the mathematical formulation of the problem that we intend to solve using a finite-element based discretization technique. Subsequently, we delve deeper into the mathematical underpinnings of the cell-matrix and matrix-free methods as applied to multivectors, specifically utilizing adaptively refined hexahedral meshes. Furthermore, we describe the various steps involved in evaluating matrix-multivector products within these frameworks, such as subdomain partitioning, the imposition of constraints to ensure continuity, extraction of FE-cell level representations and the assembly of subdomain level representations.

In \cref{sec:sec3}, we first describe the mathematical aspects of our proposed algorithm and subsequently delve into the numerical implementation strategies employed to evaluate matrix-multivector products in the matrix-free paradigm. A key consideration in this context is the adaptation of the algorithm to the specific characteristics of the underlying hardware architecture. To this end, we propose a batched algorithm in which we concurrently process a limited subset of vectors, known as a batch. The dimension of this batch is chosen based on the properties of the underlying hardware architecture. We also propose a batched layout for storing the multivectors, which significantly improves the data locality for the implementation of our proposed batched algorithm. Furthermore, we briefly describe the methods used for imposing constraints, the extraction of FE-cell level representations, and other relevant operations. We further describe the strategy employed for the implementation of the tensor contractions on both CPU-only and GPU-based architectures in \cref{sec:elecompute}.


In \cref{sec:sec4}, we benchmark the performance of our implementation using a representative FE discretized matrix on multi-node CPU architectures (NSM\footnote{National Supercomputing Mission, India} Param Pravega) and multi-node GPU architectures (ORNL\footnote{Oak Ridge National Laboratory, USA} Summit supercomputer). Specifically, as a model problem, we compute the action of the Helmholtz problem on multivectors of various sizes. We use a cell-matrix implementation and the existing matrix-free implementation from the \texttt{deal.II} library as our baselines. We begin our comparative study with the \texttt{deal.II} matrix-free approach for a single vector. Our GPU implementation outperforms the \texttt{deal.II} matrix-free baseline on a single GPU ($\sim$120k DoFs/GPU) with a speedup of 16x -- 17.5x for the single vector case with polynomial orders 6, 7 and 8. Hence, we do not consider the \texttt{deal.II} method as a baseline for the evaluation of matrix-multivector products on GPU architectures. We subsequently benchmark our matrix-free multivector implementation against the chosen baselines. Our results indicate the superior performance of our proposed implementation, demonstrating computational gains of 2x -- 2.8x on one Summit node (6 GPUs, $\sim$200k DoFs/GPU), 16\% -- 30\% on 16 Summit nodes (96 GPUs, $\sim$12k DoFs/GPU), and 2.4x -- 4.4x on 64 nodes of Param Pravega (3072 CPU \texttt{MPI} tasks, $\sim$700 DoFs per \texttt{MPI} task) for matrix-multivector products (1024 vectors) compared to the best baseline implementation for polynomial orders 6, 7 and 8. Additionally, we present the strong scaling studies of our proposed implementation on both multi-node CPU and GPU architectures. 

We further benchmark our implementation strategy by solving the eigenvalue problem involving the differential operator $-\mu\nabla^2+\kappa(\bx)$, we show speedups of  1.6x -- 2.2x on a uniform mesh for 1 Summit node (6 GPUs, $\sim$200k DoFs/GPU), 14\% -- 41\% for 4 Summit nodes (24 GPUs, $\sim$50k DoFs/GPU), and 2x -- 3x on 64 nodes of Param Pravega (3072 CPU \texttt{MPI} tasks, $\sim$700 DoFs per \texttt{MPI} task) for matrix-multivector products (1024 vectors) compared to the best baseline implementation for polynomial order 6, 7 and 8. In addition, we report benchmarks on adaptively refined meshes for our matrix-free implementation against the baselines. Finally in \cref{sec:sec5}, we present brief concluding remarks with a future outlook.

\section{Methodology}\label{sec:sec2}
\subsection{Mathematical background}\label{sec:math}
Consider a partial differential equation (PDE) defined on a bounded domain $\Omega\subset\mathbb{R}^3$ involving the differential operator $\mathcal{F}=-\mu\nabla^2+\kappa(\bx)$ with $\mu\in \mathbb{R}$ and $\kappa(\bx):\Omega \rightarrow \mathbb{R}$. Note that the operator $\mathcal{F}$ is reduced to the Laplace operator if $\mu=1,\;\kappa(\bx)=0 \;\; \forall \bx \in \Omega$ and to the Helmholtz operator if $\mu=1,\; \kappa(\bx)=k^2 \;\;\forall\bx\in\Omega$, where $k\in\mathbb{R}$ is a constant.

To elucidate our matrix-free multivector algorithmic strategies developed in the current work, we introduce the following problem of finding $u^{\beta}(\bx) \in \mathcal{V}$ with $\beta=1,2,\dots,n_v$ such that 
\begin{align}
    \mathcal{F}u^{\beta}(\bx)&=-\mu\nabla^2 u^{\beta}(\bx)+\kappa(\bx)u^{\beta}(\bx)=\begin{cases}f^{\beta}(\bx)\\\lambda^{\beta} u^{\beta}(\bx)
    \end{cases}  \forall \bx \in \Omega\nonumber
    \\ u^{\beta}(\bx)&=u_D(\bx) \;\;\;\; \forall \bx \in \partial \Omega_D \label{eqn:refprob}
\end{align}
where $\mathcal{V}$ denotes a suitable function space in which the solution of the problem in \cref{eqn:refprob} lies, and $u_D(\bx)$ in the above equation corresponds to the Dirichlet boundary condition applied on $\partial \Omega_D \subseteq \partial \Omega$ and the boundary of $\Omega$. If the choice of the RHS is a set of forcing functions $f^{\beta}(\bx):\Omega \rightarrow \mathbb{R}$, for $\beta=1,2,\dots,n_v$, the above problem represents a set of linear PDEs. If the choice of the RHS is $\lambda^{\beta}u^{\beta}(\bx)$, then \cref{eqn:refprob} represents an eigenvalue problem with $(\lambda_{\beta},u^{\beta}(\bx))$ as the eigenvalue and eigenfunction pair corresponding to the operator $\mathcal{F}$.  Eigenvalue problems of this nature with large $n_v$ are very similar to those arising in quantum-modeling of materials using Kohn-Sham density functional theory (DFT)~\cite{Das2019FastSystem,Das2022DFT-FEDiscretization}, electroelastics~\cite{Martynova2023TheMaterials} and modal analysis~\cite{Fan2014ParallelAnalysis,Fan2015SomeFramework}

We now consider the discretization of the eigenvalue problem in \cref{eqn:refprob} using finite-elements, a strictly local piecewise polynomial basis comprising of $C^{0}$ continuous Lagrange polynomials generated using Gauss Lobatto Legendre (GLL) nodal points~\cite{Brenner2008TheMethods}. To this end, we consider the finite-dimensional space $\mathbb{V}^{h}_m \subset \mathcal{V}$ with a 3D tensor-structured finite-element (FE) basis $N^{h}_J(\bx): 1 \leq J \leq m$ constructed from strictly local 1D Lagrange interpolating polynomials of order $n_p-1$, generated using the nodes of the FE triangulation $\mathcal{T}^{h}$, with the characteristic mesh size denoted by $h$. Consequently, the discretization of the solution fields in \cref{eqn:refprob} using the FE basis is given by $u^{\beta,h}(\bx)=\sum_{J=1}^{m} u_J^{\beta}\, N^{h}_J(\bx)$, where $u_{J}^{\beta}$ denotes the coefficient of the $\beta^{th}$ discretized field for $\beta=1,2,\dots,n_v$.  

Finally, the finite-element discretization of the eigenvalue problem in \cref{eqn:refprob} results in the following: 
\begin{equation}
    \begin{alignedat}{3}
        \bK\bu^{\beta}&+\bM^\kappa \bu^{\beta}=\lambda_{\beta}\bM\bu^{\beta}\\
        u^{\beta}_{J}&=\Pi u_D(\bx_J) & &\forall \bx_J \in \partial \Omega_D
    \end{alignedat}\label{eqn:matrix}
\end{equation}
to be solved for the eigenvalues $\lambda^\beta\in \mathbb{R}$ and eigenvectors $\bu^{\beta}\in \mathbb{R}^m\;\; \forall \beta=1,\dots,n_v$ comprising of the FE nodal degrees of freedom (DoFs), where $\Pi u_D(\bx_J)$ is the interpolant of $u_D(\bx)$ in $\mathbb{V}^{h}_m$ and $\bK$, $\bM$ and $\bM^\kappa$ denote the stiffness matrix, mass matrix (FE basis overlap matrix) and the weighted mass matrix respectively, and are given by: 
\begin{subequations}\label{eqn:integrals}
    \begin{align}
  K_{IJ}&=\int_{\Omega}\mu\del N^{h}_I(\bx) \cdot \del N^{h}_J(\bx) \dx \\
  M_{IJ}&=\int_{\Omega}  N^{h}_I(\bx) \, N^{h}_J(\bx) \dx \\
  M^{\kappa}_{IJ}&=\int_{\Omega} \kappa(\bx)\, N^{h}_I(\bx)\, N^{h}_J(\bx) \dx
\end{align}
\end{subequations}
Defining the multivector matrix  $\bU=\left[\bu^1\;\bu^2\;\dots\;\bu^{n_v}\right]$, we can now rewrite \cref{eqn:matrix} as

\begin{equation}
    \begin{alignedat}{3}
        \bK\bU&+\bM^\kappa \bU=\bM\bU\mathbf{\Lambda}\\
        U_{J\beta}&=\Pi u_D(\bx_J) \cn & &\forall \bx_J \in \partial \Omega_D
    \end{alignedat}\label{eqn:matrixmultivector}
\end{equation}
where $\Lambda_{\alpha \beta}=\delta_{\alpha \beta}\lambda_\alpha$.

The computational efficiency of an iterative solution strategy for solving the eigenvalue problem in \cref{eqn:matrixmultivector} relies on the efficient evaluation of matrix multivector products $\bK\bU$, $\bM^\kappa \bU$ and $\bM\bU$ on distributed heterogeneous architectures, which will be the primary focus of this work.

\subsection{Matrix multivector product}\label{sec:theory}
According to the standard prescription of finite-element (FE) discretization, we decompose
 $\Omega$ into non-overlapping volumes called finite-element cells $\Omega^{(e)}$ i.e., $\Omega=\bigcup_{e=1}^E\Omega^{(e)}$ where $E$ is the number of finite-element cells. We refer to these elements $\Omega^{(e)}$ as FE-cells, and in this work, we choose them to be hexahedra. Furthermore, we assume that a linear map exists from each FE-cell to a reference domain $\widehat{\Omega}=\left[-1,1\right]^3$ with $\bxi=[\xi_1,\xi_2,\xi_3]$ as the reference coordinate system. In this framework, the FE discretized field $u^{\beta,h}(\bx)$ for a given FE-cell ($e$) can be defined as follows:
\begin{equation}
   u^{\beta,h\,(e)}(\bx(\bxi))=\sum_{J=1}^{n_p^3}u^{\beta,h\,(e)}\left(\bx^{(e)}_{J}\right)\widehat{N}_J\left(\bxi\right) 
   \label{eqn:ufield}
\end{equation}
where $\widehat{N}_{J}$ is the 3D finite-element (FE) cell level basis function of polynomial order $(n_p-1)^3$ corresponding to the FE node $J$.

To make the problem more amenable to distributed parallelism, we partition the domain $\Omega$ into subdomains $\Omega^{(t)} \; \forall t=1,2,\dots,n_t$, where $n_t$ is the number of subdomains, and assign each subdomain $\Omega^{(t)}$ to an MPI task $t$. Let $E_t$ be the number of FE-cells, and $m_t$ be the number of basis functions in each subdomain $\Omega^{(t)}$ such that $\Omega^{(t)}=\bigcup_{e=1}^{E_t}\Omega^{(e)}$. Consequently, the matrix-multivector product $\bA \bU$, where $\bA$ denotes the FE discretized matrix (such as $\bK$, $\bM^\kappa$, $\bM$ or $\bK + \bM^{\kappa}$), can be written as follows:
\begin{equation}
\bV=\bA\bU=\left[\sum_{t}^{n_t}{\bP^{(t)}}^T{\bC^{(t)}}^T\left(\sum_e^{E_t}{\bQ^{(e,t)}}^T\bA^{\left(e\right)}\bQ^{(e,t)}\right)\bC^{(t)}\bP^{(t)}\right]\bU\label{eqn:AU}        
\end{equation}
where the multi-index $(e,t)$ denotes the FE-cell index $(e)$ associated with an MPI task $(t)$ and the Boolean sparse matrix $\bP^{(t)}$ denotes the partitioner matrix whose action on $\bU$ gives the subdomain level multivector $\bU^{(t)}$. The matrix $\bP^{(t)}$ imposes the continuity of the field $u^{\beta,h}(\bx)$ across the partitioned subdomains. Further, the Boolean sparse matrix $\bC^{(t)}$ in \cref{eqn:AU} denotes an $m_t \times m_t$ constraint matrix employed to constrain the values of the $m_t \times n_v$ matrix $\bU^{(t)}$ at certain nodal points. These constraints are used either to satisfy the necessary boundary conditions imposed on $u^{\beta,h}(\bx)$ or to deal with constraints arising from non-conforming meshes~\cite{Bangerth2009DataSoftware}. Furthermore, the imposition of the continuity condition associated with $u^{\beta,h}(\bx)$ across FE-cells within a partitioned subdomain $\Omega^{(t)}$ is accomplished by the action of $n_p^3\times m_t$ Boolean sparse matrix $\bQ^{(e,t)}$ on the constrained subdomain level multivector $\bC^{(t)}\bP^{(t)}\bU$, with $\bQ^{(e,t)}$ representing the subdomain level to FE-cell level map on the subdomain $\Omega^{(t)}$. Finally, the FE-cell level matrix $\bA^{(e)}$ arising from the finite-element discretization of the underlying PDE can be evaluated as an integral over the reference domain $\widehat\Omega$. For example, the $n_p^3 \times n_p^3$ FE-cell level matrix $\bK^{\left(e\right)}$ associated with the matrix $\bK$ in \cref{eqn:matrixmultivector} can be evaluated as  
\begin{subequations}
\begin{align}
    K^{\left(e\right)}_{IJ}&=\int_{\Omega^{\left(e\right)}}\mu \nabla  N_I\cdot\nabla  N_J d\bx \\
    &=\int_{\widehat\Omega}\mu \left({\bJ^{\left(e\right)}}^{-T}\nabla_{\bxi} \widehat N_I\right)\cdot\left({\bJ^{\left(e\right)}}^{-T}\nabla_{\bxi} \widehat N_J\right) \det{\bJ^{\left(e\right)}} d\widehat\bx \label{eqn:stiffnessmatinteg} \\
    &=\sum_{Q=1}^{n_q^3}\left(\nabla_{\bxi} \widehat N_I\right)^T{\bJ^{\left(e\right)}}^{-1}{\bJ^{\left(e\right)}}^{-T}\left(\nabla_{\bxi} \widehat N_J\right)\mu w_Q\det{\bJ^{\left(e\right)}}\Biggl|_{\widehat\bxi_Q}\label{eqn:stiffnessmatquad}
\end{align}
\end{subequations}
where $\nabla_{\bxi} \widehat N_I$ denotes the gradient of the FE-cell level basis function within reference coordinate system $\bxi$, while $\bJ^{\left(e\right)}$ denotes the Jacobian matrix of the map from $\Omega^{\left(e\right)}$ to $\widehat\Omega$. Furthermore, a tensor structured $n_q$-point quadrature rule with quadrature points $\widehat\bxi_Q$ and the quadrature weights $w_Q$ is used in \cref{eqn:stiffnessmatquad} for evaluating the integral involved in \cref{eqn:stiffnessmatinteg}. 

Defining $D^{(s)}_{QI}=\nabla_{\bxi}\widehat N_I\left(\widehat \bxi_Q\right)\cdot \widehat\bn_s$ as $n_q^3\times n_p^3$ matrices where $\widehat\bn_s, \; s=0,1,2$ represents the unit vector along the $s$ axis, we can now rewrite \cref{eqn:stiffnessmatquad} as
\begin{align}
    \bK^{\left(e\right)}=\begin{bmatrix}
        \bD^{\left(0\right)}\\
        \bD^{\left(1\right)}\\
        \bD^{\left(2\right)}
    \end{bmatrix}^T
    \begin{bmatrix}
        \bG^{\left(0,0\right)}&&\bG^{\left(0,1\right)} &&\bG^{\left(0,2\right)}\\
        \bG^{\left(1,0\right)}&&\bG^{\left(1,1\right)} &&\bG^{\left(1,2\right)}\\
        \bG^{\left(2,0\right)}&&\bG^{\left(2,1\right)} &&\bG^{\left(2,2\right)}\\
    \end{bmatrix}
    \begin{bmatrix}
        \bD^{\left(0\right)}\\
        \bD^{\left(1\right)}\\
        \bD^{\left(2\right)}
    \end{bmatrix}\label{eqn:stiffness3d}
\end{align}
where $\bG^{\left(s,d\right)}$ for $s,d = 0,1,2$ are $n_q^3\times n_q^3$ diagonal matrices with the diagonal entry $\mathcal{G}^{\left(s,d\right)}_{QQ}=\left[\left(\bJ^{\left(e\right)}\right)^{-1}\left(\bJ^{\left(e\right)}\right)^{-T}\right]_{sd}\det{\bJ^{\left(e\right)}}\mu w_Q\biggl|_{\widehat\bxi_Q}$. We can rewrite the weighted mass matrix in the same framework as 
\begin{align}
    \bM^{\kappa,\left(e\right)}=\bN^T\bG\bN \label{eqn:mass3d}
\end{align}
where $N_{QI}=\widehat N_I(\widehat\bxi_Q)$ is an $n_q^3\; \times \;n_p^3$ matrix and $\mathcal{G}_{QQ}=\kappa\det{\bJ^{\left(e\right)}}w_Q\biggl|_{\widehat\bxi_Q}$ is an $n_q^3\;\times\;n_q^3$ matrix. We obtain the unweighted mass matrix $\bM^{\left(e\right)}$ by setting $\kappa=1$.

A straightforward approach to evaluate the matrix-multivector product $\bV=\bA\bU$ as outlined in \cref{eqn:AU} is to construct the \emph{global FE discretized matrix} $\bA$ and perform the sparse-matrix dense-matrix product in a distributed setting. As demonstrated by \citet{Kronbichler2012AApplication}, this method is computationally less efficient than the alternative methods discussed herein. In the spirit of the strategies employed for FE discretized matrix-single vector multiplication, we now discuss two computationally efficient methods for evaluating the matrix-multivector products $\bV=\bA\bU$. 


\subsubsection{Evaluation -via- FE-cell level local dense matrices}\label{sec:cellmatrix}
The matrix multivector product $\bV = \bA \bU$ can be evaluated using the FE-cell level matrices $\bA^{(e)}$ and the FE-cell level multivectors~\cite{Carey1988Element-by-elementComputations,Das2022DFT-FEDiscretization}. This strategy comprises the following steps~:
\begin{itemize}
    \item [1.] \underline{Precompute} the FE-cell level operator matrices $\bA^{(e)}$\label{stp:setupele} 
    \item [2.] \underline{Extraction} of the FE-cell level multivectors $\bU^{(e,t)}$ using the \emph{subdomain level to FE-cell level map}, the \emph{constraint} and the \emph{partitioner} matrices, i.e., $\bU^{(e,t)}=\bQ^{(e,t)}\bC^{(t)}\bP^{(t)}\bU \\ \forall e=1,2,\dots,E_t$
    \item [3.] \underline{FE-cell level evaluation} of the matrix multivector product $\bV^{\left(e,t\right)}=\bA^{\left(e\right)}\bU^{\left(e,t\right)}$ using batched matrix-matrix multiplication.\label{stp:elecompute}
    \item [4.] \underline{Assembly} of the global multivector $\bV$ using the \emph{subdomain level to FE-cell level map}, the \emph{constraint} and \emph{partitioner} matrices, i.e., $\bV = \sum_{t}^{n_t}{\bP^{(t)}}^T{\bC^{(t)}}^T\sum_e^{E_t}{\bQ^{(e,t)}}^T\bV^{(e,t)}$
\end{itemize}

In the above framework, the FE-cell level evaluation (Step 3) is the computationally dominant step with the computational complexity of $O(n_p^6n_v)$. Furthermore, this method requires us to store the FE-cell level matrices and multivectors, resulting in a memory footprint of $O(n_p^6+n_p^3n_v)$.




\subsubsection{Evaluation -via- matrix-free approach}
Here, we propose a matrix-free approach to compute matrix-multivector products, inspired by the existing matrix-free matrix-vector multiplication strategies~\cite{Kronbichler2012AApplication}. In this approach, we avoid the precomputation of the FE-cell level matrices $\bA^{(e)}$ and instead, the FE-cell level matrix multivector products $\bA^{(e)}\bU^{(e,t)}$ are evaluated \emph{on-the-fly}. Using the expressions in \cref{eqn:stiffness3d,eqn:mass3d}, we observe that the first step in evaluating $\bV^{\left(e,t\right)} =\bA^{(e)}\bU^{(e,t)}=\left(\bK^{\left(e\right)}+\bM^{\kappa,\left(e\right)}\right)\bU^{(e,t)}$ involves computing the action of $\bD^{(k)}$ and $\bN$ on $\bU^{\left(e,t\right)}$. To accomplish this, we exploit the tensor-structured nature of the FE basis functions and the quadrature rules. Recalling $\widehat{N}_J(\bxi)$ and $w_Q$ denote the 3D FE-cell level basis functions and the 3D quadrature weights introduced in \cref{eqn:ufield,eqn:stiffnessmatquad} respectively, we have 
\begin{align}
\widehat{N}_J(\bxi) &= \widehat{N}^{1D}_{j_1}(\xi_1)\widehat N^{1D}_{j_2}( \xi_2)\widehat{N}^{1D}_{j_3}(\xi_3) \label{eq:NTensor}\\
w_Q &= w^{1D}_{q_1}w^{1D}_{q_2}w^{1D}_{q_3} \label{eq:quadTensor}
\end{align}
In the above \cref{eq:NTensor}, we express $\widehat{N}_J(\bxi)$ as the product of three 1D Lagrange interpolating polynomials of order \texttt{FEOrder}$=n_p-1$, defined on the Gauss Legendre Lobatto nodal points in $[-1,1]$, with $n_p$ denoting the number of nodal points in each direction. Further, \cref{eq:quadTensor} expresses 3D quadrature weights as the product of 1D quadrature weights $w^{1D}_q$ with $q=1,\dots,n_q$ denoting the quadrature weights of the 1D quadrature rule.


Now, we treat the FE-cell level multivector $\bU^{(e,t)}$ as a 4$^{th}$ order tensor $\bUt^{\left(e,t\right)}$ with its components denoted as  $\mathcal{U}^{\left(e,t\right)}_{\beta,j_1,j_2,j_3}=u^{\beta,h\,(e)}\left(\bx^{(e)}_{J}\right)=u^{\beta,h,\left(e\right)}\left(\bx^{(e)}_{j_1,j_2,j_3}\right)$, with one dimension of $\bUt^{\left(e,t\right)}$ corresponding to the vector index ($\beta$) and the other three corresponding to the spatial indices ($j_1,j_2,j_3$) of the node $J$. To this end, the action of $\bD^{(k)}$  and $\bN$ on $\bU^{\left(e,t\right)}$ is represented as 
\begin{align}
\bN\bU^{\left(e,t\right)}\equiv\left(\bN^{1D}\otimes\bN^{1D}\otimes\bN^{1D}\otimes\bI\right)\bUt^{\left(e,t\right)}\label{eqn:NU}\\
    \begin{bmatrix}
        \bD^{\left(0\right)}\\
        \bD^{\left(1\right)}\\
        \bD^{\left(2\right)}
    \end{bmatrix}\bU^{\left(e,t\right)}\equiv    \begin{bmatrix}
        \bN^{1D}\otimes\bN^{1D}\otimes\bD^{1D}\otimes\bI\\
        \bN^{1D}\otimes\bD^{1D}\otimes\bN^{1D}\otimes\bI\\
        \bD^{1D}\otimes\bN^{1D}\otimes\bN^{1D}\otimes\bI
    \end{bmatrix}\bUt^{\left(e,t\right)} \label{eqn:DU_def}
\end{align}
where $\bN^{1D}$ and $\bD^{1D}$ are $n_q\times n_p$ matrices corresponding to the one-dimensional FE basis function values and gradients, respectively, at quadrature points and $\otimes$ represents the Kronecker product. Using the well-known result of tensor algebra, $\left(\bA\otimes\bB\right)=\left(\bI\otimes\bB\right)\left(\bA\otimes\bI\right)$ we can reduce the above expressions into a series of tensor contractions as enunciated in \cref{alg:NU} below.
\begin{figure}[H]
    \removelatexerror
\begin{algorithm}[H]
    \caption{Evaluation of $\bT=\bN\bU^{\left(e,t\right)}$}\label{alg:NU}
    \KwIn{$\bU^{\left(e,t\right)}\left(\equiv \bUt^{\left(e,t\right)}\right)$}
    \KwData{$\bN^{1D}$}
    \KwResult{$\bT\left(\equiv\bTt\right)$}
    $\bTt \gets \left(\bI\otimes\bI\otimes\bN^{1D}\otimes\bI\right)\bUt^{\left(e,t\right)}$\;
    $\bTt \gets \left(\bI\otimes\bN^{1D}\otimes\bI\otimes\bI\right)\bTt$\;
    $\bTt \gets \left(\bN^{1D}\otimes\bI\otimes\bI\otimes\bI\right)\bTt$\;
    \Return{$\bTt$}
\end{algorithm}
\end{figure}

Similar in spirit to~\citet{Deville2002High-OrderFlow,Fischer2020ScalabilitySolvers}, we now evaluate $\bK^{\left(e\right)}\bU^{\left(e\right)}$ by expressing $\widehat N^{1D}_i(\xi)$ as $\widehat N^{1D}_i(\xi)=\sum_{\wh{q}}\widehat N^{1D}_i(\xi_{\wh q})\widetilde N^{1D}_{\wh q}(\xi)$ where $\widetilde N^{1D}_{\wh q}$ is the Lagrange polynomial defined at the quadrature point $\xi_{\wh q}$. This allows us to write $\frac{d\widehat N^{1D}_i(\xi)}{d\xi}=\displaystyle\sum_{\wh q}\textstyle{ \frac{d\widetilde N^{1D}_{\wh q}(\xi)}{d\xi} \widehat{N}^{1D}_i(\xi_{\wh{q}})}$. Consequently, we can now factorize $\bD^{1D}$ as $\bD^{1D}=\widetilde\bD^{1D}\bN^{1D}$ where $\widetilde D^{1D}_{\wh q q}=\frac{d\widetilde N^{1D}_q\left(\xi\right)}{d\xi}\biggr|_{ \xi_{\wh q}}$. \Cref{eqn:DU_def} can now be rewritten as
\begin{align}
    \begin{bmatrix}
        \bD^{\left(0\right)}\\
        \bD^{\left(1\right)}\\
        \bD^{\left(2\right)}
    \end{bmatrix}\bU^{\left(e,t\right)}=    \begin{bmatrix}
        \bI\otimes\bI\otimes\widetilde\bD^{1D}\otimes\bI\\
        \bI\otimes\widetilde\bD^{1D}\otimes\bI\otimes\bI\\
        \widetilde\bD^{1D}\otimes\bI\otimes\bI\otimes\bI
    \end{bmatrix}\bN\bU^{\left(e,t\right)}=\begin{bmatrix}
        \widetilde\bD^{\left(0\right)}\\
        \widetilde\bD^{\left(1\right)}\\
        \widetilde\bD^{\left(2\right)}
    \end{bmatrix}\bN\bU^{\left(e,t\right)} \label{eqn:DU}
\end{align}

Using this factorization in \cref{eqn:DU}, $\bK^{\left(e\right)}\bU^{\left(e,t\right)}$ can be evaluated with a computational complexity of $O((4(n_p^3n_q+n_p^2n_q^2+n_pn_q^3)+12n_q^4+3n_q^3)n_v)$. Note that this approach reduces the floating point operations required when $n_q=n_p$ by $\sim$30\% compared to tensor contractions in \cref{eqn:NU,eqn:DU_def}. Even in the case of $n_q>n_p$, this factorization is beneficial for evaluating the action of $\bA^{\left(e\right)}=\bK^{\left(e\right)}+\bM^{\kappa,\left(e\right)}$ as it allows us to reduce the number of required tensor contractions by factorizing out $\bN$ and $\bN^T$ as follows:
\begin{align}
    &\bA^{\left(e\right)}\bU^{\left(e,t\right)}=\left(\bK^{\left(e\right)}+\bM^{\kappa,\left(e\right)}\right)\bU^{\left(e,t\right)}\nonumber\\&=\bN^T\left(\begin{bmatrix}
        \widetilde\bD^{\left(0\right)}\\
        \widetilde\bD^{\left(1\right)}\\
        \widetilde\bD^{\left(2\right)}
    \end{bmatrix}^T
    \begin{bmatrix}
        \bG^{\left(0,0\right)}&&\bG^{\left(0,1\right)} &&\bG^{\left(0,2\right)}\\
        \bG^{\left(1,0\right)}&&\bG^{\left(1,1\right)} &&\bG^{\left(1,2\right)}\\
        \bG^{\left(2,0\right)}&&\bG^{\left(2,1\right)} &&\bG^{\left(2,2\right)}\\
    \end{bmatrix}
    \begin{bmatrix}
        \widetilde\bD^{\left(0\right)}\\
        \widetilde\bD^{\left(1\right)}\\
        \widetilde\bD^{\left(2\right)}
    \end{bmatrix}+\bG\right)\bN\bU^{\left(e,t\right)}\label{eqn:helmholtz}
\end{align}
Using \cref{eqn:helmholtz}, we describe the algorithm for the evaluation of $\bV^{\left(e,t\right)}$ in the case of $\bA^{\left(e\right)}=\bK^{\left(e\right)}+\bM^{\kappa,\left(e\right)}$ in \Cref{alg:ve}.
\begin{figure}[H]
    \removelatexerror
\begin{algorithm}[H]
    \caption{Evaluation of $\bV^{\left(e,t\right)}=\left(\bK^{\left(e\right)}+\bM^{\kappa,\left(e\right)}\right)\bU^{\left(e,t\right)}$}\label{alg:ve}
    \KwIn{$\bU^{\left(e,t\right)}\left(\equiv\bUt^{\left(e,t\right)}\right)$}
    \KwData{$\bN^{1D},\widetilde\bD^{1D},\bG,\bG^{\left(s,d\right)}$ where $s,d=0,1,2$}
    \KwTemp{$\bTt, \bTt^{\left(0\right)},\bTt^{\left(1\right)},\bTt^{\left(2\right)}$}
    \KwResult{$\bV^{\left(e,t\right)}$}
    $\bTt \gets \bN\bU^{\left(e,t\right)}$\tcp*{\Cref{alg:NU}}
    $\bTt^{\left(0\right)} \gets \left(\bI\otimes\bI\otimes\widetilde\bD^{1D}\otimes\bI\right)\bTt$\;
    $\bTt^{\left(1\right)} \gets \left(\bI\otimes\widetilde\bD^{1D}\otimes\bI\otimes\bI\right)\bTt$\;
    $\bTt^{\left(2\right)} \gets \left(\widetilde\bD^{1D}\otimes\bI\otimes\bI\otimes\bI\right)\bTt$\;
    $\bTt^{\left(s\right)} \gets \displaystyle\sum_{d=0}^{2} \bG^{\left(s,d\right)}\bTt^{\left(d\right)}$\;
    $\bTt \gets \bG\bTt$\;
    $\bTt \gets \bTt+\left(\bI\otimes\bI\otimes\widetilde\bD^{1D}\otimes\bI\right)^T\bTt^{\left(0\right)}$\;
    $\bTt \gets \bTt+\left(\bI\otimes\widetilde\bD^{1D}\otimes\bI\otimes\bI\right)^T\bTt^{\left(1\right)}$\;
    $\bTt \gets \bTt+\left(\widetilde\bD^{1D}\otimes\bI\otimes\bI\otimes\bI\right)^T\bTt^{\left(2\right)}$\;
    $\bV^{\left(e,t\right)} \gets \bN^T\bT$\tcp*{\Cref{alg:NU}}
    \Return{$\bV^{\left(e,t\right)}$}
\end{algorithm}
\end{figure}

\section{Hardware-aware implementation of the Matrix-free algorithm}\label{sec:sec3}
This section describes the proposed hardware-aware implementation procedures on multi-node CPU and GPU architectures for evaluating FE discretized matrix-multivector products using the matrix-free algorithm discussed in the previous section. The key steps involve: (i) the extraction step in which the FE-cell level multivectors $\bU^{\left(e,t\right)}$ of size $n_v \times n_p^3$  are constructed from the global multivector $\bU$ of size $n_v \times m$ using the \emph{subdomain level to FE-cell level map} and the \emph{partitioner}, (ii) the FE-cell level evaluation in the matrix-free framework involving tensor contractions (\cref{eqn:helmholtz}) and a point-wise multiplication to represent the action of $\bG$ and $\bG_{ij}$, (iii) 
the assembly of the output FE-cell level matrices $\bV^{\left(e,t\right)}$
to construct output node-level multivector $\bV$ employing the same map and partitioner used in the extraction phase. This procedure is described in more detail in the following subsections. 

\subsection{Mathematical formulation of batched algorithm}
The proposed algorithm involves processing several batches of a small number of vectors tailored to specific hardware architecture. This approach enables better data locality owing to the smaller size of each batch and permits further parallelization over multiple batches. We denote the number of vectors in each batch as `$b$' and the number of batches as `$n_b$'. We now present a mathematical description of a batched algorithm for evaluating matrix-multivector products within the matrix-free paradigm on both CPU and GPU architectures.



\subsubsection{CPU Batched Strategy}\label{sec:cpubatchstrat}
We propose a strategy for batched evaluation of matrix-multivector products in the case of CPUs. To this end, we introduce a Boolean sparse matrix $\bB^{\left(i_b\right)}$ whose action on the multivector results in the extraction of the multivector batch $\bU^{\left(i_b\right)}=\bB^{\left(i_b\right)}\bU$, where $\bU^{\left(i_b\right)}$ is the multivector batch indexed by $i_b$ (i.e. the batch comprising of vectors indexed from $i_b\times b$ to $(i_b+1)\times b$). Using this matrix, we recast \cref{eqn:AU} as
\begin{multline}
    {\bV}=\sum_{i_b}^{n_b}{\bB^{\left(i_b\right)}}^T\Biggl(\sum_{t}^{n_t}{\bP^{\left(i_b,t\right)}}^T{\bC^{\left(i_b,t\right)}}^T\Biggl(\sum_{e}^{E_t}{\bQ^{\left(i_b,e,t\right)}}^T\bA^{\left(e\right)}\\    \bQ^{\left(i_b,e,t\right)}\Biggr)\bC^{\left(i_b,t\right)}\bP^{\left(i_b,t\right)}\Biggr)\bB^{\left(i_b\right)}{\bU}\label{eqn:basempicpu}
\end{multline}

Here, $\bP^{\left(i_b,t\right)}$ represents the partitioner matrix that acts on $\bU^{\left(i_b\right)}$ and extracts its portion belonging to the subdomain $\Omega^t$ (on \texttt{MPI}~task~$t$), i.e., $\bU^{\left(i_b,t\right)}=\bP^{\left(i_b,t\right)}\bB^{\left(i_b\right)}{\bU}$.  Constraint matrix $\bC^{\left(i_b,t\right)}$ then acts on $\bU^{\left(i_b,t\right)}$ to ensure that all constraints are satisfied. This results in the constrained subdomain level multivector, $\bC^{\left(i_b,t\right)}{\bU}^{\left(i_b,t\right)}$. The \emph{subdomain level to FE-cell level map}, $\bQ^{\left(i_b,e,t\right)}$, then acts on $\bC^{\left(i_b,t\right)}\bU^{\left(i_b,t\right)}$ resulting in the FE-cell level multivector batch $\bU^{\left(i_b,e,t\right)}=\bQ^{\left(i_b,e,t\right)}\bC^{\left(i_b,t\right)}\bU^{\left(i_b,t\right)}$. We then evaluate the FE-cell level matrix multivector product $\bV^{\left(i_b,e,t\right)}=\bA^{\left(e\right)}\bU^{\left(i_b,e,t\right)}$. Subsequently, we map this FE-cell level product to the subdomain level product multivector via  ${\bQ^{\left(i_b,e,t\right)}}^{T}$ and then sum over the contributions from all the FE-cells belonging to $\Omega^t$. The transpose of $\bC^{\left(i_b,t\right)}$ then acts on the result to ensure that the constraints are satisfied, which results in the product multivector corresponding to batch $i_b$ and subdomain $\Omega^t$, i.e., $\bV^{\left(i_b,t\right)}={\bC^{\left(i_b,t\right)}}^T\left(\sum_{e}^{E_t}{\bQ^{\left(i_b,e,t\right)}}^T\bV^{\left(i_b,e,t\right)}\right)$. The global product multivector batch can then be evaluated by summing over the action of the ${\bP^{\left(i_b,t\right)}}^T$ on all the subdomain level product multivectors corresponding to batch $i_b$, i.e., $\bV^{\left(i_b\right)}=\sum_{t}^{n_t}\!{\bP^{\left(i_b,t\right)}}^T\bV^{\left(i_b,t\right)}$. This process is repeated for every batch to compute the global product vector $\bV=\sum_{i_b}^{n_b}\!{\bB^{\left(i_b\right)}}^T\bV^{\left(i_b\right)}$. 
\begin{figure*}[!hb]
    \centering
    \includegraphics[width=0.98\linewidth]{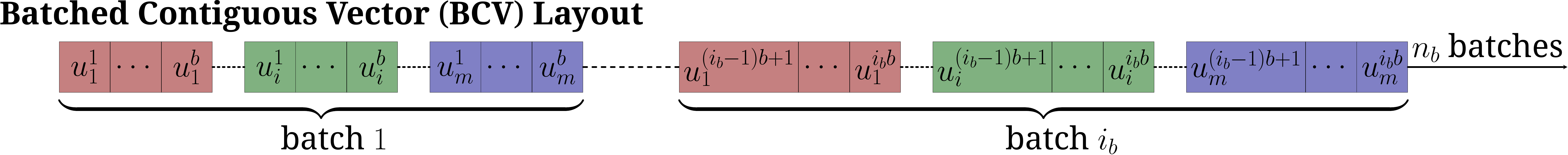}
    \caption{\small Pictorial depiction of the layout described in \cref{sec:layout}. Here $\bu^{\beta}_J=u^{\beta}\left(\bx_J\right)$ with $\beta$ representing the vector index and $J$ representing the spatial index.}\label{fig:layouts}
\end{figure*}

\subsubsection{GPU Batched Strategy}\label{sec:gpubatchstrat}
In contrast to the batched evaluation on CPU architectures discussed above, we recast \cref{eqn:AU} differently to better harness the SIMT nature of GPU architectures by further parallelizing over both FE-cells and batches. We define $\bB^{\left(i_b,t\right)}$ to represent the Boolean sparse matrix for extracting the batch $i_b$ of the subdomain multivector $\bU^{(t)}$ corresponding to $\Omega^t$ and $\bQ^{\left(i_b,e,t\right)}$ represents the \emph{subdomain level to FE-cell level map} for the FE-cell identified by $e$, batch $i_b$ and task $t$. To this end, we interchange the order of the operations involved in \cref{eqn:basempicpu} and consequently rewrite \cref{eqn:AU} as
\begin{multline}
    {\bV}=\sum_{t}^{n_t}{\bP^{\left(t\right)}}^T{\bC^{\left(t\right)}}^T\Biggl(\sum_{i_b}^{n_b}\sum_{e}^{E_t}{\bB^{\left(i_b,t\right)}}^T{\bQ^{\left(i_b,e,t\right)}}^T\bA^{\left(e\right)}\\
   \bQ^{\left(i_b,e,t\right)}\bB^{\left(i_b,t\right)}\Biggr)\bC^{\left(t\right)}\bP^{\left(t\right)}\bU\label{eqn:basempigpu}
\end{multline}
\normalsize
Recall $\bP^{\left(t\right)}$ represents the partitioner acting on $\bU$ to extract  $\bU^{(t)}=\bP^{\left(t\right)}\bU$, belonging to the subdomain $\Omega^t$ (on \texttt{MPI}~task~$t$). The constraint matrix $\bC^{\left(t\right)}$ acts on $\bU^{(t)}$, to ensure that all the constraints are satisfied, resulting in the constrained subdomain level multivector $\bC^{\left(t\right)}\bU^{(t)}$. Subsequently, we define $\bU^{\left(i_b,e,t\right)} = \bQ^{\left(i_b,e,t\right)}\bB^{\left(i_b,t\right)}\bC^{\left(t\right)}\bU^{(t)}$ and now the steps involving the action of $\bA^{(e)}$ on $\bU^{\left(i_b,e,t\right)}$ to compute FE-cell level output $\bV^{\left(i_b,e,t\right)}$ and its mapping to the subdomain level product multivector via the action of ${\bB^{\left(i_b,t\right)}}^T{\bQ^{\left(i_b,e,t\right)}}^T$ on $\bV^{\left(i_b,e,t\right)}$ are accomplished collectively. We do this for every batch `$i_b$' and FE-cell `$e$' and sum over all batches and FE-cells to compute the subdomain level product multivector $\bV^{(t)} = \sum_{i_b}^{n_b}\sum_{e}^{E_t}{\bB^{\left(i_b,t\right)}}^T{\bQ^{\left(i_b,e,t\right)}}^T\bA^{\left(e\right)}\bU^{\left(i_b,e,t\right)}$. We note that the computation of summation terms necessary for evaluation of $\bV^{(t)}$ is done concurrently for every FE-cell and batch through a single GPU kernel launch. The transpose of the constraint matrix ${\bC^{\left(t\right)}}^T$, then acts on $\bV^{(t)}$ to ensure that the constraints are satisfied. Finally, the transpose of the partitioner ${\bP^{(t)}}^T$ acts on this subdomain level product multivector $\bV^{(t)}$ to compute the global product multivector $\bV$. Further elaboration on these operations will be provided in the subsequent discussion.
\subsection{Numerical implementation strategy}
Next, we delve into the computational strategies employed on CPU and GPU architectures used for the implementation of the batched algorithm discussed above. Therefore, we propose a batched layout for storing of the subdomain level multivector.
\subsubsection{Data Layout: Storage of subdomain multivector}\label{sec:layout}
As discussed above, computations can be performed more efficiently for the matrix-free approach if the number of vectors simultaneously dealt with at a given FE node is tailored to hardware architectures, such as the SIMD vectorization width in CPUs or the shared memory size on GPUs. To this end, we propose a batched layout for storing the multivector, which we refer to as the \emph{Batched Contiguous Vector} (BCV) layout. This BCV layout stores the nodal values of a batch of $b$ vectors contiguously for all nodes in $n_b=\lceil n_v/b \rceil$ contiguous batches. We illustrate the layout in \cref{fig:layouts}.

\subsubsection{Applying the constraints: Action of \texorpdfstring{$\bC^{\left(t\right)}$}{C} and \texorpdfstring{${\bC^{\left(t\right)}}^T$}{C transpose}}\label{sec:constraints}
\begin{figure}[H]
    \centering
    \includegraphics[width=0.98\linewidth]{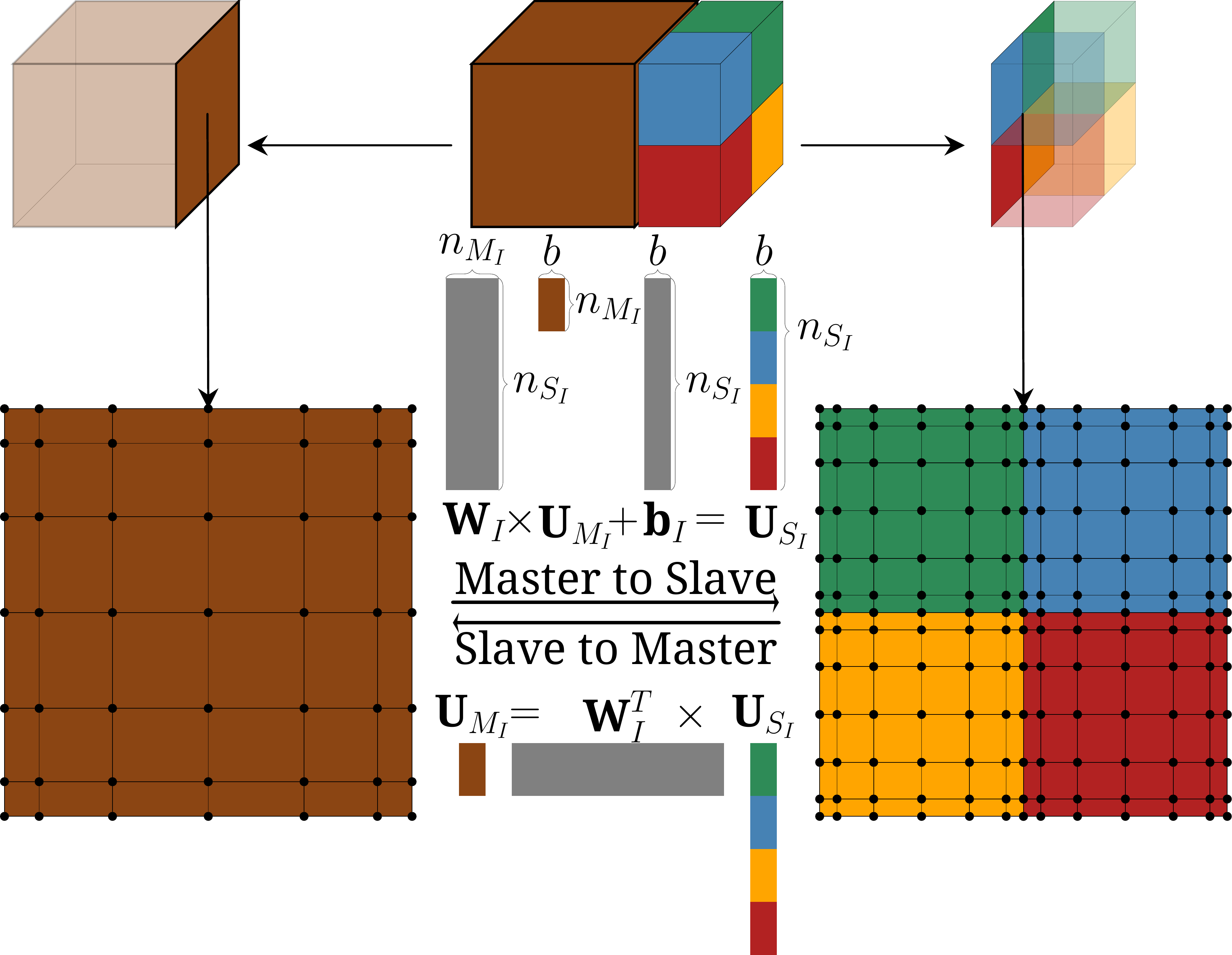}
    \caption{\small  Pictorial depiction of the constraints strategy described in \cref{sec:constraints}. Here $\bU_{S_I}$ and $\bU_{M_I}$ represent the subvectors corresponding to the slave and master nodes respectively}\label{fig:constraints}
\end{figure}

We now discuss the application of constraints, mathematically represented as the application of sparse matrices $\bC^{\left(t\right)}$ and ${\bC^{\left(t\right)}}^T$ ($\bC^{\left(i_b,t\right)}$ and ${\bC^{\left(i_b,t\right)}}^T$ in case of CPU architectures) as discussed in \cref{sec:cpubatchstrat,sec:gpubatchstrat}. We note that the most commonly encountered constraints in non-conforming adaptively refined meshes are the hanging-node constraints \cite{Bangerth2009DataSoftware}, which are locally dense, as they involve interpolation along faces/edges. Consequently, we adopt a local dense matrix approach for applying constraints which allows for utilization of optimized BLAS level 3 routines. We store the constraints as multiple sets, and each set `$I$' comprises four arrays to hold all the constraint information involving the same master nodes. The four arrays include an array containing master node indices, an array containing all the slave node indices, another consisting of the weight matrix ($\bW_I$) for this set of constraints, and finally, an array containing the inhomogeneities corresponding to the slave nodes. A pictorial depiction of the process of application of $\bC^{\left(t\right)}$ on a given batch of multivectors is shown in \cref{fig:constraints}. 
To this end, the nodal values corresponding to the master nodes for the given batch of multivector are extracted into a matrix $\bU_{M_I}$, which is multiplied by the weight matrix $\bW_I$ using BLAS \texttt{gemm} routines, and subsequently, the inhomogeneity vector $\bb_I$ is added to the result. The resulting $\bU_{S_I}$ is copied back to slave nodes of the multivector corresponding to the same batch. Hence, the application of constraints reduces to a sequence of dense matrix-matrix multiplications. The action of ${\bC^{\left(t\right)}}^T$ is evaluated in the similar manner described above. We also apply the Dirichlet boundary conditions using this framework. In this case, the master index matrix $\bU_{M_I}$ and weight matrix $\bW_I$ are empty. 


\subsubsection{Extraction and Assembly: Action of \texorpdfstring{${\bQ^{\left(i_b,e,t\right)}}$}{Q} and \texorpdfstring{${\bQ^{\left(i_b,e,t\right)}}^T$}{Q transpose}}\label{sec:extraction}
The action of the Boolean sparse matrix $\bQ^{\left(i_b,e,t\right)}$ on the subdomain level multivector batch to \emph{extract} $\bU^{\left(i_b,e,t\right)}$ is implemented as a discontiguous read from $\bU^{\left(i_b,t\right)}$ to obtain the data corresponding to the nodes within FE-cell $\Omega^e$. Similarly, we compute the action of ${\bQ^{\left(i_b,e,t\right)}}^T$ and the summation over $e$ in \cref{eqn:basempicpu,eqn:basempigpu} (assembly step) as addition into discontiguous data. The FE-cell level multivector for batch $i_b$ and FE-cell $e$ is represented by
\begin{align}
    &\texttt{U\_e\_ib[}i+bp_1+bn_pp_2+bn_p^2p_3\texttt{]}\equiv U^{\left(i_b,e,t\right)}_{i,p_1,p_2,p_3} \\&\quad i=1,\dots,b \qquad p_1,p_2,p_3=1,\dots,n_p\nonumber
\end{align}
Note that the ordering of the subscript indices represents the data contiguity in memory.
Further optimizations for this step on GPUs are discussed in \cref{sec:elecompute}.

\subsubsection{Tensor Contractions: Evaluation of \texorpdfstring{$\bA^{\left(e\right)}\bU^{\left(i_b,e,t\right)}$}{ FE-cell Level products}}\label{sec:elecompute}
We now illustrate the methodology followed for the evaluation of $\bV^{\left(i_b,e,t\right)}=\bA^{\left(e\right)}\bU^{\left(i_b,e,t\right)}$ for the specific case of $\bA^{\left(e\right)}=\bK^{\left(e\right)}+\bM^{\kappa,\left(e\right)}$ using \cref{alg:ve,alg:NU}.

Note that in both \cref{alg:ve,alg:NU} we need to evaluate products of the forms $\left(\bI\otimes\bI\otimes\bY\otimes\bI\right)\bU^{\left(i_b,e,t\right)}$, $\left(\bI\otimes\bY\otimes\bI\otimes\bI\right)\bU^{\left(i_b,e,t\right)}$ and $\left(\bY\otimes\bI\otimes\bI\otimes\bI\right)\bU^{\left(i_b,e,t\right)}$ using the tensor product vec-trick, $\left(\bY\otimes\bZ\right)\text{vec}\left(\bX\right)=\text{vec}\left(\bY^T\bX\bZ\right)$ where $\text{vec}(\bX)$ denotes the vectorization of the matrix $\bX$ by stacking the columns of $\bX$ into a single column vector, we rewrite these products as batched matrix-matrix multiplications. For instance, let $\bRt$ and $\bTt$ be fourth-order tensors and the dimension of $\bRt$ be $b\times n_p\times n_p\times n_p$. If $\bY$ is an $n_q\times n_p$ matrix, then we have 
\begin{itemize}
    \item $\bTt \gets \left(\bY\otimes\bI\otimes\bI\otimes\bI\right)\bRt$ :
    Treating $\bRt_{\beta,p_1,p_2,p_3}$ and $\bTt_{\beta,p_1,p_2,q_3}$ as matrices $\bR$ of dimensions ($bn_p^2\times n_p$) and $\bT$ of dimensions ($bn_p^2\times n_q$) respectively, we write
    \begin{align}
        \bT\gets\bR{\bY}^T \label{eqn:tc3}
    \end{align}
    \item $\bTt \gets \left(\bI\otimes\bY\otimes\bI\otimes\bI\right)\bRt$ :
    Treating $\bRt_{\beta,p_1,p_2,p_3}$ and $\bTt_{\beta,p_1,q_2,p_3}$ as sets of matrices $\bR_{p_3}$ of dimensions ($bn_p\times n_p$) and $\bT_{p_3}$ of dimensions ($bn_p\times n_q$) respectively, where $p_3=1,2,\dots,n_p$ we write
    \begin{align}
        \bT_{p_3}\gets\bR_{p_3}{\bY}^T &  & \forall p_3=1,\dots,n_p\label{eqn:tc2}
    \end{align}
    \item $\bTt\gets\left(\bI\otimes\bI\otimes\bY\otimes\bI\right)\bRt$ :
    Treating $\bRt_{\beta,p_1,p_2,p_3}$ and $\bTt_{\beta,q_1,p_2,p_3}$ as sets of matrices $\bR_{p_2,p_3}$ of dimensions ($b\times n_p$) and $\bT_{p_2,p_3}$ of dimensions ($b\times n_q$) respectively, where $p_2,p_3=1,2,\dots,n_p$ we write
    \begin{align}
        \bT_{p_2,p_3}\gets\bR_{p_2,p_3}{\bY}^T &  & \forall p_2,p_3=1,\dots,n_p\label{eqn:tc1}
    \end{align}
\end{itemize}

The other major part of \cref{alg:ve} is the evaluation of $\bG\bTt$ and $\sum_d\bG^{\left(s,d\right)}\bTt^{\left(d\right)}$. To evaluate these products we redefine the $\bN^{1D}$ and the $\widetilde\bD^{1D}$ matrices as $N^{1D}_{q,p}\gets N^{1D}_{q,p}\sqrt{w^{1D}_q}$ and $\widetilde D^{1D}_{q_1,q_2}\gets \widetilde D^{1D}_{q_1,q_2} \sqrt{w^{1D}_{q_1}/w^{1D}_{q_2}}$ where $w^{1D}_q$ are the 1D quadrature weights, as discussed in \cref{sec:theory}. This allows us to evaluate $\bG\bTt$ and $\sum_d\bG^{\left(s,d\right)}\bTt^{\left(d\right)}$ in the following manner
\begin{itemize}
    \item $\bTt \gets \bG\bTt$ :
    Considering $\bm{\kappa}^{\left(e\right)}$ to be the vector of length $n_q^3$ defined as $\kappa^{\left(e\right)}_{Q}=\kappa\left(\bx^{\left(e\right)}_{Q}\right) \; \forall Q=1,\dots,n_q^3$ we can evaluate $\bG\bTt$ as $\det{\bJ^{(e)}}\bm{\kappa}^{\left(e\right)}\circ \bTt$ where $\circ$ represents the batched Hadamard product defined as
    \begin{align}
        \bTt_{\beta,Q}\gets \det{\bJ^{(e)}}\kappa^{\left(e\right)}_{Q}\bTt_{\beta,Q} \quad \forall \beta=1,\dots,b\label{eqn:mjacobian}
    \end{align}
    Note that this reduces to matrix scaling in the case of the Helmholtz operator as $\kappa(\bx)$ is a constant.
    \item  $\bTt^{\left(s\right)} \gets \sum_d\bG^{\left(s,d\right)}\bTt^{\left(d\right)} \; \forall s=0,1,2$ :
    Defining a $bn_q^3\times 3$ matrix as $\left[{\bTt^{\left(0\right)}}{\bTt^{\left(1\right)}}{\bTt^{\left(2\right)}}\right]$ we can write this operation as a $bn_q^3\times 3$ times $3\times 3$ matrix-matrix multiplication as 
    \begin{align}
        \left[{\bTt^{\left(0\right)}}{\bTt^{\left(1\right)}}{\bTt^{\left(2\right)}}\right]\left(\left(\bJ^{\left(e\right)}\right)^{-1}\left(\bJ^{\left(e\right)}\right)^{-T}\det{\bJ^{\left(e\right)}}\mu\right)\label{eqn:sjacobian}
    \end{align}
\end{itemize}
\cn

We now discuss the implementation of the above algorithm on CPU and GPU architectures.



\subsubsection*{CPU Implementation: Evaluation of \texorpdfstring{$\bA^{\left(e\right)}\bU^{\left(i_b,e,t\right)}$}{ Cell Level products}}
\begin{figure*}[!ht]
    \centering
    \includegraphics[width=\linewidth]{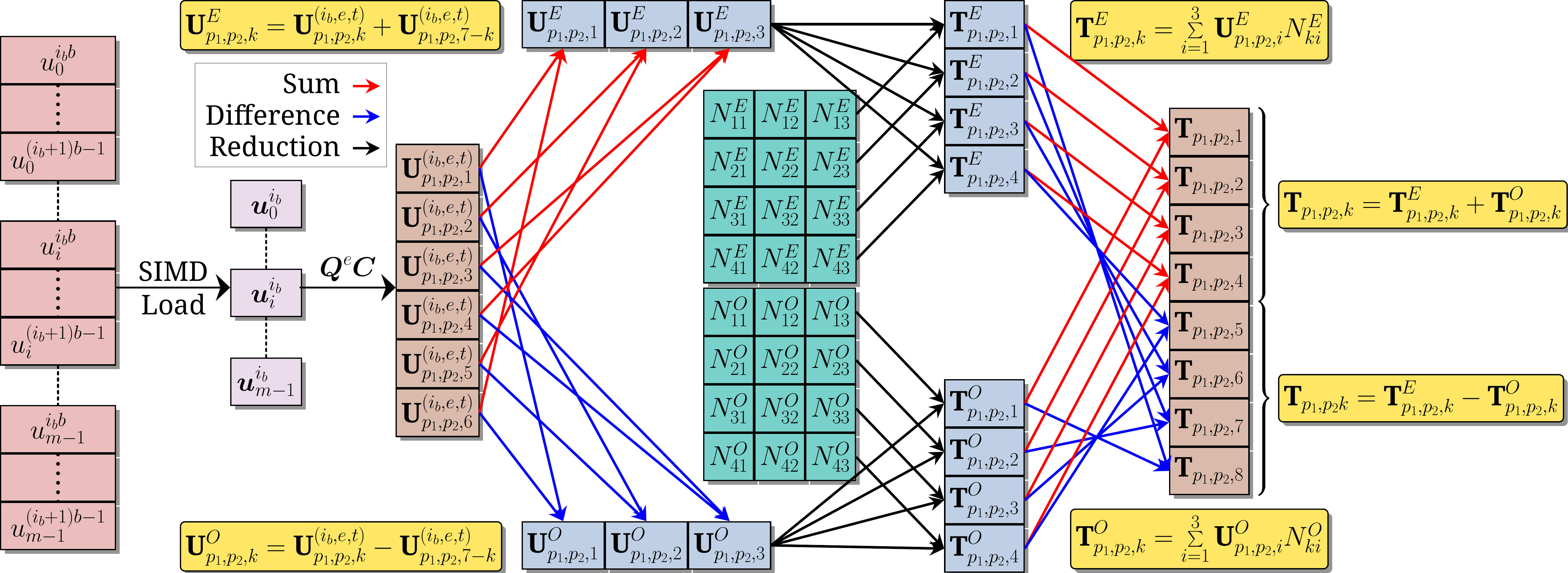}
    \caption{Evaluation of $\left(\bI\otimes\bI\otimes\bI\otimes\bN^{1D}\right)\bUt^{\left(i_b,e\right)}$ on CPU architectures using the even-odd decomposition strategy. An example with $n_p=6$ and $n_q=8$, each block in $\bU$ represents an $n_p^2$ sized array of AVX-512 doubles, which are decomposed into even and odd components to be multiplied by the corresponding shape function matrices. The results are combined to form $\bT$.}\label{fig:cpueo}
\end{figure*}

The implementation strategy used for the batch-wise evaluation of ${\bV^{\left(i_b,t\right)}}$ on CPU architectures, including constraints, extraction, and assembly, is described in \cref{alg:vibtcpu}. To perform the strided-batched matrix-matrix multiplications in \cref{alg:vibtcpu} (described by \cref{eqn:tc1,eqn:tc2,eqn:tc3}), we would need to have a function with the following signature
\begin{lstlisting}[caption={function signature for strided batched matrix-matrix multiplication},label={lst:cpusig}]
template <int m, int n, int k, int c, bool add, bool trans>
inline void 
matmul(const double |${}^{\texttt{*}}$|E, const double |${}^{\texttt{*}}$|F, double |${}^{\texttt{*}}$|C)  
\end{lstlisting}

which evaluates $\bC_i=\bE_i\texttt{op}\left(\bF\right)+\beta\bC_i \quad \forall i=1,\dots,c$ where $\bE_i$ is an $m\times k$ matrix and $\texttt{op}\left(\bF\right)$ is a $k\times n$ matrix with $\beta=1$ if \texttt{add=true} (0 otherwise) and $\texttt{op}\left(\bF\right)=\bF^T$ if \texttt{trans=true} ($\bF$ otherwise). To evaluate these batched matrix-matrix products involving  $\bN^{1D}$ and $\widetilde\bD^{1D}$ we explored three strategies:
\begin{enumerate}
    \item Employ \texttt{JIT} (Just-In-Time) modules from Intel\textsuperscript{\textregistered} MKL version 2022.1.0 \cite{Fedorov2019IntelGEMM}. For this implementation, $b=20$ yielded the best performance.
    \item Handwritten matrix-matrix multiplication code using AVX-512 intrinsics to work with 8 vectors concurrently, i.e. $b=8$.
    \item  Exploit the symmetry of the shape functions and quadrature points to reduce the floating point operations required by half via the \emph{even-odd} decomposition \cite{Kopriva2009ImplementingEquations,Solomonoff1992ADifferentiation}, and use AVX-512 intrinsics to work with eight vectors concurrently, i.e., $b=8$. An illustration of the even-odd implementation strategy to evaluate $\left(\bN^{1D}\otimes\bI\otimes\bI\otimes\bI\right)\bUt^{\left(i_b,e,t\right)}$ is provided in \cref{fig:cpueo} and sample code for the same is provided in \cref{lst:cpumatmul}.
\end{enumerate}

\begin{figure}[H]
    \centering
    \includegraphics[width=\linewidth]{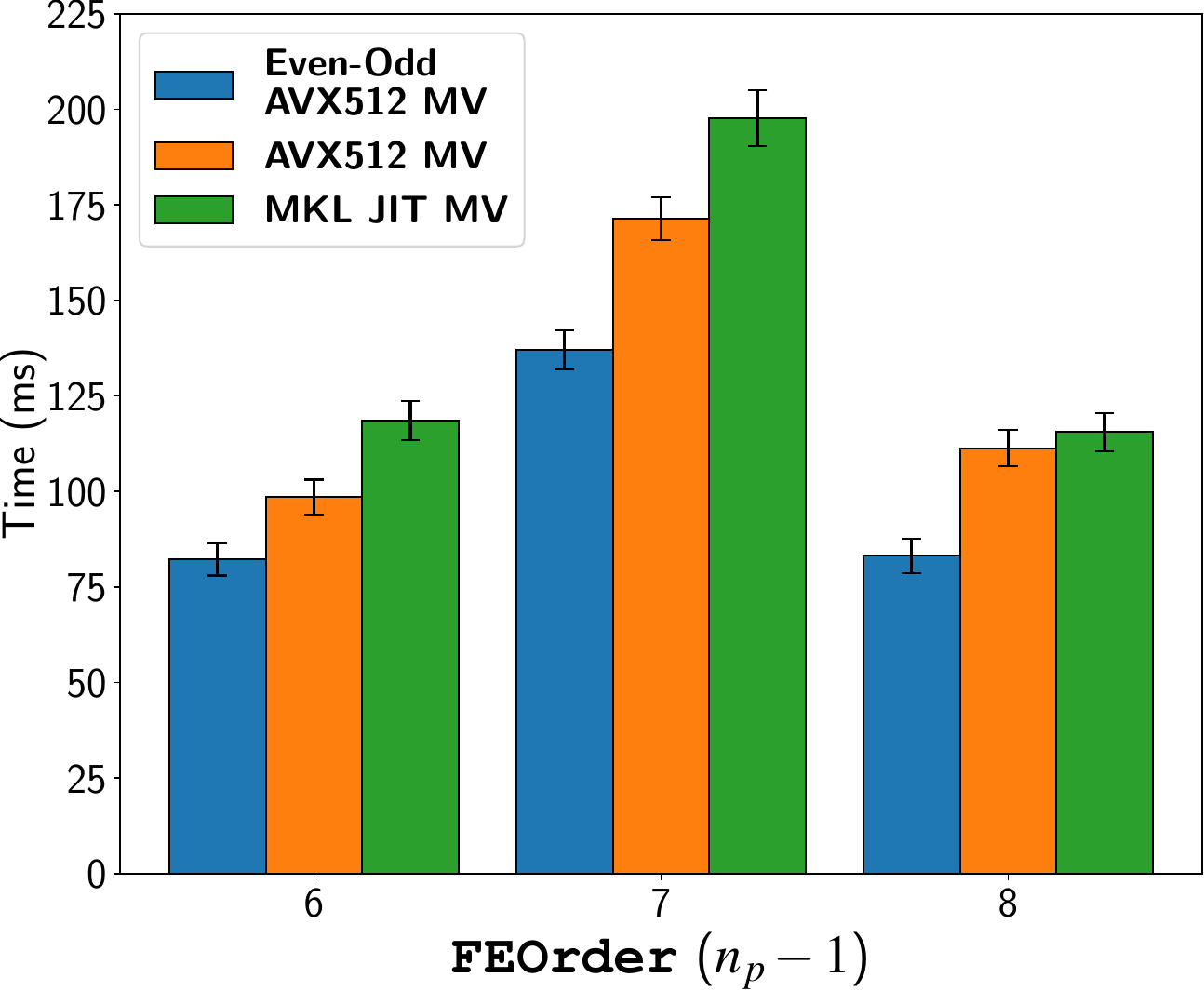}
    \caption{Performance benchmarks of our matrix-free implementation strategies using the proposed BCV layout on a single core of Intel\textsuperscript{\tiny\textregistered} Xeon\textsuperscript{\tiny\textregistered} Gold 6248R processor. Benchmark case studies: 15625 DoFs ($n_q=n_p= 7, 9$); 24389 DoFs ($n_q=n_p=8$). For AVX512 implementations $b=8$ is chosen and for the MKL JIT implementation $b=20$ is chosen.}\label{fig:layoutsdata}
\end{figure}

To compare the three strategies, we computed the action of the FE discretized Helmholtz operator obtained by setting $\mu=1$ and $\kappa\left(\bx\right)=2\pi ;\forall \bx \in \Omega$ in \cref{eqn:refprob} on randomly generated multivectors.
The results of our explorations are shown in \cref{fig:layoutsdata}. We find that the \emph{even-odd} decomposition approach yielded the best performance. On a single core, using the BCV layout, we observe speedups of up to 40\% of the \emph{Even-Odd AVX-512 Multivector} implementation over the \emph{MKL JIT Multivector}. We attribute this speedup to the fact that MKL JIT does not appear to use AVX-512 for matrices of such dimensions and instead falls back to AVX2.

\begin{figure}[H]
    \removelatexerror
\begin{algorithm}[H]
    \linespread{1.12}\selectfont
    \caption{Batchwise evaluation of ${\bV}$ on CPUs}\label{alg:vibtcpu}
    \KwIn{${\bU}$}
    \KwData{$\bB^{\left(i_b\right)},\bP^{\left(i_b,t\right)},\bC^{\left(i_b,t\right)},\bQ^{\left(i_b,e,t\right)},\bN^{1D},\bD^{1D},\bJ^{e},\bm{\kappa}$}
    \KwTemp{$\bT, \bT^{\left(0\right)},\bT^{\left(1\right)},\bT^{\left(2\right)}$}
    \KwResult{${\bV}$}
    \KwTId{$t$}
    $\bU^{\left(i_b,t\right)} \gets {\bC^{\left(i_b,t\right)}}\bU^{\left(i_b,t\right)}$\tcp*{\cref{sec:constraints}}
        \For{$e\gets1$ \KwTo $E_t$}{
            $\bT \gets \bQ^{\left(i_b,e,t\right)}\bU^{\left(i_b,t\right)}$\tcp*{\cref{sec:extraction}}
            $\showdim{\bT}{bn_p^2}{n_q} \gets \showdim{\bT}{bn_p^2}{n_p}\showdim{{\bN^{1D}}^T}{n_p}{n_q}$\tcp*{\cref{eqn:tc3}}
            \For {$q \gets 1$ \KwTo $n_q$}{
                $\showdim{\bT_{q}}{bn_p}{n_q} \gets \showdim{\bT_{q}}{bn_p}{n_p}\showdim{{\bN^{1D}}^T}{n_p}{n_q}$\tcp*{\cref{eqn:tc2}}
            }
            \For {$q \gets 1$ \KwTo $n_q^2$}{
                    $\showdim{\bT_{q}}{b}{n_q} \gets \showdim{\bT_{q}}{b}{n_p}\showdim{{\bN^{1D}}^T}{n_p}{n_q}$\tcp*{\cref{eqn:tc1}}
            }
            $\showdim{\bT^{\left(2\right)}}{bn_q^2}{n_q} \gets \showdim{\bT}{bn_q^2}{n_q}\showdim{\mbox{${\widetilde\bD}^{1D}$}^T}{n_q}{n_q}$\tcp*{\cref{eqn:tc3}}
            \For {$q \gets 1$ \KwTo $n_q$}{
                $\showdim{\bT^{\left(1\right)}_{q}}{bn_q}{n_q} \gets \showdim{\bT_{q}}{bn_q}{n_q}\showdim{\mbox{${\widetilde\bD}^{1D}$}^T}{n_q}{n_q}$\tcp*{\cref{eqn:tc2}}
            }
            \For {$q := 1$ \KwTo $n_q^2$}{
                    $\showdim{\bT^{\left(0\right)}_{q}}{b}{n_q} \gets \showdim{\bT_{q}}{b}{n_q}\showdim{\mbox{${\widetilde\bD}^{1D}$}^T}{n_q}{n_q}$\tcp*{\cref{eqn:tc1}}
            }
            $\showdim{\left[\bT^{\left(0\right)}\; \bT^{\left(1\right)}\;\bT^{\left(2\right)}\right]}{bn_q^3}{3} \gets \showdim{\left[\bT^{\left(0\right)}\;\bT^{\left(1\right)}\;\bT^{\left(2\right)}\right]}{bn_q^3}{3}\showdim{\left({\bJ^{(e)}}^{-1}{\bJ^{(e)}}^{-T}\det{\bJ^{(e)}}\mu\right)}{3}{3} $\tcp*{\cref{eqn:sjacobian}}
            $\showdimv{\bT}{bn_q^3} \gets \showdimv{\left(\det{\bJ^{(e)}}\bm{\kappa}\right)}{n_q^3}\;\circ \showdimv{\bT}{bn_q^3}$\tcp*{\cref{eqn:mjacobian}}
            $\showdim{\bT}{bn_q^2}{n_q} \gets \showdim{\bT}{bn_q^2}{n_q} \!\!\!+\!\!\!  \showdim{\bT^{\left(2\right)}}{bn_q^2}{n_q}\showdim{\mbox{${\widetilde\bD}^{1D}$}}{n_q}{n_q}$\tcp*{\cref{eqn:tc3}}
            \For {$q \gets 1$ \KwTo $n_q$}{
                $\showdim{\bT_{q}}{bn_q}{n_q} \gets \showdim{\bT_{q}}{bn_q}{n_q}\!\!\!+\!\!\! \showdim{\bT^{\left(1\right)}_{q}}{bn_q}{n_q}\showdim{\mbox{${\widetilde\bD}^{1D}$}}{n_q}{n_q}$\tcp*{\cref{eqn:tc2}}
            }
            \For {$q \gets 1$ \KwTo $n_q^2$}{
                $\showdim{\bT_{q}}{b}{n_q} \gets \showdim{\bT_{q}}{b}{n_q}\!\!\!+\!\!\!  \showdim{\bT^{\left(0\right)}_{q}}{b}{n_q}\showdim{\mbox{${\widetilde\bD}^{1D}$}}{n_q}{n_q}$\tcp*{\cref{eqn:tc1}}
            }
            $\showdim{\bT}{bn_q^2}{n_p} \gets \showdim{\bT}{bn_q^2}{n_q}\showdim{{\bN^{1D}}}{n_q}{n_p}$\tcp*{\cref{eqn:tc3}}
            \For {$p \gets 1$ \KwTo $n_p$}{
                $\showdim{\bT_{p}}{bn_q}{n_p} \gets \showdim{\bT_{p}}{bn_q}{n_q}\showdim{\bN^{1D}}{n_q}{n_p}$\tcp*{\cref{eqn:tc2}}
            }
            \For {$p \gets 1$ \KwTo $n_p^2$}{
                $\showdim{\bT_{p}}{b}{n_p} \gets \showdim{\bT_{p}}{b}{n_q}\showdim{\bN^{1D}}{n_q}{n_p}$\tcp*{\cref{eqn:tc1}}
            }
            $\bV^{\left(i_b,t\right)} \gets \bV^{\left(i_b,t\right)}+{\bQ^{\left(i_b,e,t\right)}}^T\bT$\tcp*{\cref{sec:extraction}}
        }
    $\bV^{\left(i_b,t\right)} \gets {\bC^{\left(i_b,t\right)}}^T\bV^{\left(i_b,t\right)}$\tcp*{\cref{sec:constraints}}
    \Return{${\bV}$}
\end{algorithm}
\end{figure}

\begin{lstlisting}[caption={Code snippet for the evaluation of $\left(\bN^{1D}\otimes\bI\otimes\bI\otimes\bI\right)\bU^{\left(i_b,e,t\right)}$ using the even-odd decomposition strategy. Note that this snippet is purely used to illustrate the implementation strategy, and as such \texttt{k} and \texttt{n} are assumed to be even. The actual implementation is generic. Note that if $n=k$, then $\texttt{C}=\texttt{A}$ is allowed, which results in a lower memory footprint.},label={lst:cpumatmul}]
template <int m, int n, int k>
inline void
matmul(const __m512d |${}^{\texttt{*}}$|A, const __m512d |${}^{\texttt{*}}$|B, __m512d |${}^{\texttt{*}}$|C){
/*Here |\clstcomment $\texttt{m}=n_p^2$, $\texttt{n}=n_q$, $\texttt{k}=n_p$|              */
/*and |\clstcomment $\texttt{A} \gets \bU^{\left(i_b,e,t\right)}$, $\texttt{B}\gets \left[\bN^E\;\bN^O\right]$, $\texttt{C}\gets \bU^{\left(i_b,e,t\right)}{\bN^{1D}}^T$|*/
constexpr int ko = k / 2;
constexpr int no = n / 2;
for (auto i = 0; i < m; ++i){
    /*Temporary arrays for storage of even and odd components of rows of A*/
    __m512d tempAe[ko], tempAo[ko];
    /*Evaluate even and odd components of row i (|\clstcomment$=p_1+n_pp_2 $|) of A*/
    for (auto q = 0; q < ko; ++q){
        /* tempAe[q]=|\clstcomment$U^{\left(i_b,e,t\right)}_{p_1,p_2,q}+U^{\left(i_b,e,t\right)}_{p_1,p_2,k-q}$\cn|*/
        tempAe[q] = A[i + q * m] + A[i + (k - 1 - q) * m];
        /* tempAo[q]=|\clstcomment$U^{\left(i_b,e,t\right)}_{p_1,p_2,q}-U^{\left(i_b,e,t\right)}_{p_1,p_2k-q}$\cn|*/
        tempAo[q] = A[i + q * m] - A[i + (k - 1 - q) * m];}
    for (auto j = 0; j < no; ++j){
    /*Temporary storage even and odd components of C*/
    __m512d tempCe, tempCo;
    /*tempCe=|\clstcomment$\displaystyle\sum_q^{n_p/2}$|tempAe[q]|\clstcomment $N^{E}_{j,q}$| */
    tempCe = tempAe[0] * B[j];
    for (auto q = 1; q < ko; ++q)
        tempCe += tempAe[q] * B[j + q * no];
    /*tempCo=|\clstcomment$\displaystyle\sum_q^{n_p/2}$|tempAo[q]|\clstcomment $N^{O}_{j,q}$| */
    tempCo = tempAo[0] * B[j + ko * no];
    for (auto q = 1; q < ko; ++q)
        tempCo += tempAo[q] * B[j + q * no + ko * no];
    /*Recombining tempCe and tempCo to get elements of |\clstcomment$\bC$| */
    /*|\clstcomment$T_{p_1,p_2,j}=$|tempCe+tempCo */
    C[i + m * j]           = tempCe + tempCo;
    /*|\clstcomment$T_{p_1,p_2,n-j}=$|tempCe-tempCo*/
    C[i + m * (n - 1 - j)] = tempCo - tempCe;}}}
\end{lstlisting}

\subsubsection*{GPU Implementation: Evaluation of \texorpdfstring{$\bA^{\left(e\right)}\bU^{\left(i_b,e,t\right)}$}{ Cell Level products}}
The implementation strategy, including extraction, and assembly, used for the evaluation of $\bV^{\left(t\right)}$ on GPU architectures is described in \cref{alg:vibtgpu}.
To evaluate the tensor contractions in \cref{eqn:tc3,eqn:tc2,eqn:tc1}, vendor optimized \textit{gemm} libraries (for eg. \textit{cuBLAS}, \textit{hipBLAS}) modules seem to be a natural choice at first glance for GPUs. However, the sequential library calls for each tensor contraction requires multiple reads from and writes to the device memory. Hence, to avoid such data movement,  we design a shared memory gemm implementation on GPUs by taking advantage of kernel fusion, accessing data only once from the device memory. This implementation combines the extraction, tensor contractions, and assembly steps in one kernel, performs all the computations inside the fast shared memory to finally write the data back to the device memory.

\begin{figure}[H]
    \centering
    \includegraphics[width=\linewidth]{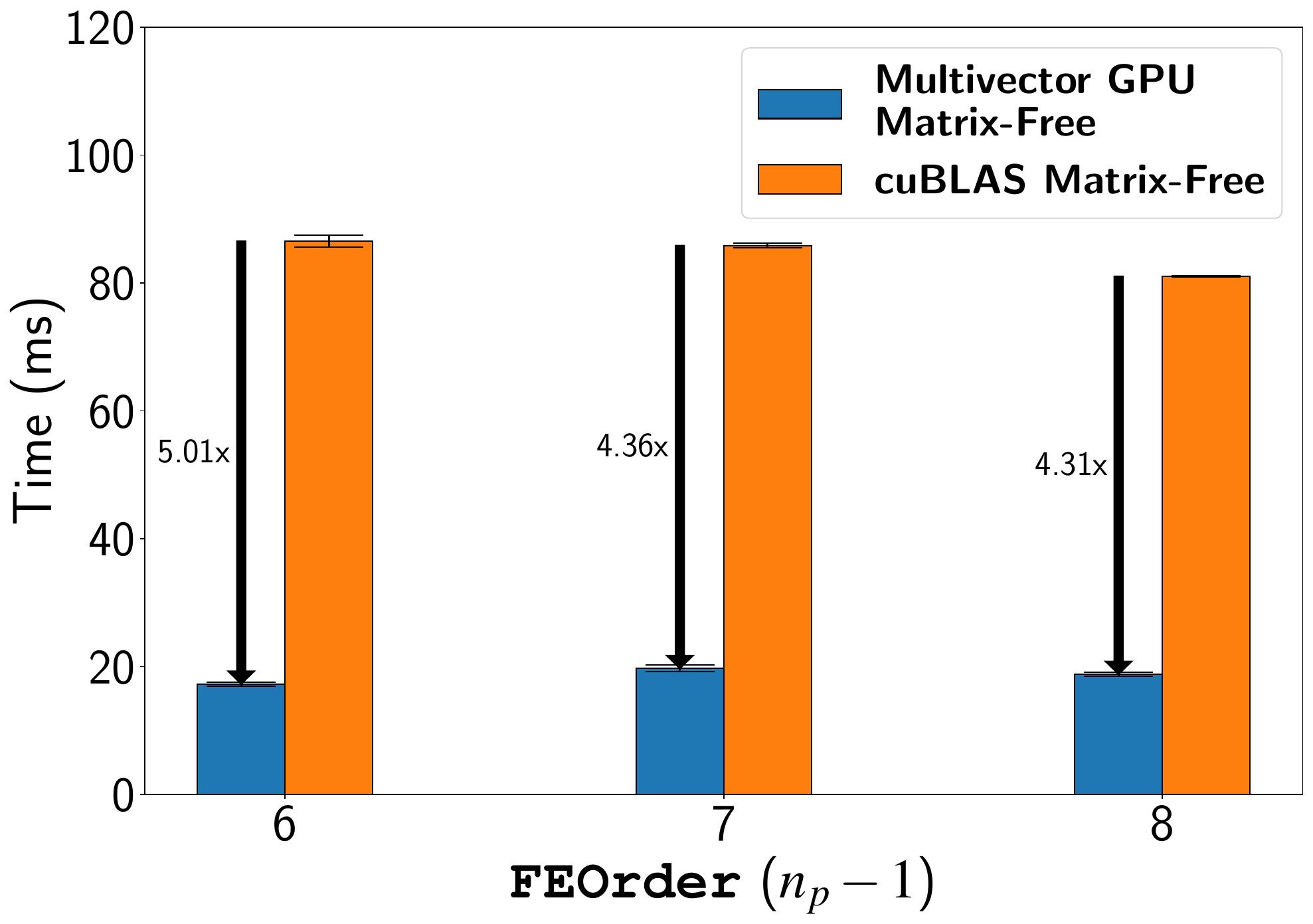}
    \caption{\small Performance benchmark of \texttt{cuBLAS dgemm} matrix-free implementation with our \emph{Multivector GPU Matrix-Free} approach for evaluating tensor contractions. Studies conducted on NVIDIA\textsuperscript{\tiny\textregistered} Tesla\textsuperscript{\tiny\textregistered} V100 SXM2 16GB (Summit Supercomputer). GPU benchmark case studies: 117649 DoFs (\texttt{FEOrder} = 6, 8); 125000 DoFs (\texttt{FEOrder} = 7) with BCV layout where $b = n_v = 1024$.} \label{fig:gpulayoutsdata}
\end{figure}

In \cref{fig:gpulayoutsdata}, we compare these two strategies, one using \texttt{cuBLAS dgemm} and the other using the shared memory implementation as discussed above. We observe speedups of 4x -- 5x for the proposed \emph{Multivector GPU Matrix-Free} (MV GPU Matrix-Free) approach compared to the \texttt{cuBLAS dgemm} approach. 


We discuss the shared memory implementation on GPUs in more detail. Unlike matrix-multivector multiplication using the FE-cell level local dense matrices approach (see \cref{sec:cellmatrix}), the shared memory kernel does not explicitly construct the cell level multivectors $\bU^{(e,t)}$ and $\bV^{(e,t)}$ in the device memory. This helps to further reduce the memory footprint. The kernel launch associated with the shared memory implementation of our MV GPU Matrix-Free approach is  as follows:
\begin{lstlisting}[mathescape=true, caption={Kernel launch for Multivector GPU Matrix-Free implementation},label={lst:gpuker}] 
compute <$b$, $n_p$, $n_q$> <<<dim3($E_t$, $n_b$), dim3($n_{t_x}$, $n_{t_y}$), smem>>> (double |${}^{\texttt{*}}$|C, const double |${}^{\texttt{*}}$|A, const double |${}^{\texttt{*}}$|B, $\cdots$)  
\end{lstlisting}

The kernel is templated with $b$, $n_p$ and $n_q$ and launched with a 2-D grid of $E_t \times n_b$ thread blocks, each with a 2-D block of $n_{t_x} = b$ threads in the x-direction and $n_{t_y} = \texttt{warpSize}\times\alpha$ threads in the y-direction where \texttt{warpSize} = 32 for NVIDIA GPUs and $\alpha$ is a tunable parameter. This choice of $n_{t_y}$ ensures that the total number of threads per thread block is a multiple of \texttt{warpSize}. The kernel is also configured with a dynamic shared memory of $\texttt{smem} = 4b{n_q}^3$. 

\begin{figure*}[!hb]
    \centering
    \includegraphics[width=\linewidth]{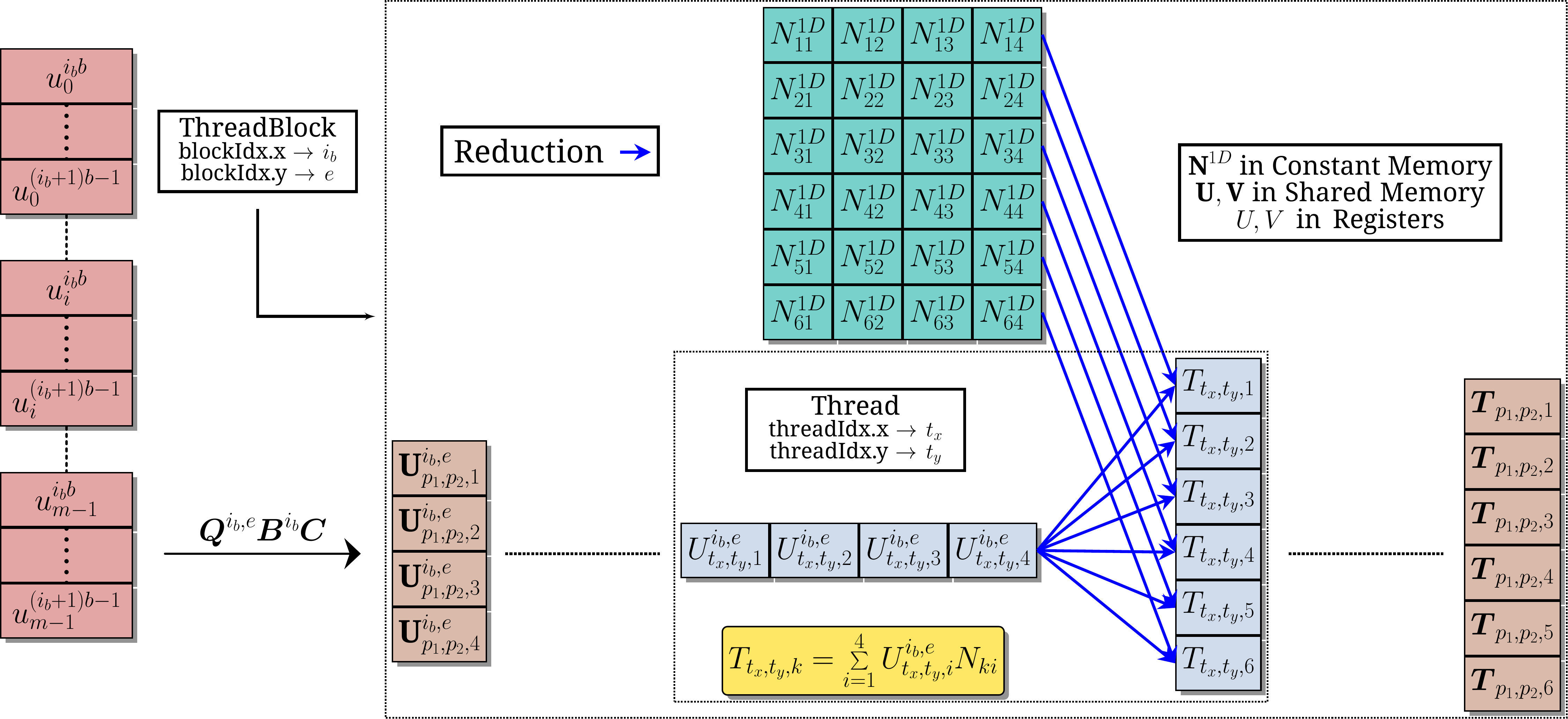}
    \caption{\small Pictorial depiction of tensor contractions done on GPUs. The extraction and first tensor contraction steps of evaluation of $\bA^{\left(e\right)}\bU^{(i_b,e,t)}$ are depicted for the case of $n_p = 4$ and $n_q = 6$. Each block in $\bU$ represents $n_p^2$ sized array of $b$ doubles.}\label{fig:TensorGPU}
\end{figure*}

The matrices $\bN^{1D}$ and $\widetilde\bD^{1D}$ are stored in constant memory as they are constant for all cells and batches. This helps reduce shared memory usage, and the matrices can be reused for all subsequent tensor contractions. We rewrite these tensor contractions as batched-matrix-matrix multiplications as discussed in \cref{sec:elecompute} and execute them as linear combinations of columns of $\bN^{1D}$ and $\widetilde\bD^{1D}$ as illustrated in \cref{fig:TensorGPU}. Thus, evaluations like \cref{eqn:tc3} can be written as
\begin{align}
    \bTt_{t_x, t_y, q}=\sum_{k = 1}^{n_p}\bRt_{t_x, t_y, k}{N_{qk}^{1D}} &  & \forall q=1,\dots,n_p\label{eqn:tcGPU}
\end{align}
where $t_x$ is \texttt{threadIdx.x} and $t_y$ is \texttt{threadIdx.y}. This execution method enables us to combine the extraction and the first tensor contraction steps. Thus, the floating point operations are performed as soon as a portion of ${\bU^{\left(t\right)}}$ is read from the device memory, without having to wait for its complete $bn_p^3$ data to be loaded inside the shared memory. Furthermore, as each thread accesses the same values from $\bN^{1D}$ and $\widetilde\bD^{1D}$, the accesses are broadcast, and because the access is from constant memory, the GPU pipelines are better utilized. To further improve performance, we utilize registers to keep the data local to each thread as much as possible. This optimization reduces data movement from shared memory and bank conflicts. Finally, in the assembly step \texttt{atomicAdd} is used to avoid race conditions and safely assemble the output ${\bV^{\left(t\right)}}$. Similar to the extraction step, we also combine the assembly step with the last tensor contraction. Unlike CPUs, we do not employ the \emph{even-odd} decomposition on GPUs because the stall wait state of the GPU warps increases due to the requirement of additional accesses to compute the even and odd components in the \emph{even-odd} decomposition approach (see \cref{fig:cpueo} and \cref{lst:cpumatmul}).

\begin{lstlisting}[caption={Code snippets for the evaluation of \cref{eqn:tc3,eqn:tc2,eqn:tc1} on GPUs.},label={lst:gpuCompute}]
    /* Snippet for |\clstcomment\cref{eqn:tc1}| */
    /* |\clstcomment $\texttt{m}=n_p^2$, $\texttt{n}=n_q$, $\texttt{k}=n_p$| */
    /* |\clstcomment $\texttt{A} \gets \bR$, $\texttt{B}\gets {\bN^{1D}}^T$, $\texttt{C}\gets \bT = \bR{\bN^{1D}}^T$|*/
    for (int i = threadIdx.y; i < m; i += blockDim.y) {
      /* Temporary arrays for storage of rows of A and C*/
      double y[n], x[k];
      for (int j = 0; j < n; j++)
        y[j] = 0.0;
      /* |\clstcomment x[q] = $R_{t_x,i,q}$| */
      for (int q = 0; q < k; q++) {
        x[q] = A[threadIdx.x + i * b + q * b * m];
        /* |\clstcomment y[j] = $\sum_q^{n_p}R_{t_x,i,q}N_{j, q}$| */
        for (int j = 0; j < n; j++)
          y[j] += B[j + q * n] * x[q]; }
      /* |\clstcomment  $T_{t_x,i,j}$ = y[j]| */
      for (int j = 0; j < n; j++)
        C[threadIdx.x + i * b + j * b * m] = y[j]; }
    
    /* Snippet for |\clstcomment\cref{eqn:tc2}| */
    /* |\clstcomment $\texttt{m}=n_p^2$, $\texttt{n}=n_q$, $\texttt{k}=n_p$| */
    /* |\clstcomment $\texttt{A} \gets \bR_v$, $\texttt{B} \gets {\bN^{1D}}^T$, $\texttt{C} \gets \bT_v = \bR_v{\bN^{1D}}^T \quad \forall  v=1,\dots,n_p$| */
    for (int i = threadIdx.y; i < m; i += blockDim.y) {
      /* Temporary arrays for storage of rows of A and C*/
      double y[n], x[k];
      int u = i % k, v = i / k;
      for (int j = 0; j < n; j++)
        y[j] = 0.0;
      /* |\clstcomment x[q] = $R_{t_x,u,q,v}$| */
      for (int q = 0; q < k; q++) {
        x[q] = A[threadIdx.x + u * b + q * b * k + v * b * k^2];
        /* |\clstcomment y[j] = $\sum_q^{n_p}R_{t_x,u,q,v}N_{j, q}$| */
        for (int j = 0; j < n; j++)
          y[j] += B[j + q * n] * x[q]; }
      /* |\clstcomment  $T_{t_x,u,j,v}$ = y[j]| */
      for (int j = 0; j < n; j++)
        C[threadIdx.x + u * b + j * b * k + v * b * k * n] = y[j]; }
    
    /* Snippet for |\clstcomment\cref{eqn:tc3}| */
    /* |\clstcomment $\texttt{m}=n_q^2$, $\texttt{n}=n_q$, $\texttt{k}=n_p$| */
    /* |\clstcomment $\texttt{A} \gets \bR_v$, $\texttt{B} \gets {\bN^{1D}}^T$, $\texttt{C} \gets \bT_v = \bR_v{\bN^{1D}}^T \quad \forall  v=1,\dots,n_p^2$| */
    for (int i = threadIdx.y; i < m; i += blockDim.y) {
      /* Temporary arrays for storage of rows of A and C*/
      double y[n], x[k];
      for (int j = 0; j < n; j++)
        y[j] = 0.0;
      /* |\clstcomment x[q] = $R_{t_x,q,i}$| */
      for (int q = 0; q < k; q++) {
        x[q] = A[threadIdx.x + q * b + i * b * k];
        /* |\clstcomment y[j] = $\sum_q^{n_p}R_{t_x,q,i}N_{j, q}$| */
        for (int j = 0; j < n; j++)
          y[j] += B[j + q * n] * x[q]; }
      /* |\clstcomment  $T_{t_x,j,i}$ = y[j]| */
      for (int j = 0; j < n; j++)
        C[threadIdx.x + j * b + i * b * n] = y[j]; }
    \end{lstlisting}

\begin{figure}[H]
    \removelatexerror
\begin{algorithm}[H]
    \SetNoFillComment
    \linespread{1.335}\selectfont
    \caption{Batchwise evaluation of ${\bV}$ on GPUs}\label{alg:vibtgpu}
    \KwIn{${\bU}$} 
\KwData{$\bC^{\left(t\right)},\bB^{\left(i_b, t\right)},\bQ^{\left(i_b,e,t\right)},\bN^{1D},\bD^{1D},,\bJ^{e},\bm{\kappa}$}
    \KwTemp{$\bT, \bT^{\left(0\right)},\bT^{\left(1\right)},\bT^{\left(2\right)}$}
    \KwResult{${\bV}$}

    \KwTId{$t$}
    \KwBIdx{$e$}
    \KwBIdy{$i_b$}

    $\bU^{\left(t\right)} \gets {\bC^{\left(t\right)}}\bU^{\left(t\right)}$\tcp*{\cref{sec:constraints}}
        \tcc{\normalsize \normalfont\textbf{Device kernel compute starts}}
            $\showdim{\bT}{bn_p^2}{n_q} \gets \showdim{\bQ^{\left(i_b,e,t\right)}\bB^{\left(i_b, t\right)}\bU^{\left(t\right)}}{bn_p^2}{n_p}\showdim{{\bN^{1D}}^T}{n_p}{n_q};$\newline\tcp{\cref{sec:layout,sec:extraction,eqn:tc3}}
                $\showdim{\bT^{\left(0\right)}_{q}}{bn_p}{n_q}\! \gets\! \showdim{\bT_{q}}{bn_p}{n_p}\showdim{{\bN^{1D}}^T}{n_p}{n_q}  \forall q=1,\dots,n_q $\tcp*{\cref{eqn:tc2}}
                    $\showdim{\bT_{q}}{b}{n_q} \gets \showdim{\bT^{\left(0\right)}_{q}}{b}{n_p}\showdim{{\bN^{1D}}^T}{n_p}{n_q} \quad \forall q=1,\dots,n_q^2$\tcp*{\cref{eqn:tc1}}
            $\showdim{\bT^{\left(2\right)}}{bn_q^2}{n_q} \gets \showdim{\bT}{bn_q^2}{n_q}\showdim{\mbox{${\widetilde\bD}^{1D}$}^T}{n_q}{n_q}$\tcp*{\cref{eqn:tc3}}
                $\showdim{\bT^{\left(1\right)}_{q}}{bn_q}{n_q} \gets \showdim{\bT_{q}}{bn_q}{n_q}\showdim{\mbox{${\widetilde\bD}^{1D}$}^T}{n_q}{n_q} \forall q=1,\dots,n_q$\tcp*{\cref{eqn:tc2}}
                    $\showdim{\bT^{\left(0\right)}_{q}}{b}{n_q} \gets \showdim{\bT_{q}}{b}{n_q}\showdim{\mbox{${\widetilde\bD}^{1D}$}^T}{n_q}{n_q}\quad \forall q=1,\dots,n_q^2$\tcp*{\cref{eqn:tc1}}
            $\showdim{\left[\bT^{\left(0\right)}\; \bT^{\left(1\right)}\;\bT^{\left(2\right)}\right]}{bn_q^3}{3} \gets \showdim{\left[\bT^{\left(0\right)}\;\bT^{\left(1\right)}\;\bT^{\left(2\right)}\right]}{bn_q^3}{3}\showdim{\left({\bJ^{(e)}}^{-1}{\bJ^{(e)}}^{-T}\det{\bJ^{(e)}}\mu\right)}{3}{3} $\;
            
            
            $\showdim{\bT}{bn_q^2}{n_q} \gets \showdimv{\left(\det{\bJ^{(e)}}\bm{\kappa}\right)}{n_q^3}\;\circ\showdimv{\bT}{bn_q^3} + \showdim{\bT^{\left(2\right)}}{bn_q^2}{n_q}\showdim{\mbox{${\widetilde\bD}^{1D}$}}{n_q}{n_q};$\newline\tcp{\cref{eqn:tc3}}
            
                $\showdim{\bT_{q}}{bn_q}{n_q} \gets \showdim{\bT_{q}}{bn_q}{n_q}+ \showdim{\bT^{\left(1\right)}_{q}}{bn_q}{n_q}\showdim{\mbox{${\widetilde\bD}^{1D}$}}{n_q}{n_q} \quad \forall q=1,\dots,n_q$\tcp*{\cref{eqn:tc2}} 
                $\showdim{\bT_{q}}{b}{n_q} \gets \showdim{\bT_{q}}{b}{n_q}+ \showdim{\bT^{\left(0\right)}_{q}}{b}{n_q}\showdim{\mbox{${\widetilde\bD}^{1D}$}}{n_q}{n_q} \quad \quad \quad \forall q=1,\dots,n_q^2$\tcp*{\cref{eqn:tc1}}
            $\showdim{\bT}{bn_q^2}{n_p} \gets \showdim{\bT}{bn_q^2}{n_q}\showdim{{\bN^{1D}}}{n_q}{n_p}$\tcp*{\cref{eqn:tc3}}
                $\showdim{\bT_{p}}{bn_q}{n_p} \gets \showdim{\bT_{p}}{bn_q}{n_q}\showdim{\bN^{1D}}{n_q}{n_p} \;\;\, \forall p=1,\dots,n_p$\tcp*{\cref{eqn:tc2}}
                $\bV^{\left(i_b,t\right)} \gets \bV^{\left(i_b,t\right)} +{\bB^{\left(i_b, t\right)}}^T{\bQ^{\left(i_b,e,t\right)}}^T \showdim{\bT_{p}}{b}{n_q}\showdim{\bN^{1D}}{n_q}{n_p} \newline \forall p=1,\dots,n_p^2$\tcp*{\cref{sec:layout,sec:extraction,eqn:tc1}}
        \tcc{\normalsize \normalfont\textbf{Device kernel compute ends}}
    $\bV^{\left(t\right)} \gets {\bC^{\left(t\right)}}^T\bV^{\left(t\right)}$\tcp*{\cref{sec:constraints}}
    \Return{${\bV}$}
\end{algorithm}
\end{figure}

\subsubsection{Distributed Parallelism: \texttt{MPI} aspects}\label{sec:mpi}
We now discuss the \texttt{MPI} communication strategies employed to reduce the communication overheads encountered when deploying on multi-node CPU and GPU architectures.
\subsubsection*{CPU Implementation: \texttt{MPI} aspects}
It is important to note that in our implementation, we do not explicitly construct $\bB^{\left(i_b\right)}{\bU}$ in memory. Instead, we evaluate the action of $\bP^{\left(i_b,t\right)}$ on it through \texttt{MPI} communication of boundary data for the multivectors across tasks that share subdomain boundaries. In addition, we evaluate the summations over $e$ and $i_b$ in \cref{eqn:basempicpu} as serial loops.
Furthermore, we overlap the communication involved in the action of $\bP^{\left(i_b,t\right)}\bB^{\left(i_b\right)}$ and ${\bB^{\left(i_b\right)}}^T{\bP^{\left(i_b,t\right)}}^T$  with the computation involved in the action of $\bC^{\left(i_b,t\right)}$, ${\bC^{\left(i_b,t\right)}}^T$, $\bQ^{\left(e,i_b,t\right)}$, ${\bQ^{\left(e,i_b,t\right)}}^T$ and $\bA^{\left(e\right)}$ as illustrated in \cref{alg:cpuoverlap}. For a given MPI task $t$, $m_{loc}^{\left(t\right)}$ represents the number of locally-owned degrees of freedom (DoFs), while $m_{ghost}^{\left(t\right)}$ denotes the number of DoFs on the shared subdomain boundary that are not owned by task $t$, commonly known as ``ghost" DoFs. The storage layout of the multivector in each process consists of a $b\times m_{loc}^{\left(t\right)}\times n_b$ sized array for storing the locally-owned data followed by a contiguous storage of $b\times m_{ghost}^{\left(t\right)}\times 2$ sized array to hold the data received from tasks that share subdomain boundaries. This allows us to store the subdomain boundary data for two batches so that we can overlap the compute of one batch with the communication involved in another batch. We note that the dimensions of these arrays are in the order of their corresponding fastest index in storage. 

    \begin{algorithm}[!htb]
        \caption{\small Overlap of computation and communication}\label{alg:cpuoverlap}
        \KwIn{${\bU}$}
        \KwData{$\bB^{\left(i_b\right)},\bP^{\left(i_b,t\right)},\bC^{\left(i_b,t\right)},\bQ^{\left(i_b,e,t\right)}$ for $e=1,\dots,E_t$ and $i_b=1,\dots,n_b$}
        \KwResult{${\bV}$}
    
        \KwTId{$t$}
        $\bU^{t,1} \gets \bP^{t,1}\bB^{1}{\bU}$\;
        \For{$i_b\gets1$ \KwTo $n_b$}{
            \tcc{\small Start communication for batch $i_b+1$ required for evaluating $\bP^{t,i_b+1}\bB^{\left(i_b+1\right)}{\bU}$ using MPI\_Isend and MPI\_Irecv}
            \If{$i_b<n_b$}{Start : $\bU^{t,i_b+1} \gets \bP^{t,i_b+1}\bB^{\left(i_b+1\right)}{\bU}$\;}
            \tcc{\small Start communication for batch $i_b-1$ required for evaluating ${\bP^{t,i_b-1}}^T{\normalfont\bB^{\left(i_b-1\right)}}^T{\normalfont\bV}^{t,i_b-1}$ using MPI\_Isend and MPI\_Irecv.}
            \If{$i_b>1$}{Start : ${\bV} \gets \bV+{\bP^{t,i_b-1}}^T{\bB^{\left(i_b-1\right)}}^T{\bV}^{t,i_b-1}$\;}
            \tcc{\small Using \cref{alg:vibtcpu} for the following evaluation.}
            $\bV^{\left(i_b,t\right)} \gets {\bC^{\left(i_b,t\right)}}^T\left(\sum_{e}^{E_t}{\bQ^{e,i_b,t}}^T\bA^{\left(e\right)}\bQ^{e,i_b,t}\right)\bC^{\left(i_b,t\right)}\bU^{\left(i_b,t\right)}$\;
            \tcc{\small MPI\_Waitall for finishing communication and processing the received data.}
            \If{$i_b<n_b$}{Finish : $\bU^{t,i_b+1} \gets \bP^{t,i_b+1}\bB^{\left(i_b+1\right)}{\bU}$\;}
            \If{$i_b>1$}{Finish : ${\bV} \gets \bV+{\bP^{t,i_b-1}}^T{\bB^{\left(i_b-1\right)}}^T{\bV}^{t,i_b-1}$\;}
                }
    ${\bV}^{t,n_b} \gets {\bP^{t,n_b}}^T{\bB^{n_b}}^T{\bV}^{t,n_b}$\;
    \Return{${\bV}$}
    \end{algorithm}

\subsubsection*{GPU Implementation: \texttt{MPI} aspects}
On GPUs, we do not explicitly construct ${\bU}$. Instead, we evaluate the action of $\bP^{\left(t\right)}$ on ${\bU}$ through \texttt{MPI} communication of boundary data for the multivectors across \texttt{MPI} tasks that share subdomain boundaries to extract the subdomain level multivector ${\bU^{(t)}}$. We denote the number of \emph{locally-owned} DoFs on task $t$ as $m_{loc}^{\left(t\right)}$ and the number of DoFs on the shared subdomain boundary of task $t$ that are not owned by it but \emph{locally-owned} by task $\hat{t}$ as $m_{ghost}^{\left(t, \hat{t}\right)}$. Further, let $\hat{n}_t$ denote the number of such tasks $\hat{t}$ for a given task $t$. The storage layout of the multivector comprises of $b \times m_{loc}^{\left(t\right)} \times n_b$ sized array for the locally-owned data followed by $b \times m_{ghost}^{\left(t, \hat{t}\right)} \times n_b \times \hat{n}_t$ sized array for the data received from the tasks $\hat{t}$ that share subdomain boundaries with task $t$. Note that the dimensions indicated above are in the order of their corresponding fastest index. This storage layout helps in evaluating $\bA^{\left(e\right)}\bU^{\left(i_b,e,t\right)}$ for each $i_b$, $e$ and $t$ concurrently. We use CUDA-Aware MPI to optimize communications which pipelines message transfers and uses NVIDIA\textsuperscript{\tiny\textregistered}  GPUDirect\textsuperscript{\tiny\textregistered} for various inter-rank communications like intra-node, inter-node, and RDMA inter-node communication. We further explore a mixed precision strategy to communicate data on the shared subdomain boundary where the boundary data communicated is recast as FP32 floats, which reduces the amount of data that needs to be communicated.


\section{Performance Benchmarks}\label{sec:sec4}
We now assess the performance of the proposed matrix-free algorithm for multivectors using representative benchmark problems. To this end, we first consider the action of the finite-element (FE) discretized Helmholtz operator on randomly generated multivectors using multi-node CPU and GPU architectures. We begin by examining the sustained performance and strong scaling efficiencies of our implementation for various higher-order FE interpolating polynomial orders. Subsequently, we benchmark our performance against established baselines. The first baseline chosen for benchmarking our performance on both multi-node CPUs and GPUs involves a cell-matrix approach suited naturally for multivectors, as discussed in \cref{sec:cellmatrix}, and has also been employed in previous works \cite{Carey1988Element-by-elementComputations,Hughes1987Large-scaleGradients,Cantwell2011FromElements} and particularly in recent works \cite{Das2019FastSystem,Das2022DFT-FEDiscretization} that have been nominated as one of the 2019 ACM Gordon Bell Prize finalists~\cite{2019GordonACM}. We also consider a second baseline for benchmarking the performance of matrix multivector products on multi-node CPUs, which involves the single-component matrix-free framework of \texttt{deal.II} by looping over the constituent vectors. We consider two benchmark problems to test and evaluate our implementation: (a) The evaluation of the Helmholtz operator action on randomly generated multivectors. In this case, we set $\mu=1$ and $\kappa\left(\bx\right)=2\pi ;\forall \bx \in \Omega$ in \cref{eqn:refprob}, and (b) the solution of Helmholtz eigenvalue problem using the Chebyshev Filtered Subspace Iteration (ChFSI) method. Here, we set $\mu=1/2$ and $\kappa$ as a precomputed potential in \cref{eqn:refprob}.

For our evaluations, we selected the number of nodes for the 1D base mesh to be $n_p = 7, 8, 9$, resulting in Lagrange interpolating polynomial orders $\texttt{FEOrder}=6, 7, 8$. This selection is motivated by the potential of the proposed methods to accelerate the eigensolvers employed to solve FE discretized large-scale eigenvalue problems arising in the domain of quantum modeling of materials. The chosen \texttt{FEOrder} offers a balanced approach between the reduction in the number of DoFs required to achieve the desired accuracy and the increased cost per DoF associated with a higher \texttt{FEOrder}, as discussed by \citet{Motamarri2013Higher-orderTheory,Motamarri2020DFT-FECalculations,Das2022DFT-FEDiscretization}. It is worth noting that the desired accuracy of evaluating the integrals in \cref{eqn:integrals} may not always be achieved by using the quadrature rule of order $n_q=n_p$. Therefore, we benchmark the cases involving $n_q=n_p$ and $n_q>n_p$.

To conduct these benchmarks, we employ the computing clusters, Param Pravega (for benchmarking on CPUs) and Summit supercomputer (for benchmarking on GPUs), the configurations of which are described in \cref{tbl:sysconf}. We also report GPU performance benchmarks conducted on the Selene supercomputer in \cref{sec:selene}. 
\begin{table}[H]
\rowcolors{2}{gray!25}{white}
\begin{tabular}{M{0.189\linewidth} M{0.335\linewidth} M{0.335\linewidth}}
\rowcolor{gray!50}
\textbf{System Config} & \small\textbf{Summit Supercomputer} & \small\textbf{Param Pravega (CPU only nodes)} \\
Processor         & \small IBM\textsuperscript{\textregistered} POWER9 & \small Intel\textsuperscript{\textregistered} Xeon\textsuperscript{\textregistered} Platinum 8268  \\
GPU        & \small NVIDIA\textsuperscript{\textregistered} Tesla\textsuperscript{\textregistered} V100 SXM2 16GB & -  \\
Nodes             & \small 4608 & \small 428 + 156 (High Memory)  \\
\small CPU cores/Node    & 32 & 48  \\
GPUs/Node         & 6 & -  \\
\small Node Performance  & \small 42 TFLOP/s \quad \quad (V100 FP64) & \small 1.459 TFLOP/s (AVX-512 FP64)  \\
\small Memory/Node       & \small 512 GB DDR4 + 96 GB HBM2 & \small 192 GB or 768 GB (High Memory) DDR4 \\
Interconnect      &\footnotesize Mellanox\textsuperscript{\textregistered} EDR 100G InfiniBand & \footnotesize Mellanox\textsuperscript{\textregistered} ConnectX\textsuperscript{\textregistered}-6 MT28908 \\
OS & \small RHEL 8.2 & \small CentOS 7
\end{tabular}
\caption{System configurations for the benchmark architectures.}\label{tbl:sysconf}
\end{table}

The compilers, MPI and BLAS libraries used are listed in \cref{tbl:libconf}.
\begin{table}[H]
    \rowcolors{2}{gray!25}{white}
    \begin{tabular}{M{0.189\linewidth} M{0.335\linewidth} M{0.335\linewidth}}
    \rowcolor{gray!50}
    Library  & GPU Benchmarks & CPU Benchmarks \\
    Compiler         &\;\;\;\;\; gcc 9.1.0 \;\;\;\;\; nvcc 11.0 & gcc 12.2.0  \\
    Compiler Flags & \texttt{-O3 -arch=sm\_70 -lcublas} &\texttt{-O3 -fopenmp-simd -march=native}  \\
    MPI    & IBM Spectrum MPI 10.4 & Intel\textsuperscript{\textregistered} oneAPI MPI 2021.9.0  \\
    BLAS  & \small cuBLAS 11.0 & Intel\textsuperscript{\textregistered} oneAPI MKL 2023.1.0  \\
    \end{tabular}
    \caption{External libraries and compiler flags used for compilation.}\label{tbl:libconf}
\end{table}
    

\subsection{Helmholtz Operator action}
We use \texttt{deal.II} library version 9.4.2~\cite{Arndt2022The9.4} with the \texttt{p4est}~\cite{Burstedde2011P4estOctrees} backend to perform the MPI-parallel meshing and domain decomposition. We consider a uniform FE mesh with homogeneous Dirichlet boundary conditions.

\begin{figure*}[!ht]
    \centering
    \includegraphics[width=\linewidth]{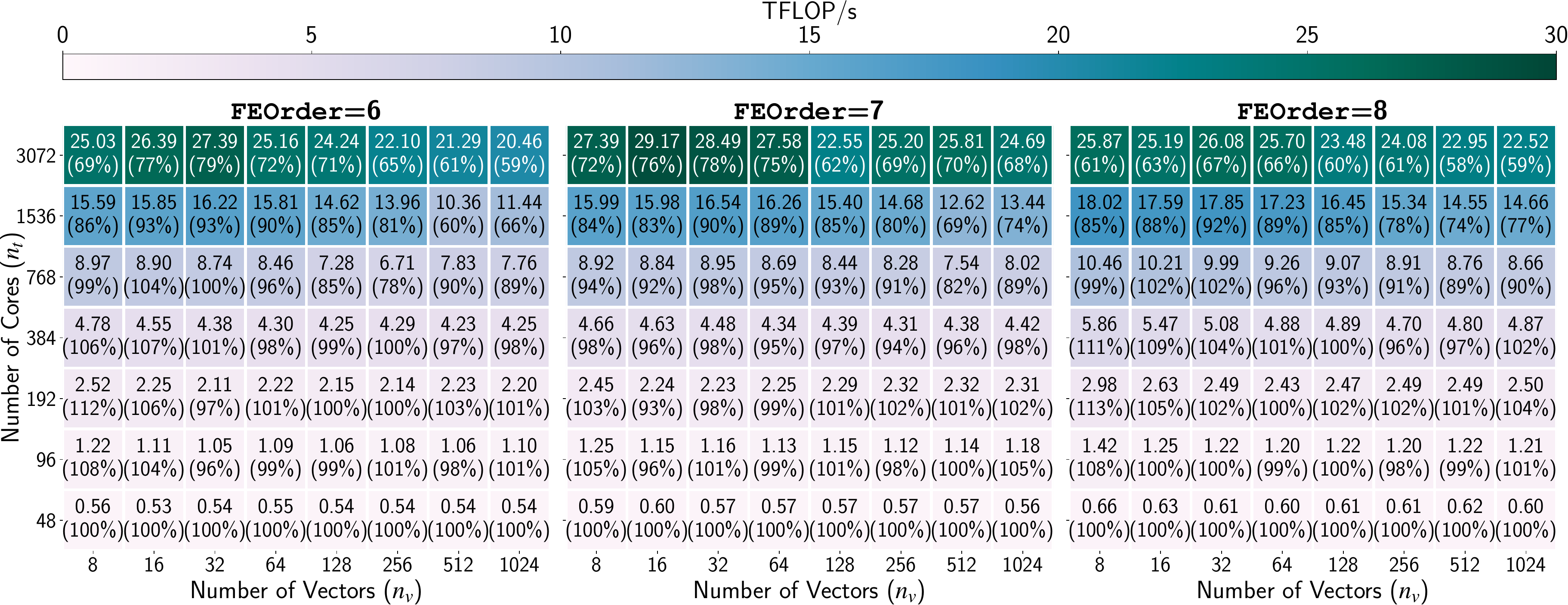}
    \caption{Scaling study of our matrix-free implementation. Case studies: 2048383 DoFs (\texttt{FEOrder}=6, 7); 2146689 DoFs (\texttt{FEOrder}=8).}\label{fig:cpuheatmap}
\end{figure*}


\subsubsection{CPU Benchmarks}\label{sec:cpubench}
We use the marker API of the LIKWID tool~\cite{Gruber2022LIKWID} with the \texttt{perf\_event} backend to obtain performance metrics on CPU architectures. To this end, we executed the MPI executable using the command :
\begin{lstlisting}[caption={\texttt{MPI} execution call},label={lst:cpusigmpi}]
likwid-mpirun -np $NTASKS -g MEM_DP -m \ $EXECUTABLE 
\end{lstlisting}

In \cref{fig:cpuheatmap}, we show the sustained performance and strong scaling efficiencies of our implementation for $\texttt{FEOrder}=6, 7, 8$ and the number of vectors $n_v=8,16,32,64,128,256,512,1024$ with $n_q=n_p$ until 3072 \texttt{MPI} tasks. This scaling study ranges from $\sim$43,000 DoFs per \texttt{MPI} task to $\sim$670 DoFs per \texttt{MPI} task in the case of \texttt{FEOrder}$=6,7$ and $\sim$45,000 DoFs per \texttt{MPI} task to $\sim$700 DoFs per \texttt{MPI} task in the case of \texttt{FEOrder}$=8$. We note that even in the extreme scaling regime of a few hundred DoFs per \texttt{MPI} task, our implementation maintains strong scaling efficiencies of about $60\%-70\%$. The high scaling efficiency observed in our experiments can be attributed to two key factors. First, compute and communication overlap, as described in Section \ref{sec:mpi}, allows for concurrent execution of computation and communication tasks, thereby minimizing idle time and maximizing resource utilization. Second, using SIMD parallelism and MPI parallelism over different indices, namely multivector batches and subdomains, further enhances the scaling efficiency. We achieve sustained performance of $24.69$ TFLOP/s for \texttt{FEOrder}=7 and $n_v$=1024, which is $\sim$26\% of the theoretical peak performance. It is imperative to note that despite achieving a lower percentage of the theoretical peak performance compared to the baseline cell-matrix implementations, the matrix-free approach, as will be demonstrated, achieves a lower time to solution. 

We benchmark our implementation against the baselines of the cell-matrix and \texttt{deal.II} matrix-free implementation. On CPU architectures, we employ the BCV layout with a batchsize of 128, which gives the best performance for the cell-matrix implementation (see \cref{sec:cpucellmatrix}). We also implement the extraction/assembly operations in the same manner as we do for the matrix-free implementation (discussed in \cref{sec:extraction}). We evaluate the FE-cell level products $\bV^{\left(e,t\right)}=\bA^{\left(e\right)}\bU^{\left(e,t\right)}$ using the \texttt{dgemm} module from Intel\textsuperscript{\textregistered} oneAPI MKL version 2023.1.0 and compute the matrix multivector products in batches of 128 vectors. Further, we note that many of the optimizations done in the case of the matrix-free implementation are not transferable to the cell-matrix implementation. For instance, the cost of construction of the intermediate data structures required for the application of constraints is prohibitive in case of the cell-matrix implementation due to the larger batchsize employed. Additionally, the overlap of MPI communication with compute across batches leads to cache pollution due to the larger batchsize and causes performance degradation. Thus, in the case of cell-matrix implementation, we utilize level-1 BLAS modules for the application of constraints and do not overlap compute and communication. Choosing a smaller batchsize to mitigate these issues leads to performance degradation in the computation of FE-cell level matrix-multivector products using \texttt{dgemm} modules due to a reduction in arithmetic intensity. 

For the second baseline, the \texttt{deal.II} matrix-free implementation, we find that the multi-component vector implementation is not very efficient when the number of components is in the order of hundreds. Instead, we compute the FE discretized matrix-multivector product using \texttt{deal.II}'s single-component matrix-free implementation by looping over the constituent vectors (see \cref{sec:dealiiCPU} for details). Note that \texttt{deal.II} also utilizes SIMD vectorization, but unlike our approach, they treat multiple FE-cells concurrently using hardware intrinsics.

\begin{figure}[!hb]
    \centering
    \includegraphics[width=\linewidth]{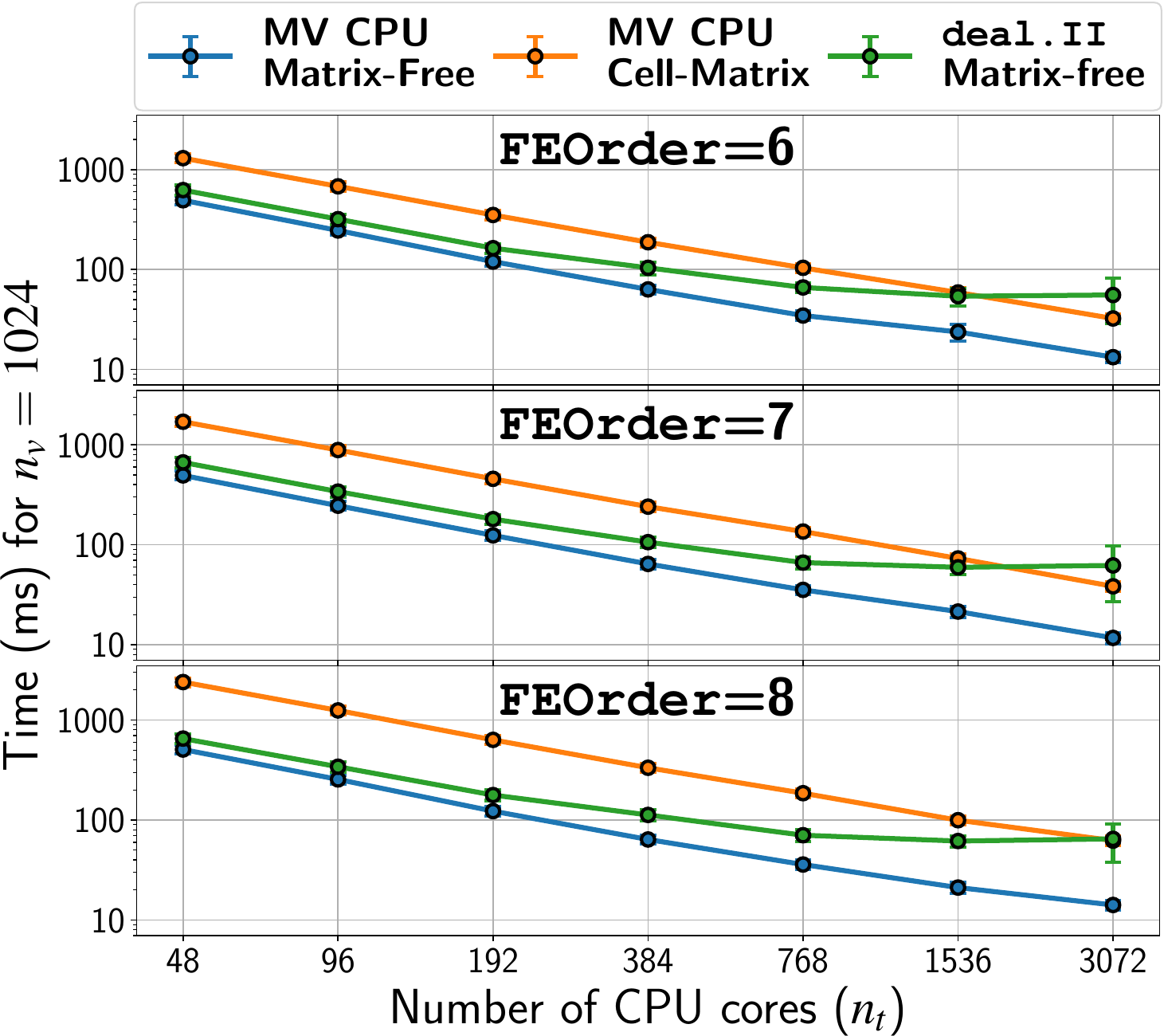}
    \caption{Comparative scaling study of our matrix-free implementation with respect to the cell-matrix method and the \texttt{deal.II} matrix-free implementation for $n_v=1024$ with $n_q=n_p$ and for uniform meshes. Case studies: 2048383 DoFs (\texttt{FEOrder}=6, 7); 2146689 DoFs (\texttt{FEOrder}=8).}\label{fig:cpumpiscaling}
\end{figure}
\begin{figure}[!ht]
    \centering
    \includegraphics[width=\linewidth]{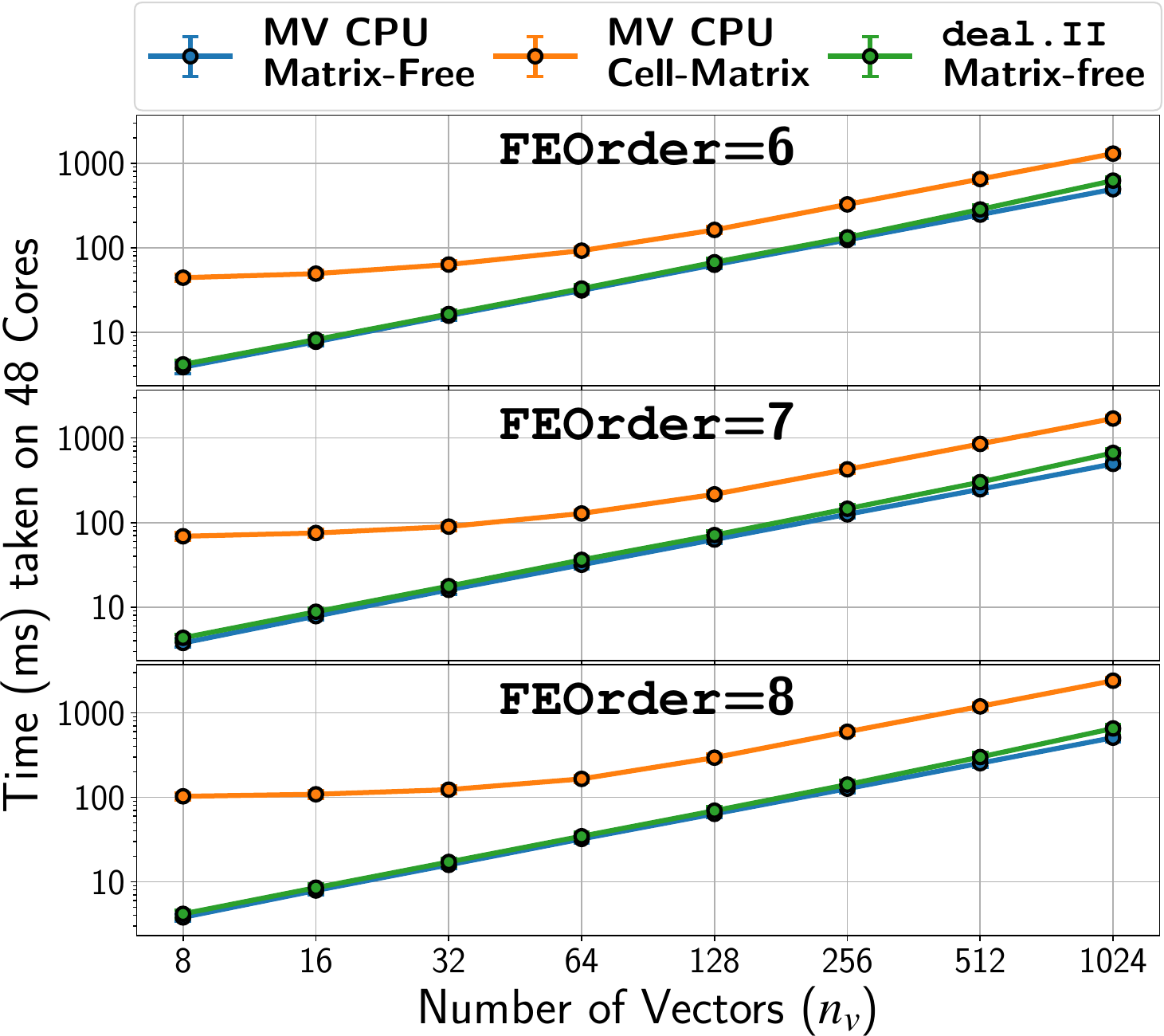}
    \caption{Performance benchmarks of our matrix-free implementation compared to the cell-matrix and \texttt{deal.II} matrix-free baseline implementations on 48 \texttt{MPI} tasks with $n_q=n_p$ and for uniform meshes. Case studies: 2048383 DoFs (\texttt{FEOrder}=6, 7); 2146689 DoFs (\texttt{FEOrder}=8).}\label{fig:cpuvecscalingn1}
\end{figure}
\begin{figure}[!hb]
    \centering
    \includegraphics[width=\linewidth]{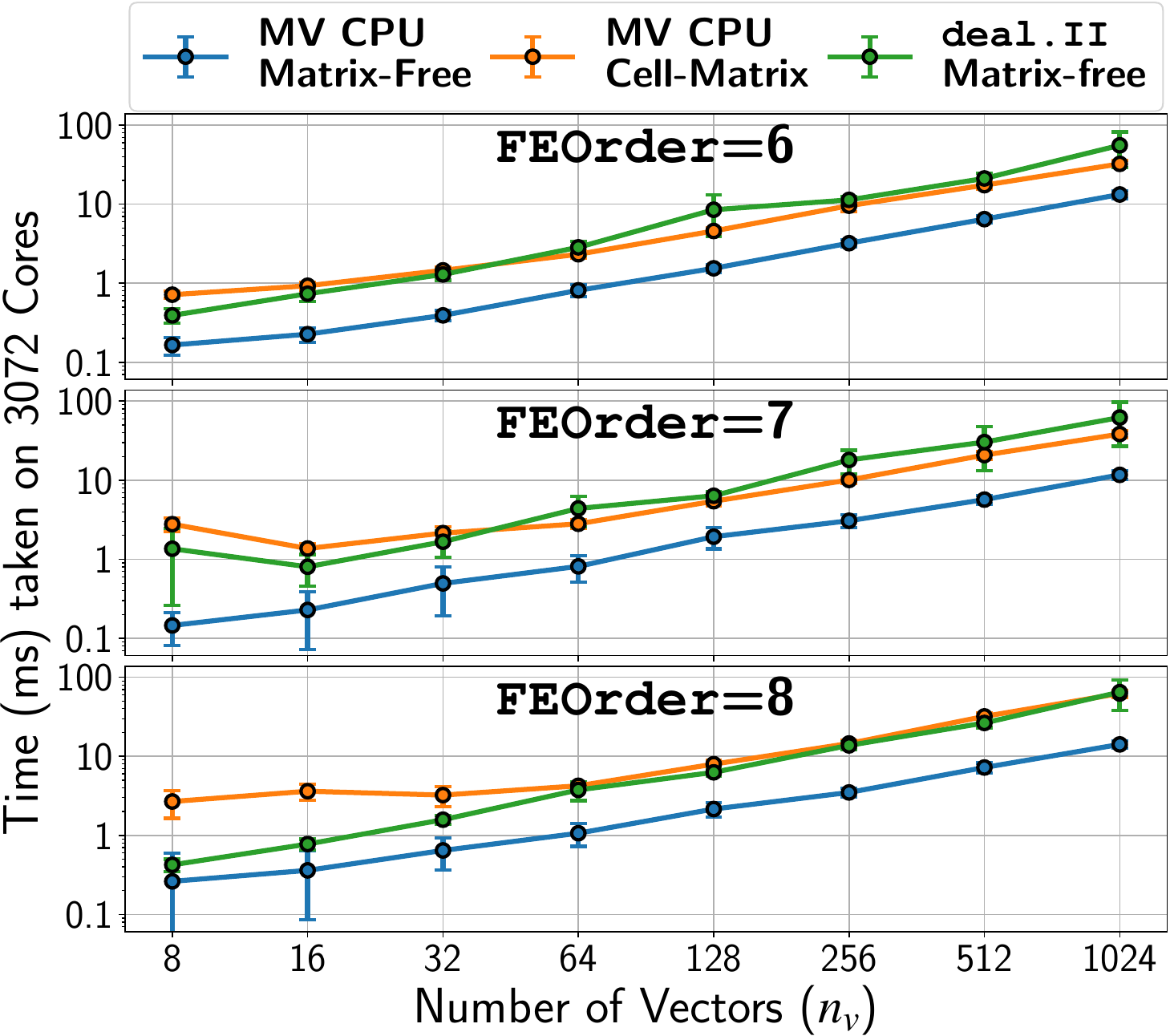}
    \caption{Performance benchmarks of our matrix-free implementation compared to the cell-matrix and \texttt{deal.II} matrix-free baseline implementations on 3072 \texttt{MPI} tasks with $n_q=n_p$ and for uniform meshes. Case studies: 2048383 DoFs (\texttt{FEOrder}=6, 7); 2146689 DoFs (\texttt{FEOrder}=8).}\label{fig:cpuvecscalingn32}
\end{figure}

In \cref{fig:cpumpiscaling}, we show the scaling data of the proposed implementation compared to the cell-matrix and the \texttt{deal.II} matrix-free implementations for $n_q=n_p$. Our implementation has a clear and noticeable performance advantage over the cell-matrix and the \texttt{deal.II} matrix-free implementations across various \texttt{MPI} tasks. The quantitative performance advantage over both the baseline implementations varies with \texttt{MPI} tasks. In particular, we show comparisons in more detail (with varying $n_v$) for 48 and 3072 \texttt{MPI} tasks in \cref{fig:cpuvecscalingn1,fig:cpuvecscalingn32} respectively.

From \cref{fig:cpuvecscalingn1}, we observe that \texttt{deal.II} matrix-free implementation is the closest competitor to our proposed approach at all values of $n_v$ in the regime of $\sim43k-45k$ DoFs per \texttt{MPI} task. Our implementation shows a performance improvement ranging from 5\% -- 35\% over the \texttt{deal.II} matrix-free implementation and achieves a speedup of 2.6x -- 27.1x over the cell-matrix implementation in this scaling regime.

\cref{fig:cpuvecscalingn32} shows benchmark comparisons in the extreme scaling regime with $\sim670-700$ DoFs per \texttt{MPI} task. To this end, we observe poor scaling behavior of the \texttt{deal.II} matrix-free implementation, attributed to the inefficient utilization of SIMD vectorization for FE cells because there are fewer FE cells per \texttt{MPI} task in this regime. However, the proposed matrix-free implementation does not suffer from this drawback. Furthermore, our implementation shows a performance improvement ranging from 2.9x -- 5.9x over the \texttt{deal.II} matrix-free implementation and 2.4x -- 4.4x over the cell-matrix implementation in this scaling regime (for $n_v\geq64$).

We further benchmark our implementation for the case $n_q>n_p$ and, to that end, choose $n_q=n_p+2$ for our investigations. We achieve a performance improvement ranging from 3\% -- 29\% over the \texttt{deal.II} matrix-free implementation, and speedups ranging from 1.8x -- 19.9x over the cell-matrix implementation in the regime of $\sim43k-45k$ DoFs per \texttt{MPI} task. On the other extreme, in the regime of $\sim670-700$ DoFs per \texttt{MPI} task, we achieve speedups ranging from 2.8x -- 7.3x over the \texttt{deal.II} matrix-free implementation and 1.2x -- 5.4x over the cell-matrix implementation (for $n_v\geq64$). A discussion of these results is provided in the Appendix (see \cref{fig:cpunq})

\subsubsection{GPU Benchmarks}

We use NVIDIA\textsuperscript{\tiny\textregistered} Tesla\textsuperscript{\tiny\textregistered} V100 SXM2 16GB GPUs, available on the Summit supercomputer, to analyze the performance of our proposed approach on multi-node GPUs. The computational times are measured using the \texttt{clock\_gettime} function with the \texttt{CLOCK\_MONOTONIC} argument as it has a nanosecond resolution. Appropriate barriers such as \texttt{MPI\_Barrier} and \texttt{cudaDeviceSynchronize} are used around the code of interest. To reduce the noise in our reported timings, the collected data is averaged over 100 repetitions.   
NVIDIA\textsuperscript{\tiny\textregistered} Nsight\textsuperscript{\texttrademark} Compute 2021.2 profiler is used to obtain the total floating point operations.       \texttt{cudaProfilerStart} and \texttt{cudaProfilerStop} are used to mark the code of interest, and the following wrapper script is used in conjunction with \texttt{mpirun} to profile:
\begin{lstlisting}[caption={Wrapper script for profiling with Nsight Compute for multi-node GPUs},label={lst:gpuncu}, basicstyle=\ttfamily\footnotesize]
metrics+="
sm__sass_thread_inst_executed_op_dadd_pred_on.sum,\ 
sm__sass_thread_inst_executed_op_dfma_pred_on.sum,\ 
sm__sass_thread_inst_executed_op_dmul_pred_on.sum"

ncu --metrics $metrics --profile-from-start off --target-processes all $EXECUTABLE
\end{lstlisting}

The \texttt{compute} kernel (\cref{lst:gpuker}) is launched with a 2-D grid of $E_t \times n_b$ thread blocks, each with a 2-D block of $n_{t_x}= b$ threads in the x-direction and $n_{t_y} = \texttt{warpSize}\times\alpha$ threads in the y-direction where \texttt{warpSize} = $32$ for NVIDIA GPUs and $\alpha$ is a tunable parameter. The optimal batchsize $b$ and parameter $\alpha$ are determined for each \texttt{FEOrder} by benchmarking for various values within the limits allowed by the GPU hardware. For instance, V100 GPUs have a default shared memory limit of 48 $kB$, which can be increased to a maximum of 96 $kB$ by the user. Hence, $b$ is limited by available shared memory. The value of $\alpha$ is limited by the maximum number of threads per thread block and the maximum number of registers per thread block. Furthermore, $n_{t_y}$ threads in each thread block are used to loop over an index of size $n_q^2$ (or $n_p^2$ or $n_pn_q$) which in turn affects the optimal values of $n_{t_y}$ to be used for each \texttt{FEOrder} (see \cref{lst:gpuCompute}). A sustained performance analysis is performed to obtain the optimal values of $n_{t_x}$ and $n_{t_y}$ (see \cref{fig:gpuBlockSizeMF}) and are tabulated in Table~\ref{tbl:batchsizeStudy}. We note that the subsequent GPU benchmarking studies in this section employ these tabulated optimal values. Furthermore, the above analysis (see \cref{fig:gpuBlockSizeMF}) indicates that our Multivector GPU Matrix-Free implementation achieves a sustained performance of $\sim$2.99 TFLOP/s on a single GPU involving 1024 vectors and $\sim$120k DoFs which is about 38\% of the peak performance of a NVIDIA\textsuperscript{\tiny\textregistered} V100 GPU.


\begin{table}[H]
\rowcolors{2}{gray!25}{white}
\begin{tabular}{M{0.189\linewidth} M{0.335\linewidth} M{0.335\linewidth}}
\rowcolor{gray!50}
\texttt{FEOrder} & \small$b$ & \small$\alpha$ \\
6 & 8 & 2  \\
7 & 4 & 2  \\
8 & 2 & 4  \\
\end{tabular}
\caption{Optimal values of $b$ and $\alpha$ for various \texttt{FEOrder} to decide the values of $n_{t_x}$ and $n_{t_y}$}\label{tbl:batchsizeStudy}
\end{table}


\begin{figure*}[!ht]
    \centering
    \includegraphics[width=\linewidth]{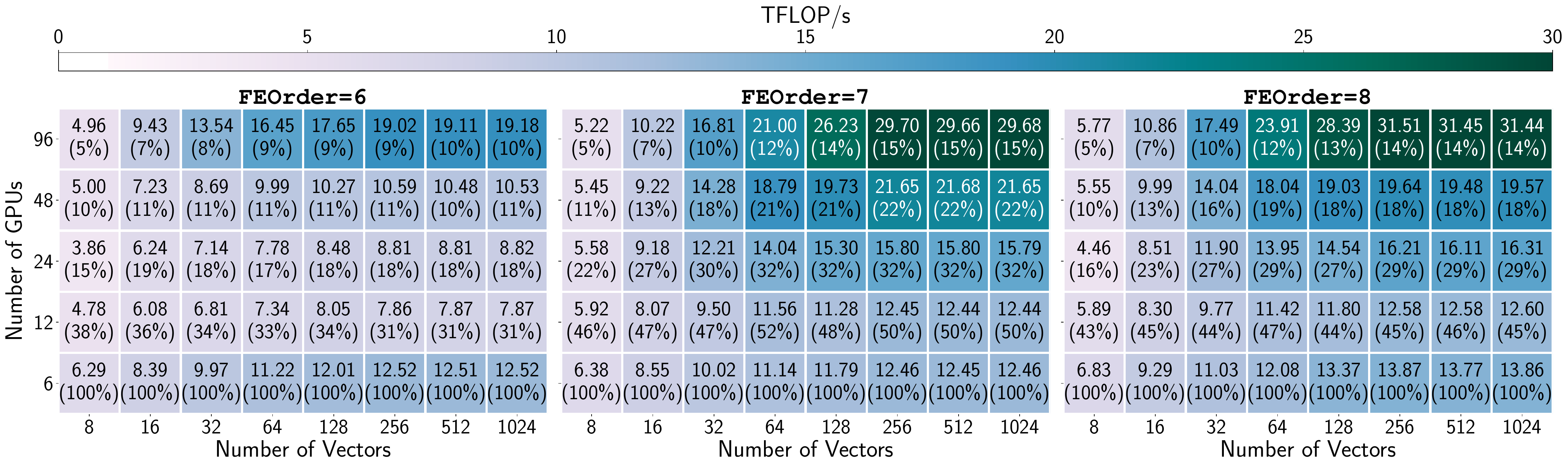}
    \caption{\small Scaling study of our matrix-free implementation on 6 to 96 V100 GPUs employing the number of vectors $n_v$ = 8, 16, 32, 64, 128, 256, 512, 1024. For a large number of vectors (512-1024), our implementation results in parallel scaling efficiencies of $\sim$30-50\% for 12 GPUs ($\sim$90k DoFs/GPU) and $\sim$10-15\% for 96 GPUs ($\sim$12k DoFs/GPU). Case studies: 1092727 DoFs (\texttt{FEOrder}=6); 1191016 DoFs (\texttt{FEOrder}=7); 1157625 DoFs (\texttt{FEOrder}=8).}\label{fig:heatGPU}
\end{figure*}

We subsequently evaluate the performance of our matrix-free implementation by conducting a strong scaling study on number of GPUs ranging from 6 to 96 employing the number of vectors $n_v$ = 8, 16, 32, 64, 128, 256, 512, 1024. \Cref{fig:heatGPU} shows the heatmap corresponding to this study. We observe that for a large number of vectors (512-1024), our implementation results in parallel scaling efficiencies of $30\%-50\%$ for 12 GPUs ($\sim$90k DoFs/GPU) and $10\%-15\%$ for 96 GPUs ($\sim$12k DoFs/GPU). The matrix-free method has a reduced arithmetic complexity compared to the cell-matrix approach; hence, inter-node communication quickly becomes the dominant cost on GPUs as the number of nodes increases, resulting in a lower scaling efficiency. It is important to note that the matrix-free multivector approach proposed here achieves lower solution times than the cell-matrix approach, as will be demonstrated, despite the lower percentage of theoretical peak performance. This can be attributed to the reduced arithmetic complexity and the proposed hardware-aware implementation strategies for the matrix-free approach minimizing the data movement costs during matrix-multivector multiplication.

We now compare our matrix-free implementation with the cell-matrix approach as a baseline (\cref{sec:cellmatrix}) for matrix-multivector products. For the cell-matrix approach, we follow the method described in \cite{Das2019FastSystem,Das2022DFT-FEDiscretization}, i.e., after extraction of the global nodal vector $\bU$ to a cell-level vector $\bU^{\left(e,t\right)}$ in device memory, we evaluate the FE-cell level products $\bV^{\left(e,t\right)}=\bA^{\left(e\right)}\bU^{\left(e,t\right)}$ using the \texttt{cublasDgemmStridedBatched} module from NVIDIA\textsuperscript{\tiny\textregistered} CUDA 11.0 and compute these dense matrix multivector products sequentially over batches with a batchsize $b = 256$ vectors in BCV layout (\cref{sec:layout}) when $n_v > 256$. This batchsize is chosen after conducting a performance study with varying batchsizes in the case of the cell-matrix approach (see \cref{fig:gpuheatmapCM}). Finally, an assembly operation is performed to build the global product vector $\bV$ as discussed in \cref{sec:cellmatrix}. Currently, state-of-the-art FE libraries such as \texttt{deal.II} do not have a multivector matrix-free implementation on GPUs. Hence, we compare our proposed matrix-free approach in the case of FE discretized matrix-single vector multiplication against \texttt{deal.II}'s single vector matrix-free implementation and observe speedups up to $\sim$17x for \texttt{FEOrder} = 6, 7 and 8 for the Helmholtz operator (see \cref{fig:dealii}). Thus, we compare our matrix-free implementation with only the cell-matrix approach for multivectors.

\begin{figure}[!htb]
    \centering
    \includegraphics[width=\linewidth]{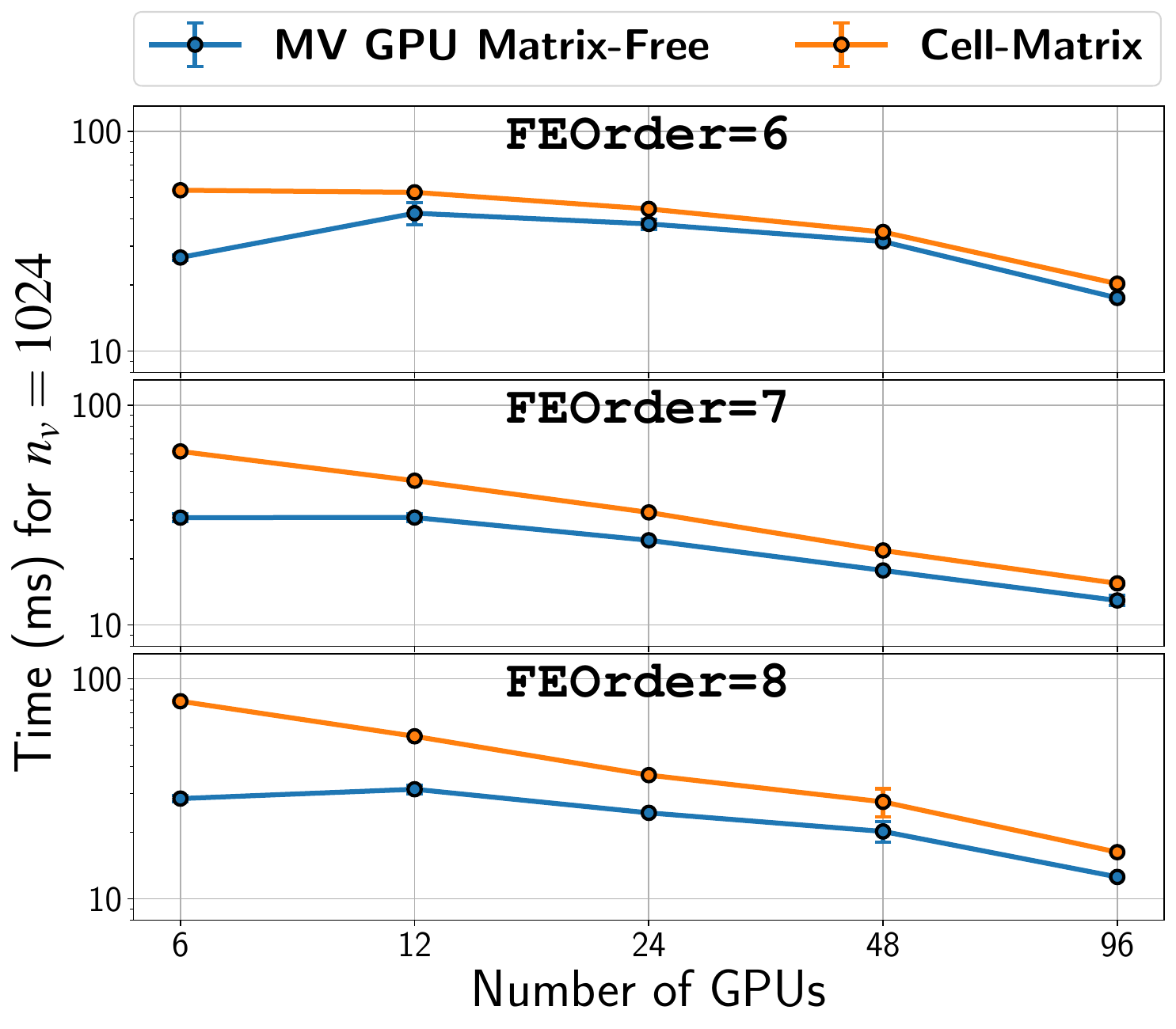}
    \caption{Scaling study comparisons of the proposed matrix-free multivector implementation with cell-matrix baseline for 1024 vectors. Case studies: 1092727 DoFs (\texttt{FEOrder}=6); 1191016 DoFs (\texttt{FEOrder}=7); 1157625 DoFs (\texttt{FEOrder}=8) for the Helmholtz problem on V100 GPUs.}\label{fig:gpuscalingvec1024}
\end{figure}

\begin{figure}[!b]
    \centering
    \includegraphics[width=\linewidth]{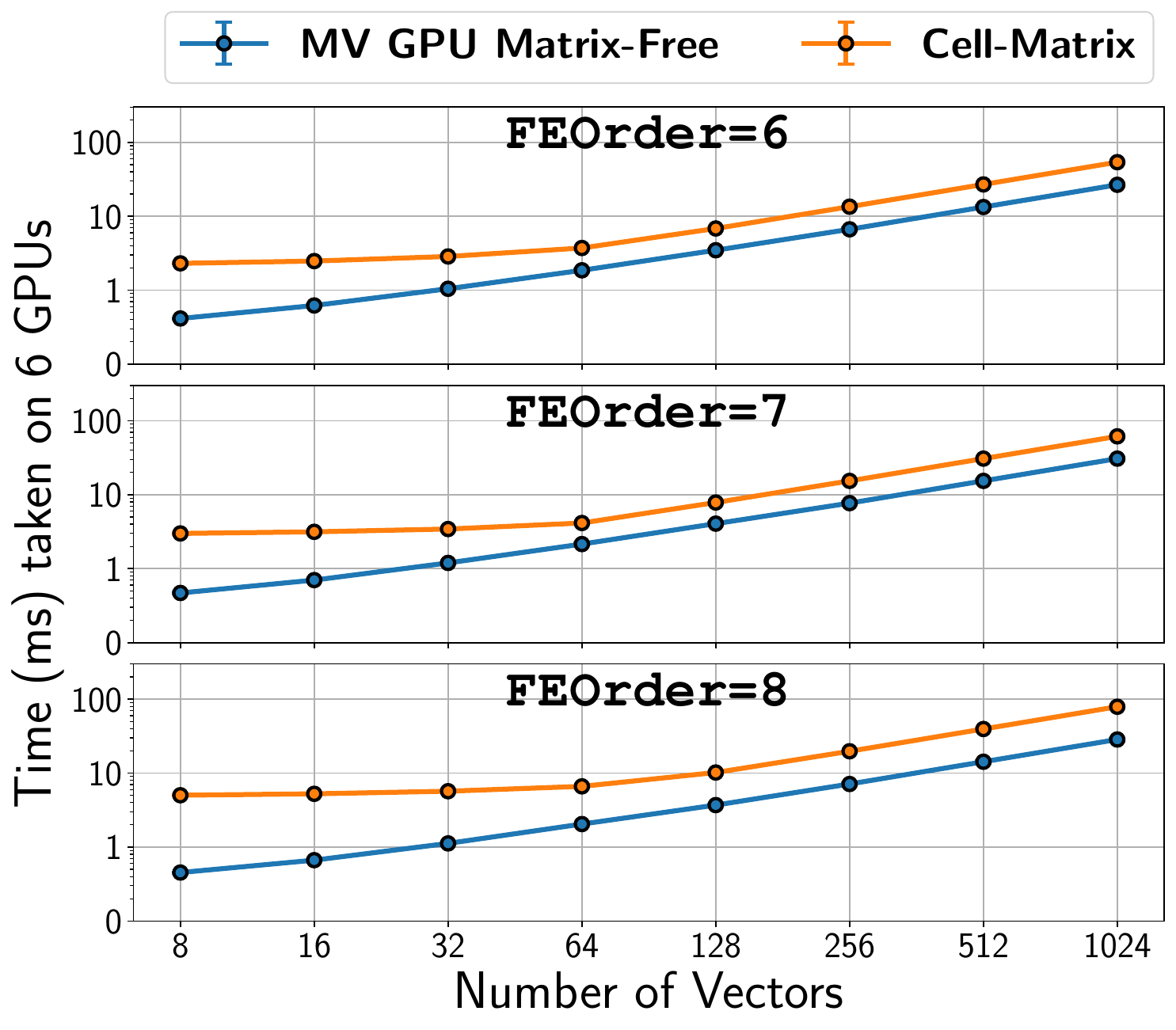}
    \caption{Performance benchmarks of our matrix-free implementation compared to the cell-matrix method on 1 node. Case studies: 1092727 DoFs (\texttt{FEOrder}=6); 1191016 DoFs (\texttt{FEOrder}=7); 1157625 DoFs (\texttt{FEOrder}=8) for the Helmholtz problem on V100 GPUs.}\label{fig:gpuscalingN1}
\end{figure}

\begin{figure}[!htb]
    \centering
    \includegraphics[width=\linewidth]{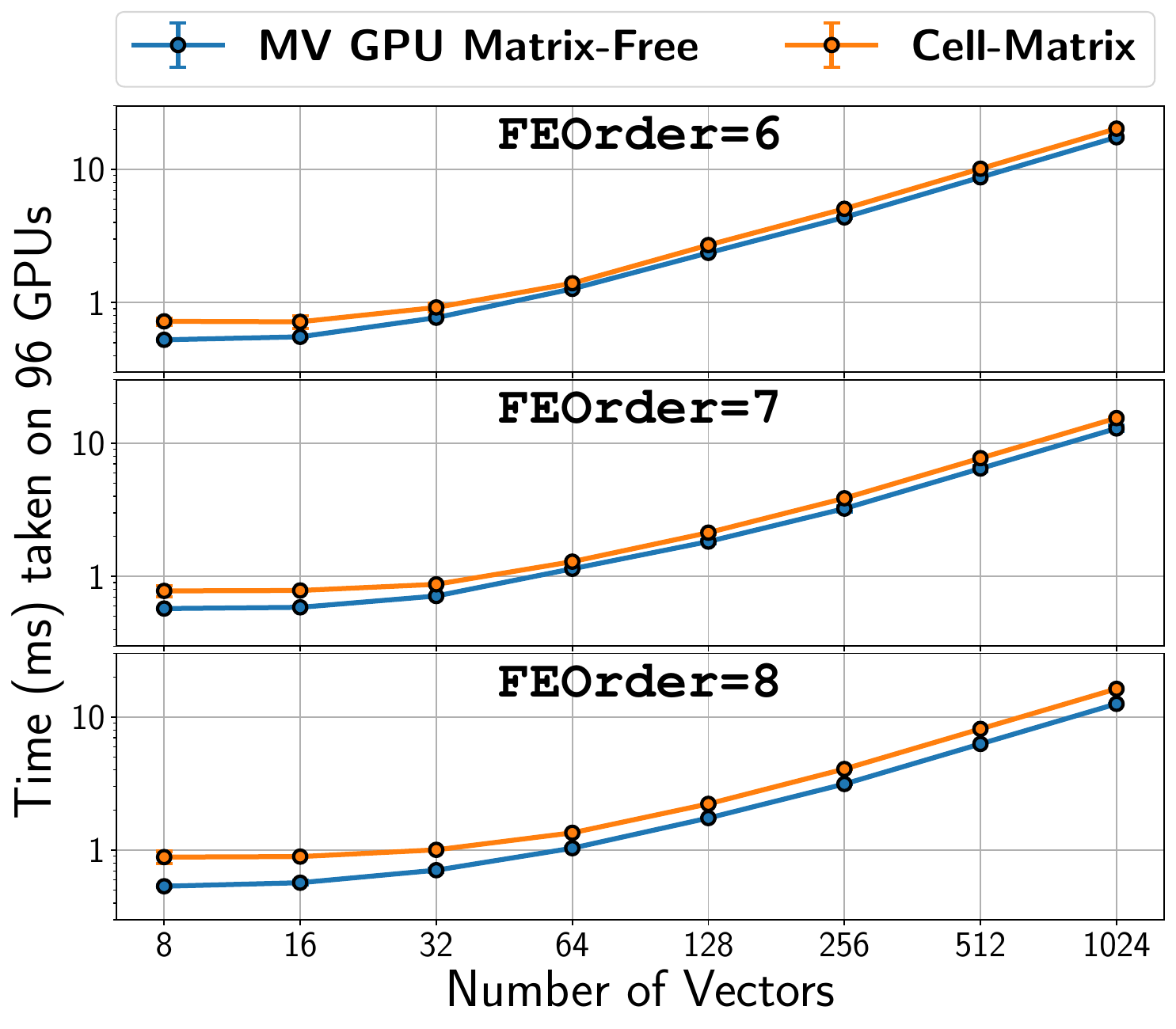}
    \caption{Performance benchmarks of our matrix-free implementation compared to the cell-matrix method on 16 nodes. Case studies: 1092727 DoFs (\texttt{FEOrder}=6); 1191016 DoFs (\texttt{FEOrder}=7); 1157625 DoFs (\texttt{FEOrder}=8) for the Helmholtz problem on V100 GPUs.}\label{fig:gpuscalingN16}
\end{figure}

To this end, \cref{fig:gpuscalingvec1024} shows a comparative strong scaling study with the two approaches for 1024 vectors on a problem involving $\sim$1.2m DoFs, and we note that GPU matrix-free implementation has a noticeable performance advantage over the cell-matrix method across all MPI tasks for \texttt{FEOrder} = 6, 7 and 8. In \cref{fig:gpuscalingvec1024}, we observe a slight increase in computational wall time in the case of matrix-free approach from one Summit node to two Summit nodes in contrast to the cell-matrix approach. We attribute this increase in time to the increased cost of inter-node communication, which is more apparent in the case of matrix-free approach because of its reduced arithmetic complexity compared with the cell-matrix approach. We also note that the timings begin to decrease with an increase in the number of nodes beyond two, since the degrees of freedom per GPU involved in communication reduce. \cref{fig:gpuscalingN1,fig:gpuscalingN16} illustrate the timing comparisons for 1 Summit node (6 GPUs) and 16 Summit nodes (96 GPUs), respectively for varying number of vectors ($n_v$).

On 1 Summit node (6 GPUs, $\sim$200k DoFs/GPU), we observe speedups of close to 2.0x for \texttt{FEOrder} = 6 and 7 and a 2.8x speedup for \texttt{FEOrder} = 8 in comparison to the cell-matrix approach for the case of 1024 vectors. In the case of 8 vectors, we observe speedups close to 6x for \texttt{FEOrder}=6 and 7 and close to 11x for \texttt{FEOrder} = 8 for 1024 vectors on 1 Summit node. 

\cref{fig:gpuscalingN16} shows the performance comparisons in the case of 16 Summit nodes (96 GPUs, $\sim$12k DoFs/GPU), and we observe computational gains of 16\%, 20\% and 30\%  for \texttt{FEOrder} = 6, 7 and 8 respectively against the cell-matrix method for 1024 vectors. On 4 Summit nodes (24 GPUs, $\sim$45k DoFs/GPU), we observe better performance improvements of up to 50\% for \texttt{FEOrder} = 8 in this case of 1024 vectors.

\subsection{Helmholtz Eigenvalue Problem}
We now present an important benchmark involving the solution of the FE discretized eigenvalue problem (EVP), leveraging the proposed matrix-free implementation to evaluate matrix multivector products arising during the course of an iterative procedure adopted to solve the EVP.  Consequently, we consider the FE discretized EVP corresponding to the Helmholtz operator, which can be expressed as follows:
\begin{equation}
    \bH\bU=\bM\bU\mathbf{\Lambda} \label{eqn:ghep}
\end{equation}
where $\bH=\bK+\bM^\kappa$ and $\bM$ is the FE basis overlap matrix (mass matrix), as discussed in \cref{eqn:matrix}. We employed the Chebyshev Filtered Subspace Iteration (ChFSI) algorithm \cite{Zhou2006Self-consistent-fieldIteration} to solve for $n_{ev}$ smallest eigenvalue/eigenvector pairs. To this end, the generalized eigenvalue problem is converted into a standard eigenvalue problem by defining $\widetilde{\bH}=\bM^{-1/2}\bH\bM^{-1/2}$ and $\widetilde{\bU}=\bM^{1/2}\bU$, resulting in
\begin{equation}
    \widetilde{\bH}\widetilde{\bU}=\widetilde{\bU}\mathbf{\Lambda} \label{eqn:shep}
\end{equation}
To efficiently compute $\bM^{-1/2}$ in \cref{eqn:shep}, the overlap integral involved in $\bM$ is evaluated using a Gauss-Lobatto-Legendre quadrature rule of order $\texttt{FEOrder}$ resulting in a diagonal matrix $\bM$ \cite{Motamarri2020DFT-FECalculations}. \Cref{alg:ChFSi} below describes the ChFSI procedure that is employed to solve the standard eigenvalue problem in \cref{eqn:shep}.
\begin{figure}[H]
    \removelatexerror
\begin{algorithm}[H]
    \caption{Chebyshev Filtered Subspace Iteration}\label{alg:ChFSi}
    \KwIn{Initial Guess of $\bU$}
    \KwData{Chebyshev polynomial order $m$, estimates of the bounds of the eigenspectrum $\lambda_{max},\lambda_{min}$, estimate of the upper bound of the wanted spectrum $\lambda_{u}$ and the tolerance for the residual $\tau$}
    \KwTemp{$e,c,\sigma,\sigma_1,\gamma,\alpha_1,\alpha_2,\bX$ and $\bY$}
    \KwResult{$\bU$ and $\mathbf{\Lambda}$}
    \While{$\norm{\widetilde{\bH}\bU-\bU\mathbf{\Lambda}}>\tau$}{$e \gets \frac{\lambda_{max}-\lambda_{u}}{2}$\;
    $c \gets \frac{\lambda_{max}+\lambda_{u}}{2}$\;
    $\sigma \gets \frac{e}{\lambda_{min}-c}$\;
    $\sigma_1 \gets \sigma$\;
    $\gamma \gets \frac{2}{\sigma_1}$\;
    $\alpha_1 \gets \frac{\sigma_1}{e}$\;
    $\alpha_2 \gets -c$\;
    $\bY \gets 0$\;
    $\bX \gets \bU$\;
    $\bY \gets \alpha_1\widetilde{H}\bX+\alpha_1\alpha_2\bX$\;
    \For{$d\gets 2$ \KwTo $m$}{
        $\sigma_2\gets \frac{1}{\gamma-\sigma}$\;
        $\alpha_1\gets \frac{2\sigma_2}{e}$\;
        $\alpha_2\gets -\sigma\sigma_2$\;
        $\bX\gets \alpha_1\widetilde{\bH}\bY+\alpha_2\bX-c\alpha_1\bY$\;
        swap$\left(\bX,\bY\right)$\;
    }
    $\bX \gets\bY$\;
    $\bX_o \gets$ orthogonalize$(\bX)$\;
    solve $\bX_o^T\widetilde{\bH}\bX_o\bQ=\bQ\mathbf{\Lambda}$\;
    $\bU\gets\bX_o\bQ$\;
    }
    \Return{$\bU$ and $\mathbf{\Lambda}$}
\end{algorithm}
\end{figure}

We use the \texttt{deal.II} library version 9.4.2~\cite{Arndt2022The9.4} with the \texttt{p4est}~\cite{Burstedde2011P4estOctrees} backend to perform MPI-parallel meshing and domain decomposition. We compute $n_{ev}=1024$ smallest eigenvalue/eigenvector pairs driving the eigenvalue problem residual to a value $\tau=5\times10^{-5}$, and we consider a buffer of 25\%, resulting in the trial subspace $\bU$ comprising of 1280 vectors to be employed in the ChFSI \cref{alg:ChFSi}. We employ our baselines described earlier to compute matrix-multivector products during the ChFSI procedure and conduct comparative studies with our proposed matrix-free implementation on both multi-node CPUs and GPUs as described subsequently.

\subsubsection{CPU Benchmarks}
\begin{figure}[H]
    \centering
    \includegraphics[width=\linewidth]{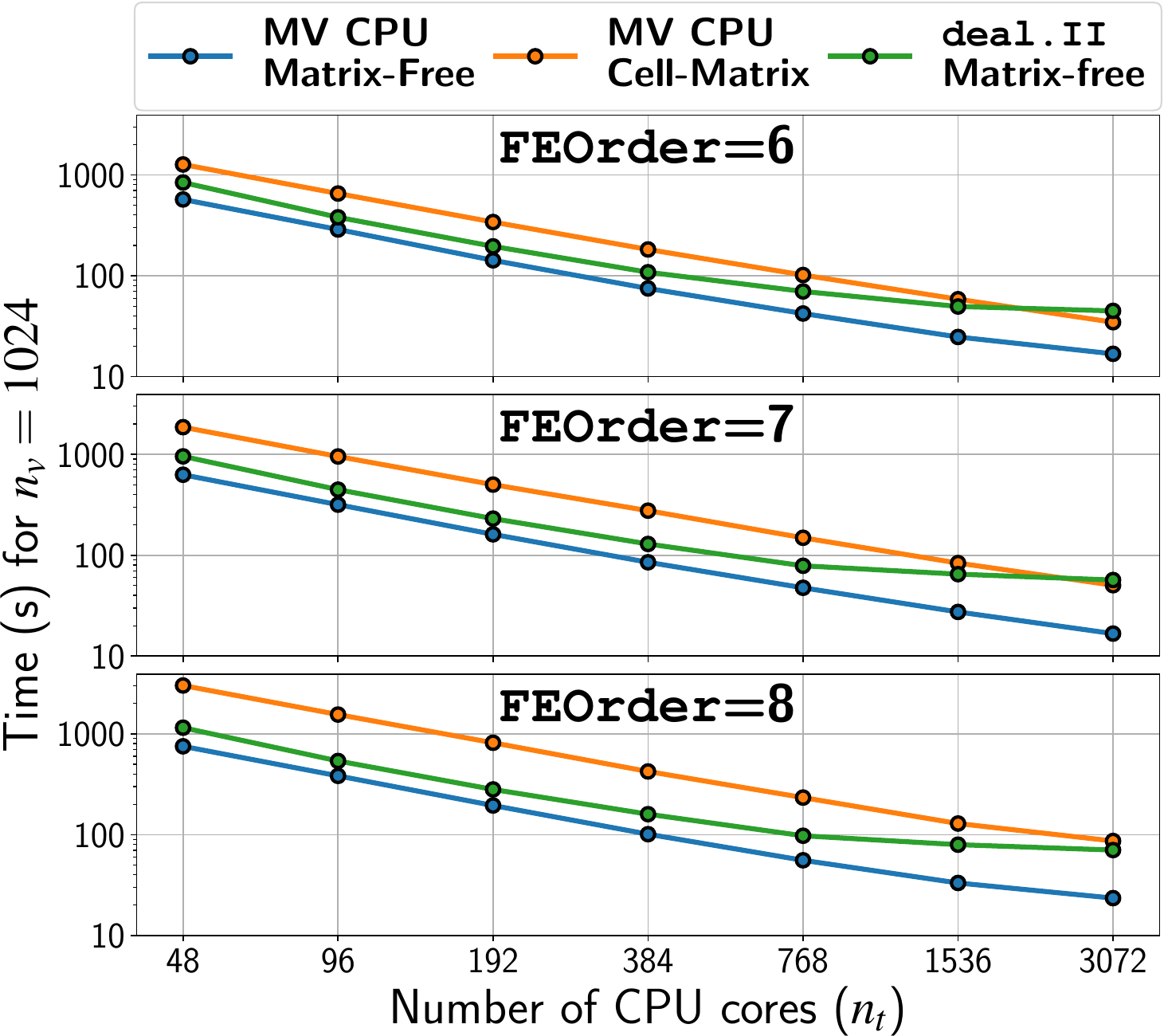}
    \caption{Performance benchmarks of our matrix-free implementation compared to the cell-matrix and \texttt{deal.II} matrix-free baseline implementations for the eigenvalue problem on uniform meshes. Case studies: 2048383 DoFs (\texttt{FEOrder}=6, 7); 2146689 DoFs (\texttt{FEOrder}=8). Chebyshev polynomial orders 67, 76 and 83 were chosen for \texttt{FEOrder}=6, 7 and 8 respectively.}\label{fig:eigenucpu}
\end{figure}

\cref{fig:eigenucpu} shows the strong scaling data of our implementation compared to the cell-matrix and the \texttt{deal.II} matrix-free implementations for uniform meshes. Our implementation has a clear and noticeable performance advantage over the cell-matrix and the \texttt{deal.II} matrix-free implementations across various \texttt{MPI} tasks. On 48 \texttt{MPI} tasks, we achieve speedups of about 2.2x, 3.0x, and 4.0x compared to the cell-matrix implementation and around 1.5x compared to the \texttt{deal.II} implementation for \texttt{FEOrder} values of 6, 7, and 8. Similarly, on 3072 \texttt{MPI} tasks, our implementation yields speedups of about 2.1x, 3.0x, and 3.7x over the cell-matrix implementation, and about 2.7x, 3.4x, and 3.0x over the \texttt{deal.II} implementation for \texttt{FEOrder} values of 6, 7, and 8, respectively.

\begin{figure}[!hb]
    \centering
    \includegraphics[width=0.7\linewidth]{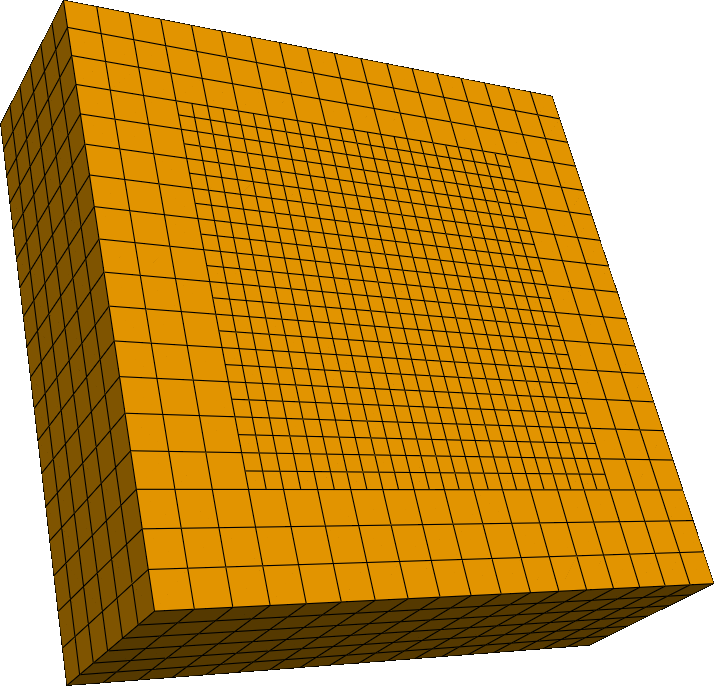}
    \caption{Example of an adaptively refined mesh with a single level of refinement}\label{fig:refmesh}
\end{figure}

We also consider the case of an adaptively refined FE mesh with hanging node constraints (see \cref{fig:refmesh}) to benchmark the performance of our matrix-free implementation within the eigenvalue solver framework using ChFSI.

\begin{figure}[!ht]
    \centering
    \includegraphics[width=\linewidth]{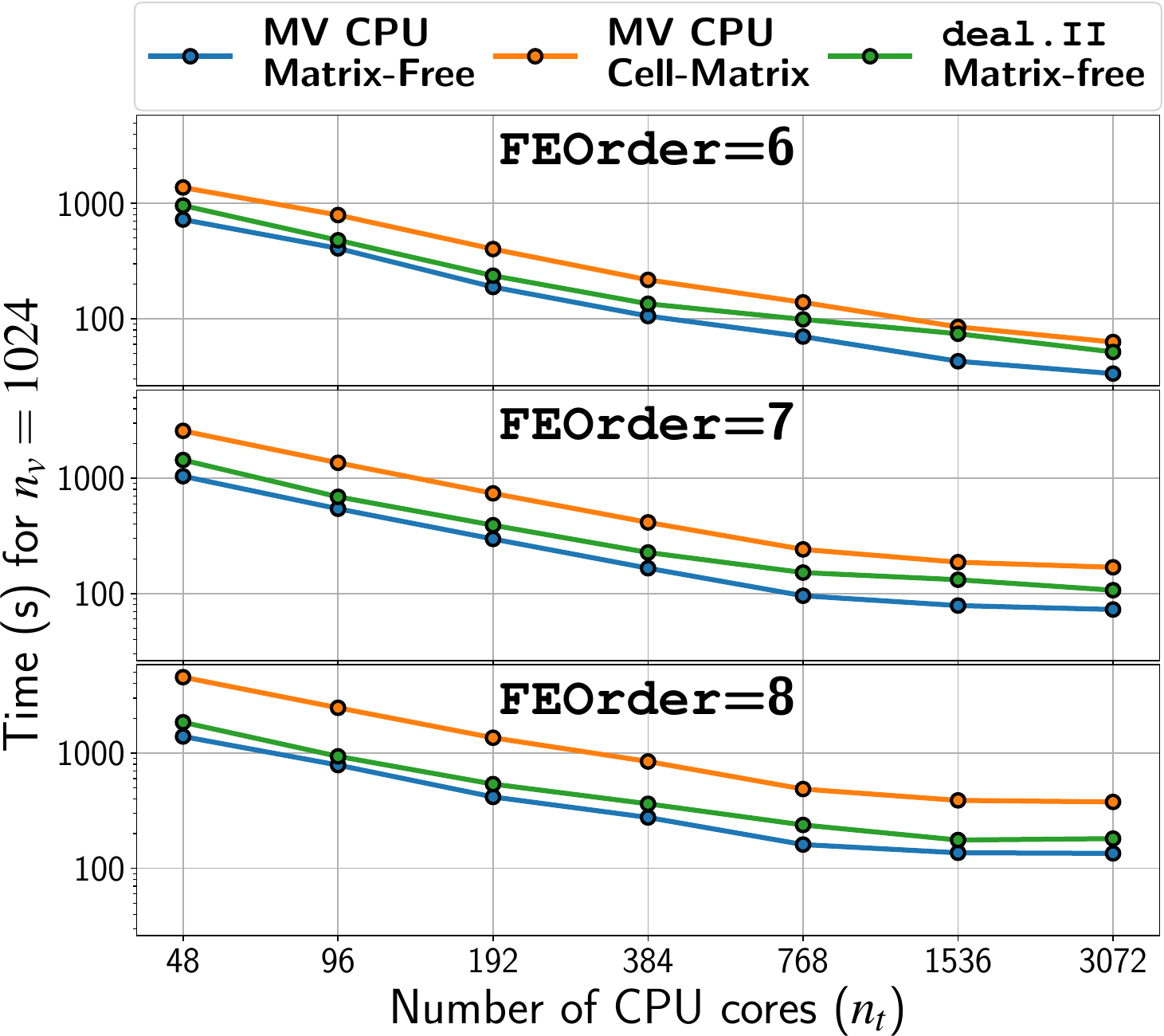}
    \caption{Performance benchmarks of our matrix-free implementation compared to the cell-matrix and \texttt{deal.II} matrix-free baseline implementations for the eigenvalue problem on adaptively refined meshes (1 level of refinement). Case studies: 2032609 DoFs (\texttt{FEOrder}=6); 2018187 DoFs (\texttt{FEOrder}=7); 2054553 DoFs (\texttt{FEOrder}=8). Chebyshev polynomial orders 67, 71 and 83 were chosen for \texttt{FEOrder}=6, 7 and 8 respectively.}\label{fig:eigencpu}
\end{figure}
The results plotted in \cref{fig:eigencpu} indicate speedups of 1.9x, 2.3x, and 2.8x, for \texttt{FEOrder}=6, 7 and 8 respectively, over the cell-matrix implementation, and 1.5x for \texttt{FEOrder}=6, 7 and 1.3x for \texttt{FEOrder}=8, over the \texttt{deal.II} implementation in the extreme-scaling regime (3072 \texttt{MPI} tasks with $\sim$700 DoFs per \texttt{MPI} task). We attribute the drop in speedups in comparison with the uniform mesh to the increase in the time taken for the application of hanging-node constraints as described in \cref{fig:constraints}.

\subsubsection{GPU Benchmarks}

\begin{figure}[!htb]
    \centering
    \includegraphics[width=\linewidth]{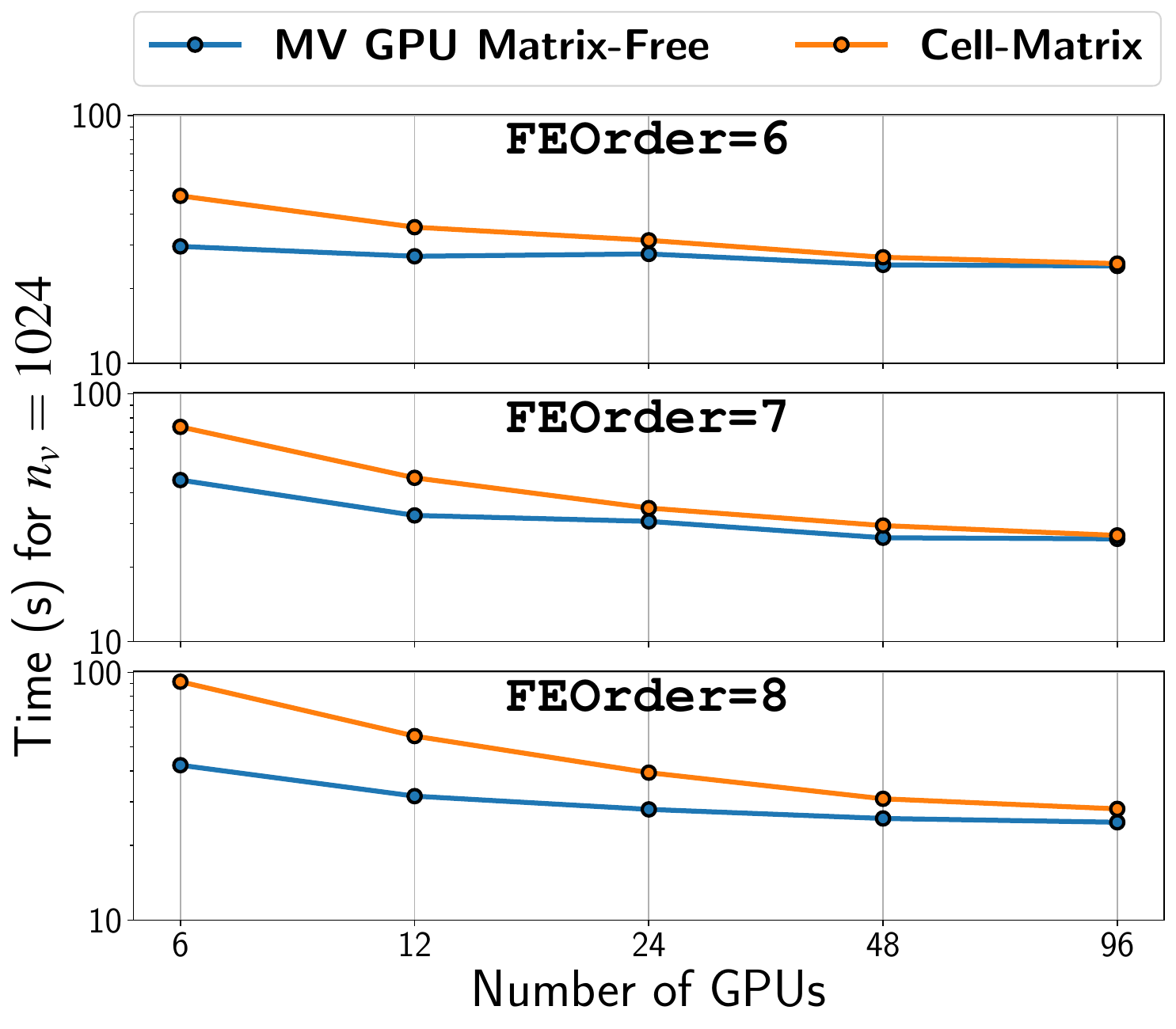}
    \caption{Performance benchmarks of our matrix-free implementation compared to the cell-matrix implementation for the eigenvalue problem on uniform meshes. Case studies: 1092727 DoFs (\texttt{FEOrder}=6); 1191016 DoFs (\texttt{FEOrder}=7); 1157625 DoFs (\texttt{FEOrder}=8). Chebyshev polynomial orders 67, 76 and 83 were chosen for \texttt{FEOrder}=6, 7 and 8 respectively.}\label{fig:eigenUgpu}
\end{figure}

\Cref{fig:eigenUgpu} shows the performance benchmark of our Multivector GPU Matrix-Free implementation compared with the cell-matrix implementation in the case of uniform meshes for the solution of the Helmholtz eigenvalue problem.  We also explore a mixed precision strategy to communicate data on the shared subdomain boundary of MPI task 't'. To this end, the boundary data communicated is recast as FP32 floats, which reduces the amount of data that needs to be communicated. The results indicate that our implementation has a clear and noticeable performance advantage over the cell-matrix implementation across varying \texttt{MPI} tasks, and this advantage improves with increase in \texttt{FEOrder}.  For instance, on one node ($\sim$200k DoFs/GPU), we obtain performance improvements of up to 60\% for \texttt{FEOrder}=6, 64\% for \texttt{FEOrder}=7 over the cell-matrix approach. Furthermore, a speedup of 2.2x is obtained in the case of \texttt{FEOrder}=8. On four nodes ($\sim$45k DoFs/GPU), we obtain performance improvements of around 14\% for \texttt{FEOrder}=6, 13\% for \texttt{FEOrder}=7, and 41\% for \texttt{FEOrder}=8  over the cell-matrix implementation. For 16 nodes ($\sim$12k DoFs/GPU), speedups of $\sim$10\% are observed for all \texttt{FEOrders}. This drop in speedup with an increase in the number of nodes can be attributed to communication costs becoming dominant compared to the cost of floating point operations due to the reduced arithmetic complexity of matrix-free multivector products. Further performance on multi-node GPUs can be obtained by overlapping compute on GPUs, MPI communication between GPUs, and data movement from device memory, which will be a part of future investigations. 
\begin{figure}[!b]
    \centering
    \includegraphics[width=\linewidth]{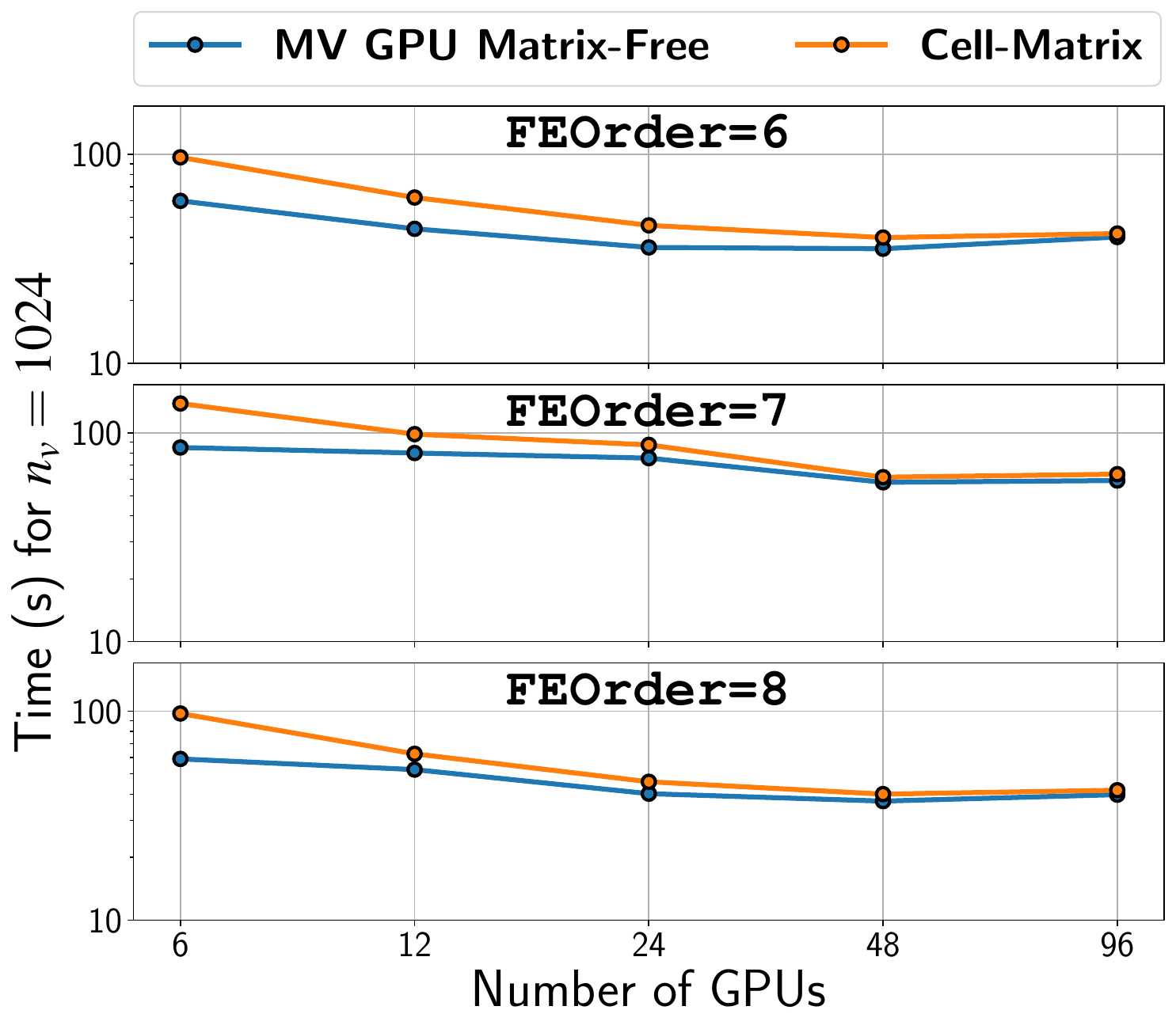}
    \caption{Performance benchmarks of our matrix-free implementation compared to the cell-matrix implementation for the eigenvalue problem on adaptively refined meshes (1 level of refinement). Case studies: 1185321 DoFs (\texttt{FEOrder}=6); 1177963 DoFs (\texttt{FEOrder}=7); 1226673 DoFs (\texttt{FEOrder}=8). Chebyshev polynomial orders 67, 76, and 83 were chosen for \texttt{FEOrder}=6, 7 and 8 respectively.}\label{fig:eigenHgpu}
\end{figure}

Similar to CPU benchmarks, we also consider the case of an adaptively refined (FE) mesh with hanging node constraints to report our performance benchmarks on GPUs. The results plotted in \cref{fig:eigenHgpu} demonstrate a performance advantage over the cell-matrix implementation across various \texttt{MPI} tasks.  On one node, we obtain a performance advantage of 62\% for \texttt{FEOrder}=6 and 7 and that of 65\% for \texttt{FEOrder}=8 over the cell-matrix implementation. For 16 nodes, a performance advantage of $\sim$10\% is observed for all \texttt{FEOrders}. These results in the case of adaptively refined mesh indicate a similar performance advantage over the cell-matrix approach as in the case of a uniform FE mesh.




\section{Conclusion and future work}\label{sec:sec5}

In conclusion, this work presents an efficient hardware-aware algorithm and implementation strategies for computing FE discretized matrix-multivector products in the matrix-free paradigm on multi-node CPU and multi-node GPU architectures. The proposed method addresses a significant gap in the currently available implementations of matrix-free methods, which are neither optimal nor directly applicable for the action of an FE discretized operator on a large number of FE discretized fields. We propose a batched layout for storing the multivector whose batchsize can be tuned to the underlying hardware architectures. Our implementation employs different batched evaluation strategies to compute the matrix-multivector products depending on the architecture to achieve the best possible performance. We also employ architecture-specific implementation strategies to evaluate the tensor contractions encountered in the matrix-free approach. For CPU architectures, we use even-odd decomposition to reduce computation, SIMD vectorization to exploit thread-level parallelism, and overlapping computation and communication to increase scaling efficiency. On GPU architectures, we employ GPU shared memory and kernel fusion for GPU architectures to reduce accesses to and from device memory and registers to reduce bank conflicts. Furthermore, we utilize constant memory on GPUs to broadcast accesses and reduce shared memory usage and bank conflicts. We also design an algorithm to overlap computation and data movement in conjunction with the proposed batched layout on GPUs. These techniques have allowed us to achieve significant performance gains. Our results indicate that this implementation outperforms the closest benchmark, achieving computational gains of 2.77x on 1 Summit node (6 GPUs, $\sim$200k DoFs/GPU), 30\% on 16 Summit nodes (96 GPUs, 12k DoFs/GPU), and 4.43x on 64 nodes of Param Pravega (3072 CPU cores, $\sim$700 DoFs/core) for matrix-multivector products (1024 vectors) for polynomial order 8. Additionally, the strong scaling studies and performance benchmarks we showed on both multi-node CPU and GPU architectures demonstrate the effectiveness of this implementation in solving large-scale problems over existing matrix-free implementations. We also demonstrated that the proposed method is particularly suitable for solving large-scale nonlinear eigenvalue problems. To this end, we have performed benchmark studies for the solution of an eigenvalue problem using the Chebyshev Filtered Subspace Iteration (ChFSI) \cite{Zhou2006Self-consistent-fieldIteration} approach and achieved speedups of 1.6x -- 2.17x on a uniform mesh for 1 Summit node (6 GPUs, $\sim$200k DoFs/GPU), 14\% -- 41\% for 4 Summit nodes (24 GPUs, $\sim$50k DoFs/GPU), $\sim$10\% on 16 Summit nodes (96 GPUs, $\sim$12k DoFs/GPU), and 2.0x -- 3.0x on 64 nodes of Param Pravega (3072 CPU cores, $\sim$700 DoFs/core) for matrix-multivector products (1024 vectors) compared to the best baseline implementation for polynomial order 6, 7, and 8. 

The methodologies discussed in this work can be straightforwardly extended to other blocked iterative eigensolvers. Furthermore, these methodologies can also be utilized for solving linear systems of equations arising from FE discretizations with multiple RHS vectors and can accelerate algorithms such as Block Krylov subspace methods~\cite{OLeary1980TheMethods} employed to solve these problems.

Although we observe significant performance improvements of our matrix-free implementation compared to the baselines, we note that the performance advantage of our implementation decreases with an increase in the number of nodes on multi-node GPU architectures. We attribute this drop in performance advantage to inter-node communication becoming the dominant cost due to the reduction in floating point operations in the matrix-free approach. This necessitates the optimization of communication involved in the action of the Helmholtz operator on the multivector to achieve further performance. Hence, in this regard, as part of future work, strategies like CUDA streams will be employed to overlap computations with communication. Extensions of the proposed algorithm to more complicated FE discretized operators such as Kohn-Sham DFT~\cite{Hohenberg1964InhomogeneousGas,Kohn1965Self-consistentEffects} Hamiltonian will be part of future investigations.


\section{Acknowledgments}
The authors gratefully acknowledge the seed grant from Indian Institute of Science (IISc) and the SERB Startup Research Grant from the Department of Science and Technology (DST), India (Grant Number:SRG/2020/002194) for the purchase of a GPU cluster, which provided computational resources for this work. The research used the resources of PARAM Pravega at the Indian Institute of Science, supported by National Supercomputing Mission (NSM) R\&D for exa-scale grant (DST/NSM/R\&D\_Exascale/2021/14.02). This research also used resources of the Oak Ridge Leadership Computing Facility (OLCF) at the Oak Ridge National Laboratory (ORNL), supported by the Office of Science of the U.S. Department of Energy (DoE) under Contract No. DE-AC05-00OR22725. We also acknowledge financial support in the form of the Prime Minister's Research Fellowship (PMRF) from the Ministry of Education (MoE), India, and the Junior Research Fellowship (JRF) from the Council of Scientific and Industrial Research (CSIR), Ministry of Science and Technology, India. We also thank Prathu Tiwari, Vinay Deshpande and Bharatkumar Sharma, all from NVIDIA, India, for fruitful discussions and for helping us run a few of our benchmarks on NVIDIA Selene.
\section{Declaration of Generative AI and AI-assisted technologies in the writing process}
During the preparation of this work the authors used Paperpal in order to proofread. After using this tool/service, the authors reviewed and edited the content as needed and take full responsibility for the content of the publication.
\onecolumn
\appendix
\textbf{\LARGE Appendix}
\section{Matrix multivector products -- CPU implementations}
\subsection{Multivector CPU matrix-free implementation}
\begin{figure}[H]
    \includegraphics[width=\linewidth]{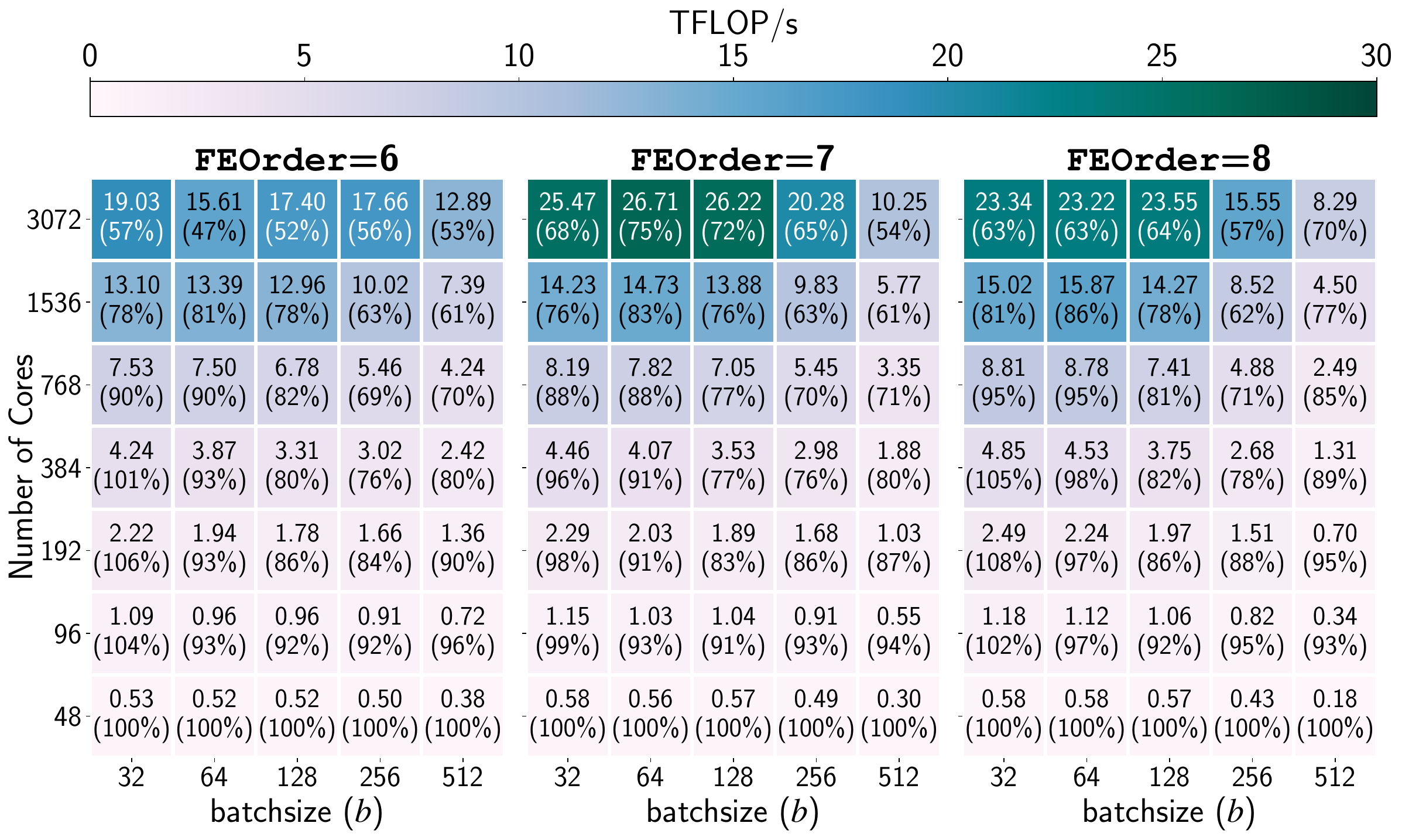}
    \caption{Performance study of our multivector CPU implementation for varying batchsizes. Case studies: 2048383 DoFs (\texttt{FEOrder}=6, 7); 2146689 DoFs (\texttt{FEOrder}=8).}\label{fig:cpuheatmapmf}
\end{figure}

We discuss the effect of varying the batchsize in our CPU multivector matrix-free implementation and its sustained performance as shown in \cref{fig:cpuheatmapmf}. We observe no appreciable gain upon increasing the batchsize to be greater than the SIMD width, and hence, the SIMD width has been chosen for all our CPU studies reported in this work. 

\subsection{Cell-matrix CPU implementation}\label{sec:cpucellmatrix}
\begin{figure}[H]
    \includegraphics[width=\linewidth]{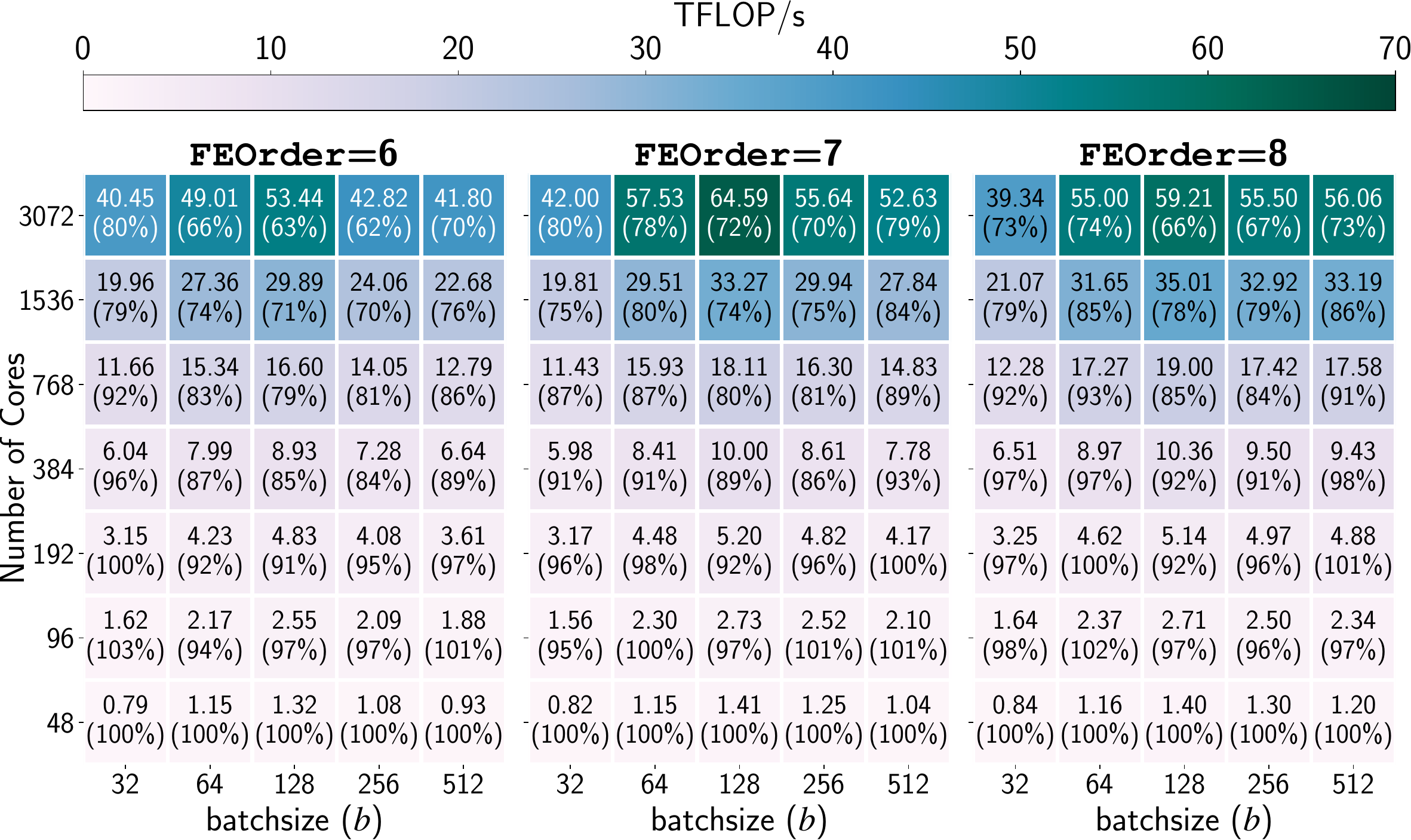}
    \caption{Performance study of our cell-matrix CPU implementation for varying batchsizes. Case studies: 2048383 DoFs (\texttt{FEOrder}=6, 7); 2146689 DoFs (\texttt{FEOrder}=8).}\label{fig:cpuheatmapc}
\end{figure}

We also vary the batchsize in our CPU multivector cell-matrix implementation, and the resulting sustained performance is shown in \cref{fig:cpuheatmapc}. We note that the best performance is obtained using a batchsize of 128 and is used as a baseline to compare our matrix-free implementations with all the results reported in this work.

\subsection{\texttt{deal.II} matrix-free CPU implementation}\label{sec:dealiiCPU}
In the case of \texttt{deal.II} matrix-free implementation, we implement the FE discretized matrix multivector product using the \texttt{deal.II}'s single-component matrix-free implementation by looping over the constituent vectors, which is equivalent to setting $b=1$ in our framework.
\lstset{style=mystyle,escapeinside={$$}}
\begin{lstlisting}[caption={Implementation of multivector using the \texttt{deal.II} matrix-free framework},label={lst:dealiimf}]
template <unsigned int FEOrder>
void
SolverProblem<FEOrder>::vmult(
    dealii::LinearAlgebra::distributed::BlockVector<double> &Ax,
    dealii::LinearAlgebra::distributed::BlockVector<double> &x)
{
    for (auto i = 0; i < d_blocksize; ++i)
        d_matrixFreeDataPtr->cell_loop(
            &SolverProblem<FEOrder>::AX,
            this,
            Ax.block(i),
            x.block(i),
            true);
}
template <unsigned int FEOrder>
void
SolverProblem<FEOrder>::AX(
  const dealii::MatrixFree<3, double>                      &matrixFreeData,
  dealii::LinearAlgebra::distributed::Vector<double>       &y,
  const dealii::LinearAlgebra::distributed::Vector<double> &x,
  const std::pair<unsigned int, unsigned int>              &cell_range) const
{
const dealii::VectorizedArray<double> tpi =
dealii::make_vectorized_array((2.0 * M_PI));
dealii::FEEvaluation<3, FEOrder, FEOrder + 3> fe_eval(
matrixFreeData,
d_matrixFreeVectorComponent,
d_matrixFreeQuadratureComponentAX);

for (unsigned int cell = cell_range.first; cell < cell_range.second; ++cell)
{
    fe_eval.reinit(cell);
    fe_eval.gather_evaluate(x,
                            dealii::EvaluationFlags::gradients |
                                dealii::EvaluationFlags::values);
    for (unsigned int q = 0; q < fe_eval.n_q_points; ++q)
    {
        fe_eval.submit_gradient(fe_eval.get_gradient(q), q);
        fe_eval.submit_value(fe_eval.get_value(q) * tpi, q);
    }
    fe_eval.integrate_scatter(dealii::EvaluationFlags::gradients |
                            dealii::EvaluationFlags::values,
                            y);
}
}
\end{lstlisting}
\lstset{style=mystyle}
\subsection{Peformance comparisons for \texorpdfstring{$n_q=n_p+2$}{nq=np+2}}

\begin{figure}[!hbt]
    \centering
    \begin{subfigure}[t]{.32\textwidth}
        \centering
        \includegraphics[width=\textwidth]{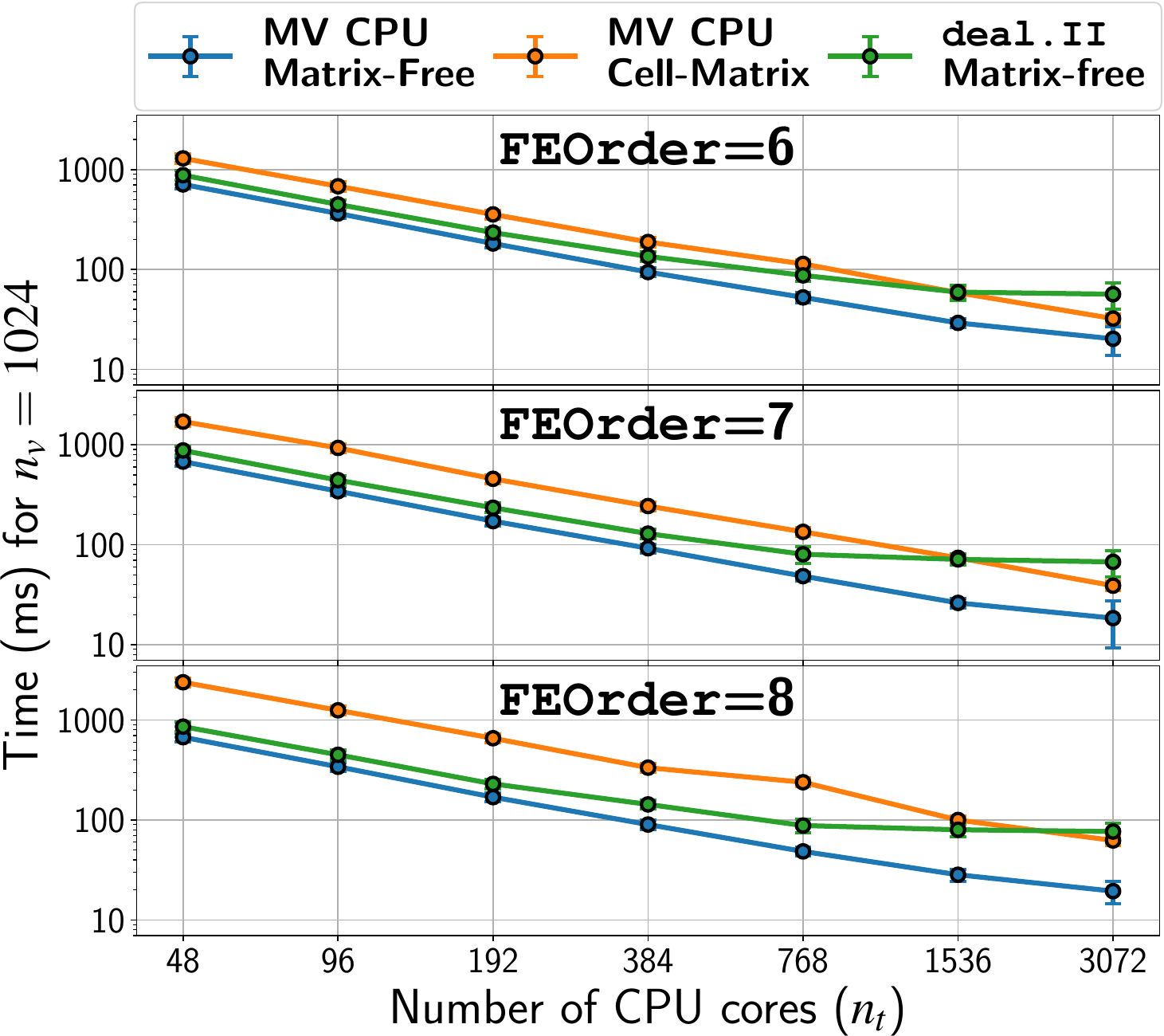}
        \caption{Comparative scaling study of our implementation with respect to the cell-matrix method and \texttt{deal.II} matrix-free implementation for $n_v=1024$.}\label{fig:cpumpiscalingq}
    \end{subfigure}\hfill
    \begin{subfigure}[t]{.32\textwidth}
        \centering
        \includegraphics[width=\textwidth]{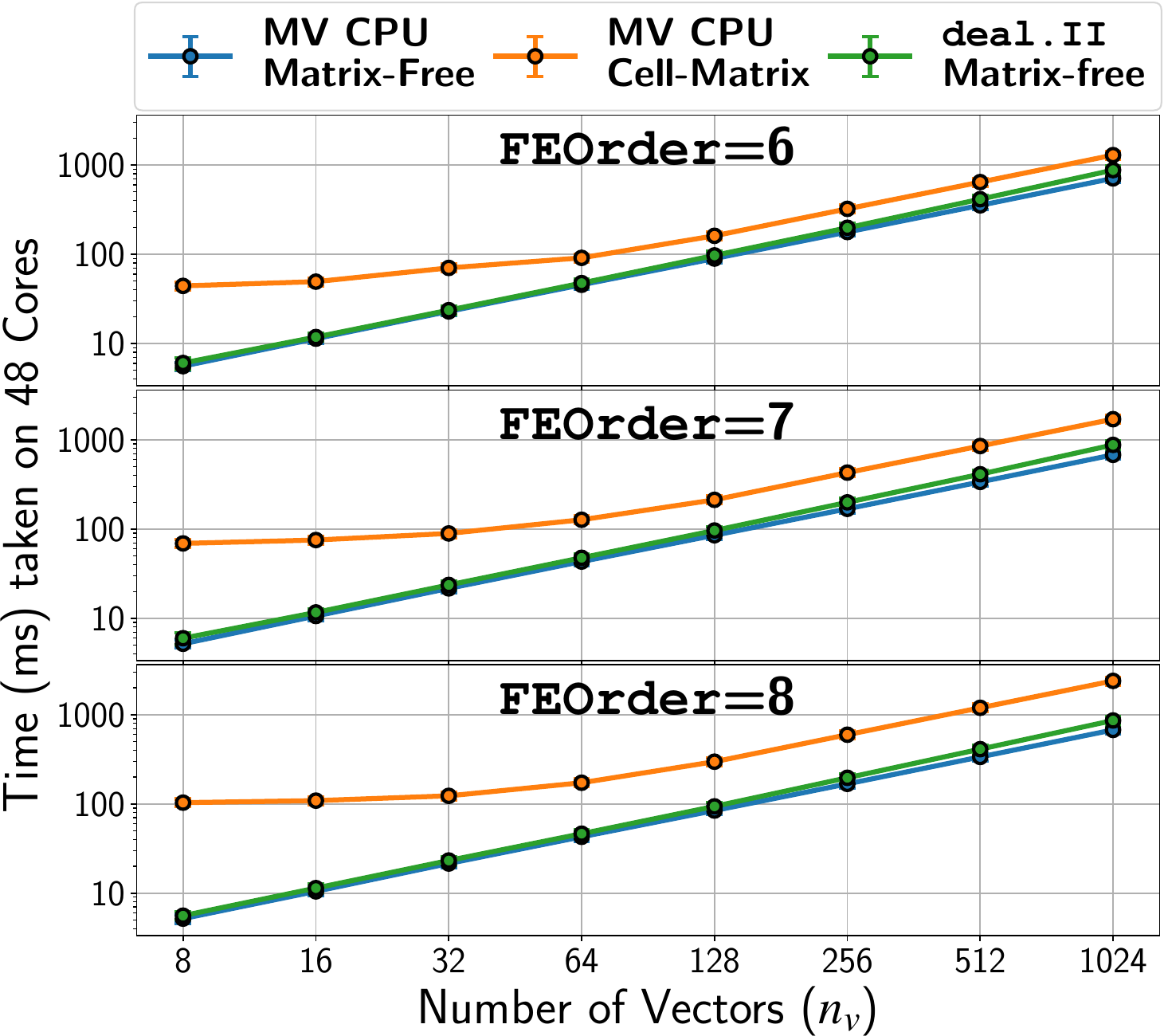}
        \caption{Performance benchmark of our implementation against the cell-matrix and \texttt{deal.II} matrix-free baseline implementations on 48 \texttt{MPI} tasks.}\label{fig:cpuvecscalingn1q}
    \end{subfigure}\hfill
    \begin{subfigure}[t]{.32\textwidth}
        \centering
        \includegraphics[width=\textwidth]{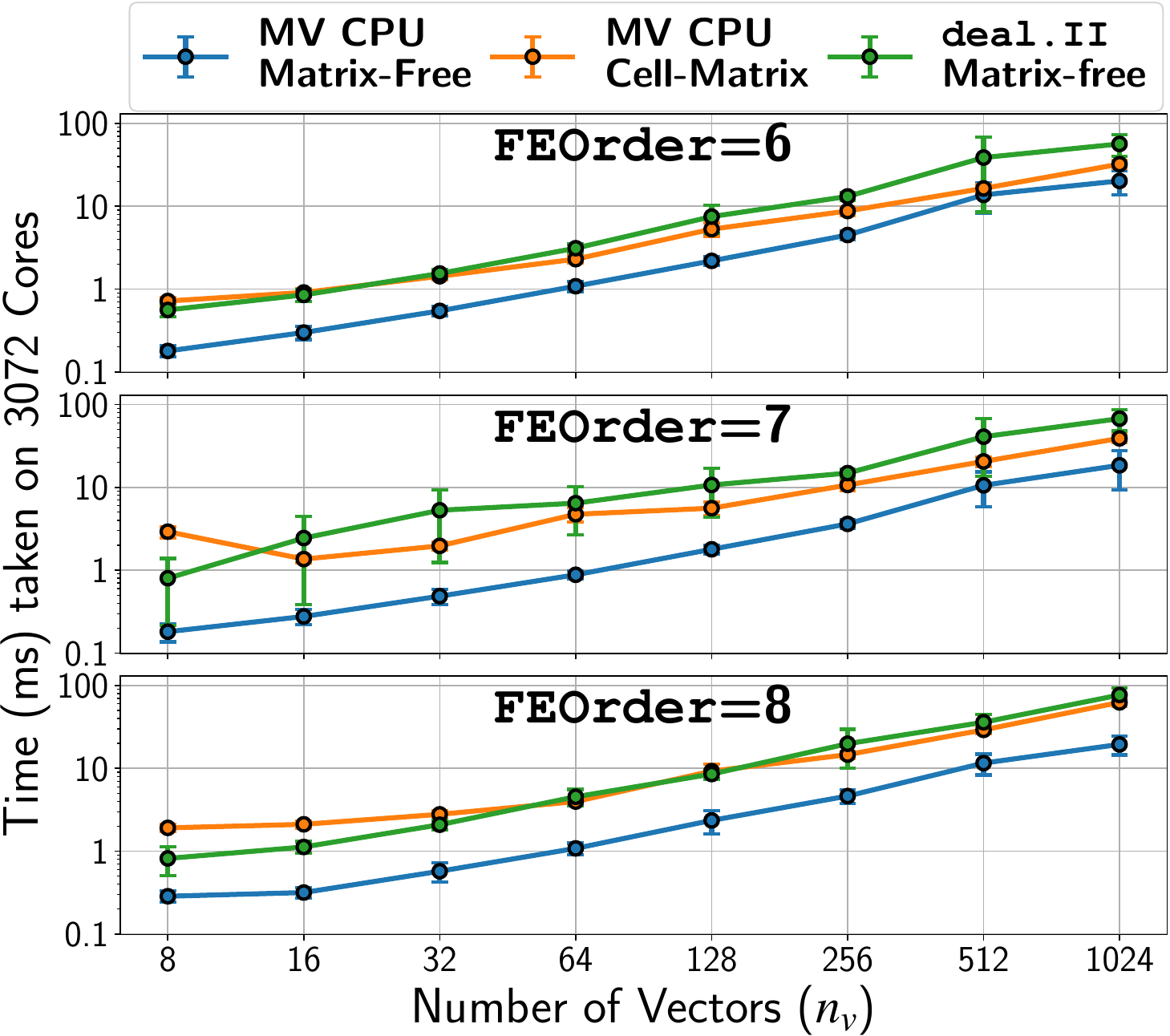}
        \caption{Performance benchmark of our implementation against the cell-matrix and \texttt{deal.II} matrix-free baseline implementations on 3072 \texttt{MPI} tasks.}\label{fig:cpuvecscalingn32q}
    \end{subfigure}%
    \caption{Benchmarks of our implementation with cell-matrix and \texttt{deal.II} matrix-free implementations for the case of $n_q=n_p+2$ and uniform mesh. Case studies: 2048383 DoFs (\texttt{FEOrder}=6, 7); 2146689 DoFs (\texttt{FEOrder}=8).}\label{fig:cpunq}
\end{figure}

In this section, we discuss the comparative studies conducted in the case of $n_q=n_p+2$.  To this end, \cref{fig:cpumpiscalingq} shows the scaling data of our implementation compared with that of the cell-matrix and the \texttt{deal.II} matrix-free implementations. Our implementation has a clear and noticeable performance advantage over the cell-matrix and the \texttt{deal.II} matrix-free implementations across various \texttt{MPI} tasks. We show the comparisons in more detail (with varying $n_v$) for 48 and 3072 \texttt{MPI} tasks in \cref{fig:cpuvecscalingn1q,fig:cpuvecscalingn32q} respectively. From \cref{fig:cpuvecscalingn1q}, we see that the closest competitor to our implementation at every value of $n_v$ in the regime of $\sim 43k-45k$ DoFs per core is the \texttt{deal.II} matrix-free implementation. We note that our implementation shows a performance improvement ranging from 1.03x to 1.29x over the \texttt{deal.II} matrix-free implementation and 1.80x to 19.94x over the cell-matrix implementation in this scaling regime. However, as shown in \cref{fig:cpuvecscalingn32q}, we see that the closest competitor to our implementation in the regime of $\sim 670-700$ DoFs per MPI task is not well-defined in the case of $n_q=n_p+2$ just as in the case of $n_q=n_p$ reported in \cref{sec:cpubench}. Our implementation shows a performance improvement ranging from 2.79x to 7.32x over the \texttt{deal.II} matrix-free implementation and 1.20x to 5.37x over the cell-matrix implementation in this scaling regime (for $n_v\geq64$).

\section{Eigensolver implementations using ChFSI on CPUs}
\subsection{Multivector matrix-free CPU implementation}\label{sec:cpuchfsi}
To implement matrix-free multivector products in CPUs, the core compute kernel was modified to compute

\begin{align}
    \bY=a\bM^{-1/2}\bH\bM^{-1/2}\bX+b\bX+c\bY
\end{align}

Scaling of the data structures $\bX$ and $\bY$ with the diagonal matrix $\bM^{-1/2}$ and scalar constants $a,b$ and $c$ is performed during extraction and assembly, as this allows us to reuse cached data more often as opposed to scaling $\bX$ and $\bY$ in entirety before/after the extraction/assembly. The constraint matrices are modified appropriately to allow for the computation of $\bM^{-1/2}\bX$ at the cell-level.

\begin{lstlisting}[caption={Extraction of cell-level multivector combined with the action of $\bM^{-1/2}$ for the cell indexed by \texttt{iCell} and batch indexed by \texttt{iBatch}},label={lst:m1b2extraction}]
for (unsigned int iDoF = 0; iDoF < d_ndofsPerCell; ++iDoF)
{
    unsigned int l2g =
    singleVectorGlobalToLocalMap[iDoF + d_ndofsPerCell * iCell];
    temp10v[iDoF] =
    x[getMultivectorIndex(l2g, iBatch)] *
    d_invSqrtElementalMassVector[iDoF + d_ndofsPerCell * iCell];
}
\end{lstlisting}
\begin{lstlisting}[caption={Assembly from cell-level multivector combined with the action of $\bM^{-1/2}$ and scaling with $a,b$ and $c$ for the cell indexed by \texttt{iCell} and batch indexed by \texttt{iBatch}},label={lst:m1b2assembly}]
for (auto i = 0; i < d_ndofsPerCell; ++i)
{
    unsigned int l2g =
    singleVectorGlobalToLocalMap[i + d_ndofsPerCell * iCell];
    if (dofEncountered[l2g])
    y[getMultivectorIndex(l2g, iBatch)] +=
        a * (temp10v[i] *
        d_invSqrtElementalMassVector[i + d_ndofsPerCell * iCell]);
    else
    {
        dofEncountered[l2g] = true;
        if (isConstrained[l2g] || l2g >= d_nLocalDofs)
        y[getMultivectorIndex(l2g, iBatch)] =
            scalar1 *
            (temp10v[i] *
            d_invSqrtElementalMassVector[i +
                                        d_ndofsPerCell * iCell]);
        else
        y[getMultivectorIndex(l2g, iBatch)] =
            scalar1 *
            (temp10v[i] *
                d_invSqrtElementalMassVector[i + d_ndofsPerCell *
                                                iCell]) +
            c * y[getMultivectorIndex(l2g, iBatch)] +
            b * x[getMultivectorIndex(l2g, iBatch)];
    }
}
\end{lstlisting}

\subsection{Cell-matrix CPU implementation}
The extraction and assembly operations in the cell-matrix implementation are also modified to account for the scaling of the data structures $\bX$ and $\bY$ with the diagonal matrix $\bM^{-1/2}$ and scalar constants $a,b$ and $c$ using a methodology similar to that used for the matrix-free implementation.

\subsection{\texttt{deal.II} matrix-free CPU implementation}
For the \texttt{deal.II} implementation we utilize the \emph{pre-} and \emph{post-} operations as described in \citet{Kronbichler2022EnhancingImplementations}.
\begin{lstlisting}[caption={Implementaion of the $\bY=a\bM^{-1/2}\bH\bM^{-1/2}\bX+b\bX+c\bY$ operation using \texttt{deal.II} matrix-free},label={lst:dealiiEigen}]
const double ratio = c / a;
for (auto i = 0; i < numberWaveFunctions; ++i)
    {
    const auto &pre = [&](const unsigned int start_range,
                            const unsigned int end_range) {
        for (int j = start_range; j < end_range; ++j)
        {
            x.block(i).local_element(j) *=
            d_invSqrtMassVector.local_element(j);
        }
        for (int j = start_range; j < end_range; ++j)
        {
            y.block(i).local_element(j) *=
            ratio * d_sqrtMassVector.local_element(j);
        }
    };
    const auto &post = [&](const unsigned int start_range,
                            const unsigned int end_range) {
        for (int j = start_range; j < end_range; ++j)
        {
            y.block(i).local_element(j) *=
            a * d_invSqrtMassVector.local_element(j);
        }
        for (int j = start_range; j < end_range; ++j)
        {
            x.block(i).local_element(j) *=
            d_sqrtMassVector.local_element(j);
            y.block(i).local_element(j) +=
            b * x.block(i).local_element(j);
        }
    };

    dftPtr->matrix_free_data.cell_loop(
        &kohnShamDFTOperatorClass<FEOrder, FEOrderElectro>::
        computeLocalHamiltonianTimesXdealii,
        this,
        y.block(i),
        x.block(i),
        pre,
        post
        );
    }
\end{lstlisting}

\section{Matrix multivector products -- GPU implementations}
\subsection{Multivector matrix-free GPU implementation}

\begin{figure}[!ht]
    \includegraphics[width=\linewidth]{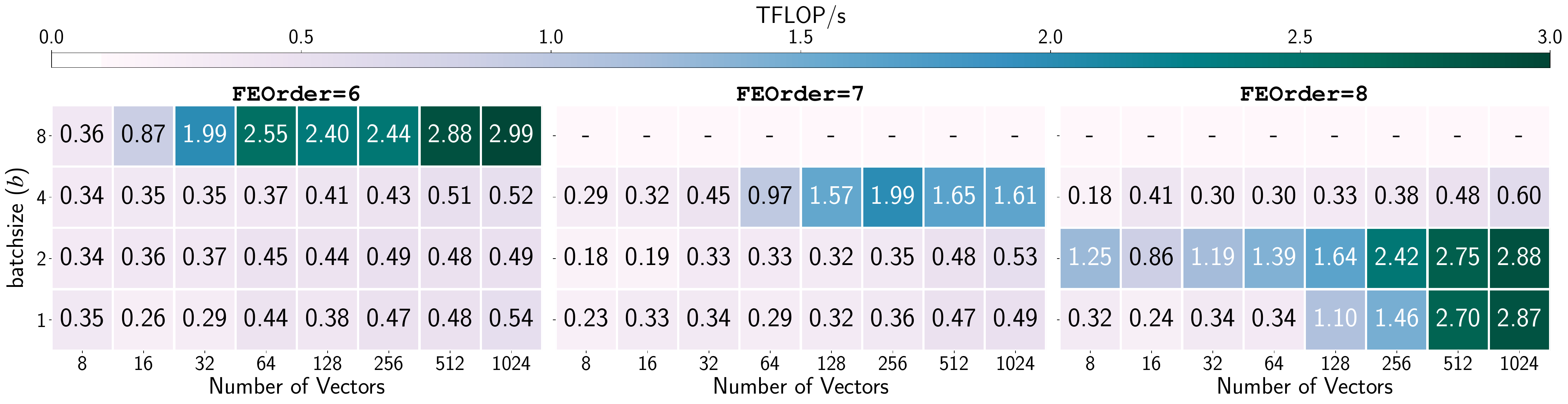}
    \caption{Performance study of our multivector matrix-free GPU implementation for varying batchsizes on a NVIDIA\textsuperscript{\tiny\textregistered} Tesla\textsuperscript{\tiny\textregistered} V100 SXM2 16GB (Summit Supercomputer). Case studies: 117649 DoFs (\texttt{FEOrder}=6 and 8); 125000 DoFs (\texttt{FEOrder}=7).}\label{fig:gpuBlockSizeMF}
\end{figure}

A sustained performance analysis for the multivector matrix-free GPU implementation on an NVIDIA\textsuperscript{\tiny\textregistered} Tesla\textsuperscript{\tiny\textregistered} V100 SXM2 16GB varying the total number of vectors $n_v$ and the batchsize $b$ (or $n_{t_x}$) shows the optimal batchsize for each \texttt{FEOrder}. We observe that $b = 8$ for \texttt{FEOrder}=6, $b = 4$ for \texttt{FEOrder}=7 and $b = 2$ for \texttt{FEOrder}=8 exhibits the best performance. Note that for \texttt{FEOrder}=7 and 8, batchsize $b = 8$ is represented with ``-'' in the above table as the kernel launch fails due to exceeding the maximum dynamic shared memory of V100 GPU. The threads launched in the y-direction $n_{t_y}$ for each thread block are used to loop over an index of size $n_q^2$ (or $n_p^2$ or $n_pn_q$) which in turn affects the optimal values of $n_{t_y}$ for each \texttt{FEOrder}. The optimal value of $n_{t_y}$ is determined by varying $n_{t_y}$ in multiples of \texttt{warpSize} (32 for NVIDIA GPUs), and the optimal values are found to be as follows: $n_{t_y}$ = 64 for \texttt{FEOrder}=6 and 7, and $n_{t_y}$ = 128 for \texttt{FEOrder}=8. These values of $n_{t_x}$ and  $n_{t_y}$ are used to launch the GPU kernel (see \cref{lst:gpuker}).

\subsection{Comparison between matrix-free GPU implementation, cell-matrix and \texttt{deal.II}'s matrix-free implementations for single-vector}
\begin{figure}[!ht]
    \centering
    \includegraphics[scale=0.3]{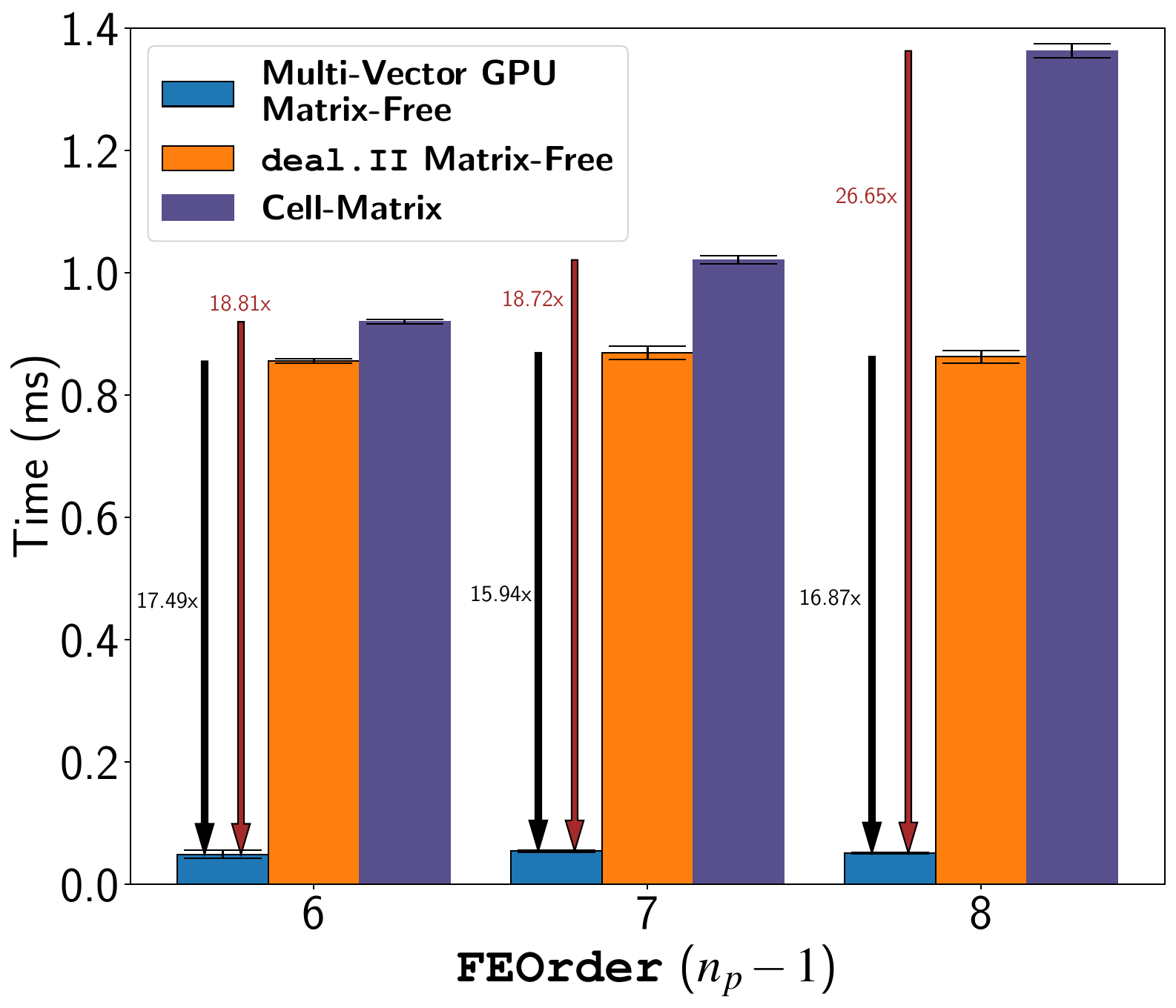}
    \caption{Comparison of our single-vector matrix-free implementation against \texttt{deal.II}'s matrix-free method and the cell-matrix method on a NVIDIA\textsuperscript{\tiny\textregistered} Tesla\textsuperscript{\tiny\textregistered} V100 SXM2 16GB (Summit Supercomputer). Case studies: 117649 DoFs (\texttt{FEOrder}=6 and 8); 125000 DoFs (\texttt{FEOrder}=7).}\label{fig:dealii}
\end{figure}

We note that \texttt{deal.II} does not have a multivector matrix-free implementation on GPUs; hence, we compare our single vector matrix-free implementation against \texttt{deal.II}'s single vector matrix-free implementation and results are illustrated in \cref{fig:dealii}. Speedups of about 16x-18x are observed for our single vector matrix-free implementation for the Helmholtz operator compared to deal.II's matrix-free baseline on a V100 GPU. Our implementation results in even larger speedups of about 19x-25x compared with the cell-matrix approach for a single vector on a V100 GPU.

\subsection{Cell-matrix GPU implementation}

\begin{figure}[H]
    \includegraphics[width=\linewidth]{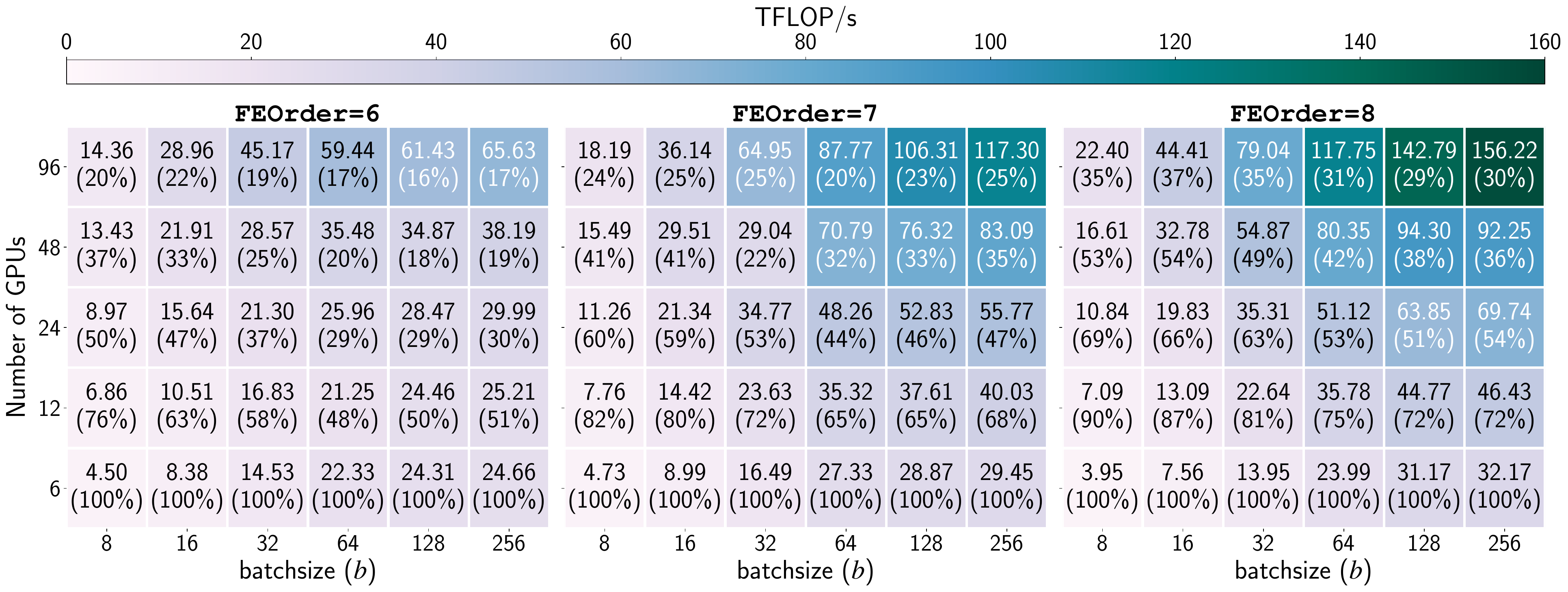}
    \caption{Performance study of the cell-matrix GPU implementation for various batchsizes on 1 to 16 nodes of Summit supercomputer. Case studies: 1092727 DoFs (\texttt{FEOrder}=6); 1191016 DoFs (\texttt{FEOrder}=7); 1157625 DoFs (\texttt{FEOrder}=8) for the Helmholtz problem and $n_v = 1024$ on GPUs.}\label{fig:gpuheatmapCM}
\end{figure}

We adopt the BCV layout in the cell-matrix implementation to compute the Helmholtz operator action on a total number of vectors $n_v = 1024$. To this end, a performance study is conducted where the Helmholtz action is evaluated sequentially over batches with varying batchsizes $b = 8, 16, 32, 64, 128, 256$ on 1 to 16 GPU nodes (Summit supercomputer). The resulting sustained performance is shown in \cref{fig:gpuheatmapCM}. We note that the time taken for computing the Helmholtz operator action on multivectors with $n_v = 1024$, does not vary appreciably from batchsize $b = 128$ to $b = 256$.  We choose $b = 256$ as the batchsize for performing all the benchmark studies since it gives the best sustained performance out of other batchsizes considered in the study.

\subsection{Performance comparisons for \texorpdfstring{$n_q=n_p+2$}{nq=np+2}}

\begin{figure}[!ht]
    \centering
    \begin{subfigure}[t]{.32\textwidth}
        \centering
        \includegraphics[width=\textwidth]{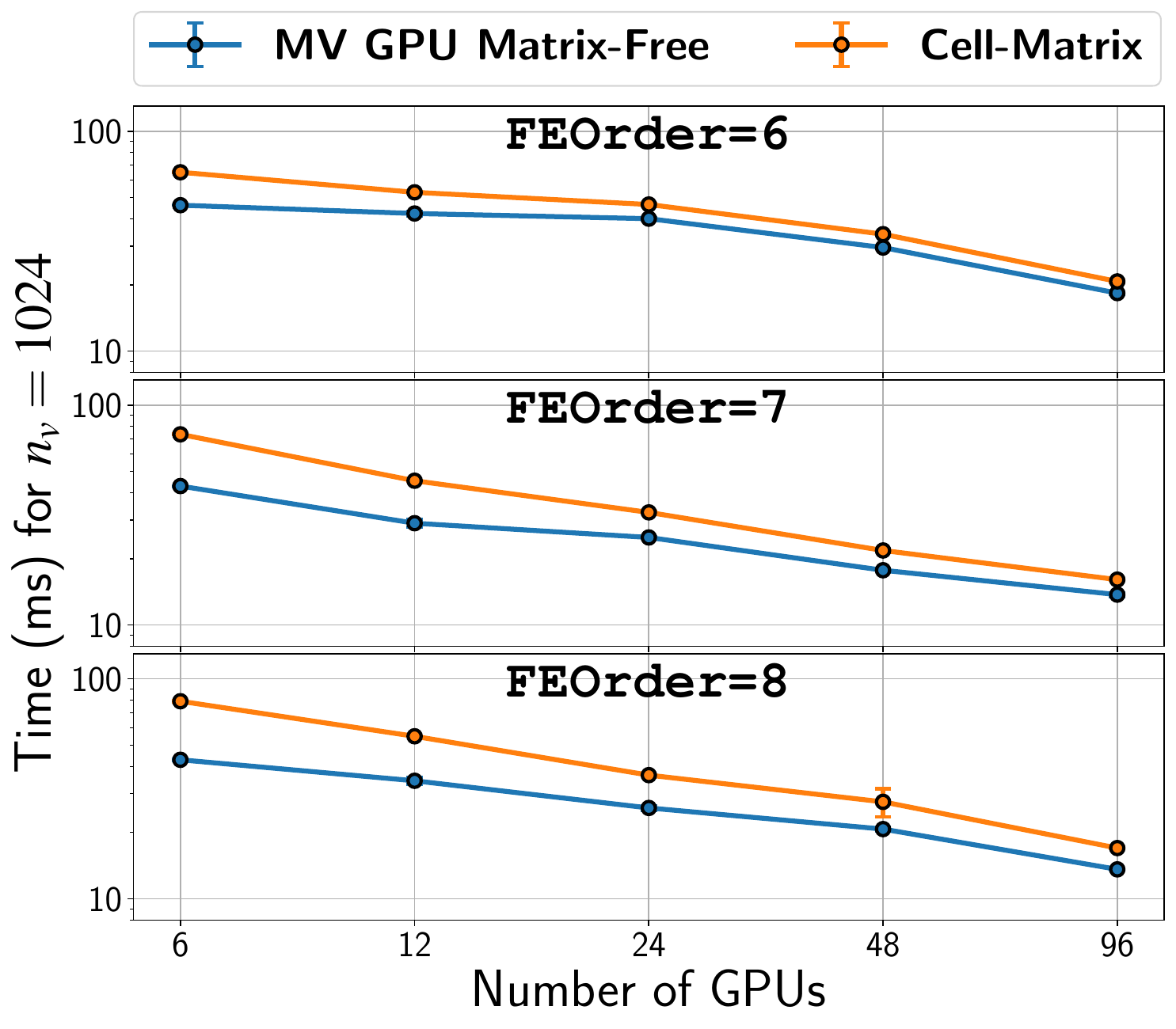}
        \caption{Comparative scaling study of our implementation against the cell-matrix method for 1024 vectors.}
        \label{fig:gpuscalingvec1024pq}
    \end{subfigure}\hfill
    \begin{subfigure}[t]{.32\textwidth}
        \centering
        \includegraphics[width=\textwidth]{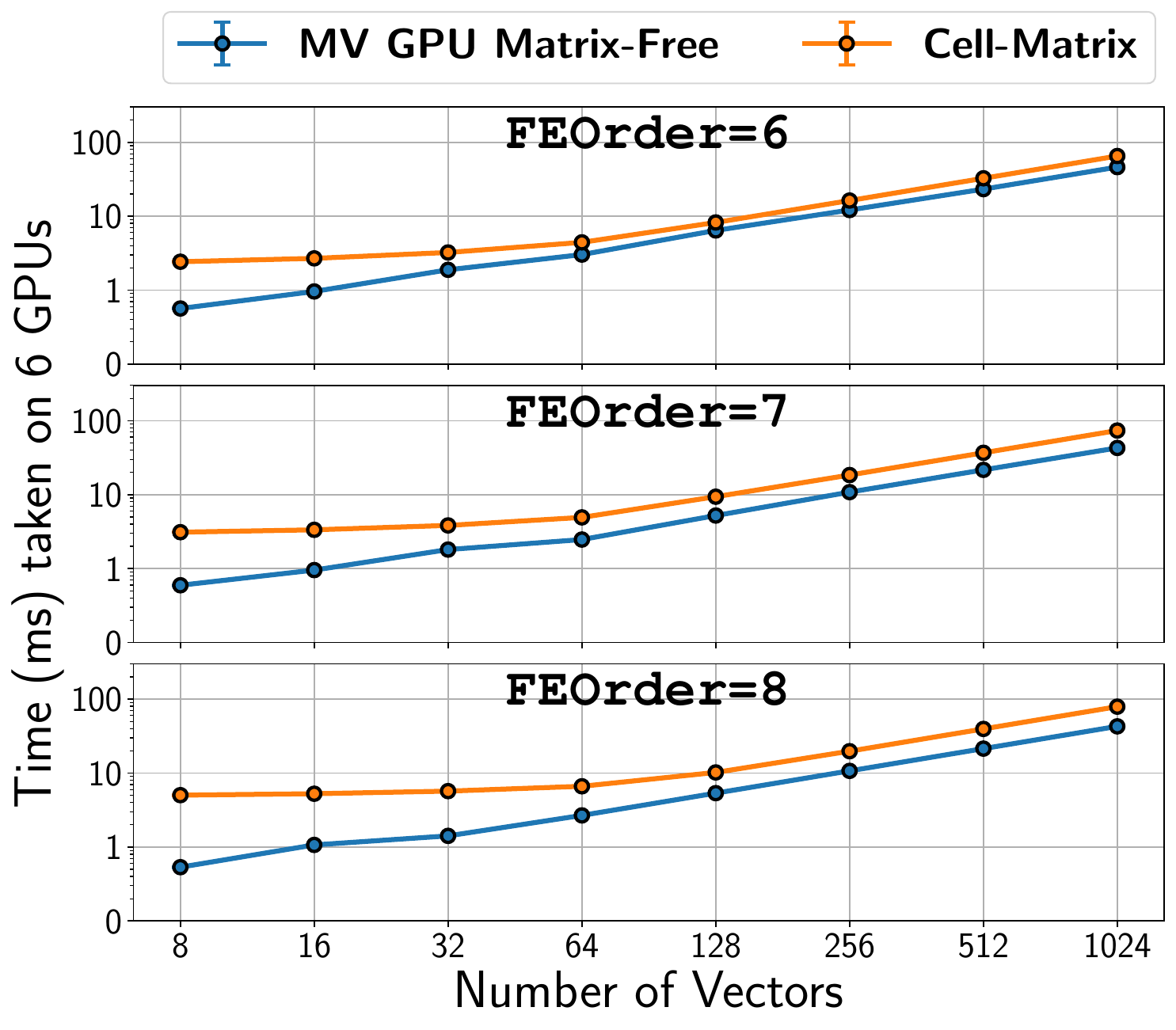}
       \caption{Performance benchmark of our implementation against the cell-matrix method on 1 node.}
       \label{fig:gpuscalingN1pq}
    \end{subfigure}\hfill
    \begin{subfigure}[t]{.32\textwidth}
        \centering
        \includegraphics[width=\textwidth]{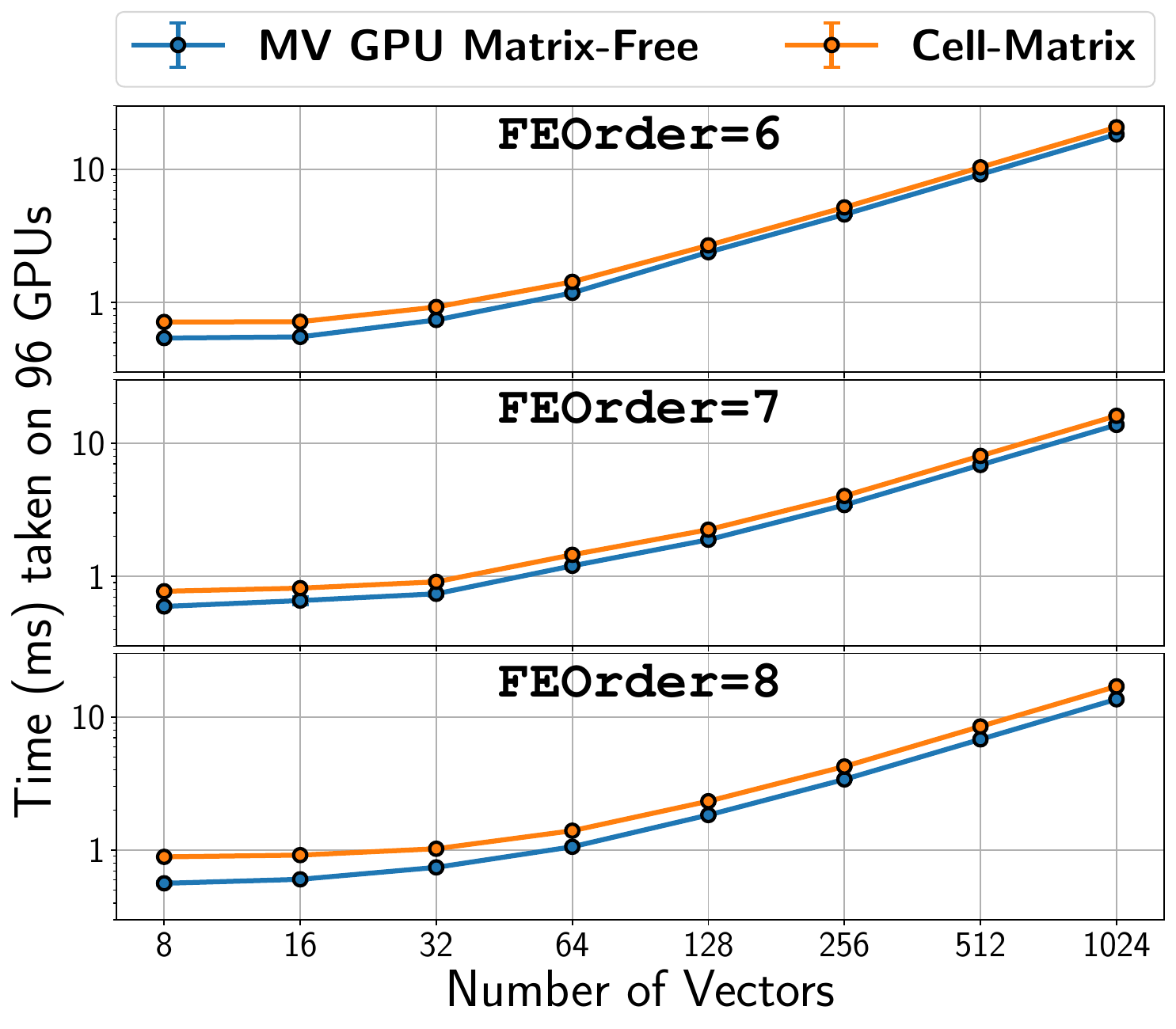}
        \caption{Performance benchmark of our implementation against the cell-matrix method on 16 nodes.}
        \label{fig:gpuscalingN16pq}
    \end{subfigure}%
    \caption{Benchmarks of our matrix-free implementation with cell-matrix implementation for the case of $n_q = n_p + 2$ and uniform mesh. Case studies: 1092727 DoFs (\texttt{FEOrder}=6); 1191016 DoFs (\texttt{FEOrder}=7); 1157625 DoFs (\texttt{FEOrder}=8) for the Helmholtz problem on NVIDIA\textsuperscript{\tiny\textregistered} Tesla\textsuperscript{\tiny\textregistered} V100 SXM2 16GB (Summit Supercomputer).}
\end{figure}


Choosing a FE mesh of around $1.2m$ DoFs, we conducted a strong scaling study of the proposed matrix-free multivector implementation in the case of $n_q = n_p + 2$ for 1024 vectors and compared it with the cell-matrix approach. \cref{fig:gpuscalingvec1024pq} shows the time to solution for this comparative study. Our GPU matrix-free implementation has a noticeable performance advantage over the cell-matrix method across all MPI tasks for \texttt{FEOrder} = 6, 7 and 8. In particular, we show the comparisons (with varying $n_v$) for 6 and 96 GPUs in \cref{fig:gpuscalingN1pq,fig:gpuscalingN16pq} respectively. On a single node we observe speedup of 41\% for \texttt{FEOrder} = 6, a speedup of 72\% for \texttt{FEOrder} = 7 and a 85\% speedup for \texttt{FEOrder} = 8 over the cell-matrix method in the case of 1024 vectors. In the case of 8 vectors on 1 node (6 GPUs, $\sim$200k DoFs/GPU), we observe a speedup of 4.3x for \texttt{FEOrder} = 6, a speedup of 5.2x for \texttt{FEOrder} = 7, and a 9.4x speedup for \texttt{FEOrder} = 8 over the cell-matrix method. On the other extreme, benchmarks for various numbers of vectors on 16 nodes (96 GPUs, $\sim$12k DoFs/GPU) show performance gains of 13\% for \texttt{FEOrder} = 6, a speedup of 17\% for \texttt{FEOrder} = 7, and around 25\% for \texttt{FEOrder} = 8 against the cell-matrix method for 1024 vectors. In the case of 8 vectors, we observe improvements of up to 30\% for \texttt{FEOrder} = 6, 7 and around 58\% for \texttt{FEOrder} = 8 against the cell-matrix method on 96 GPUs.

\subsection{Multivector matrix-free GPU implementation on Selene supercomputer}
\label{sec:selene}
\begin{figure}[!ht]
    \centering
    \begin{subfigure}[t]{.32\textwidth}
        \centering
        \includegraphics[width=\textwidth]{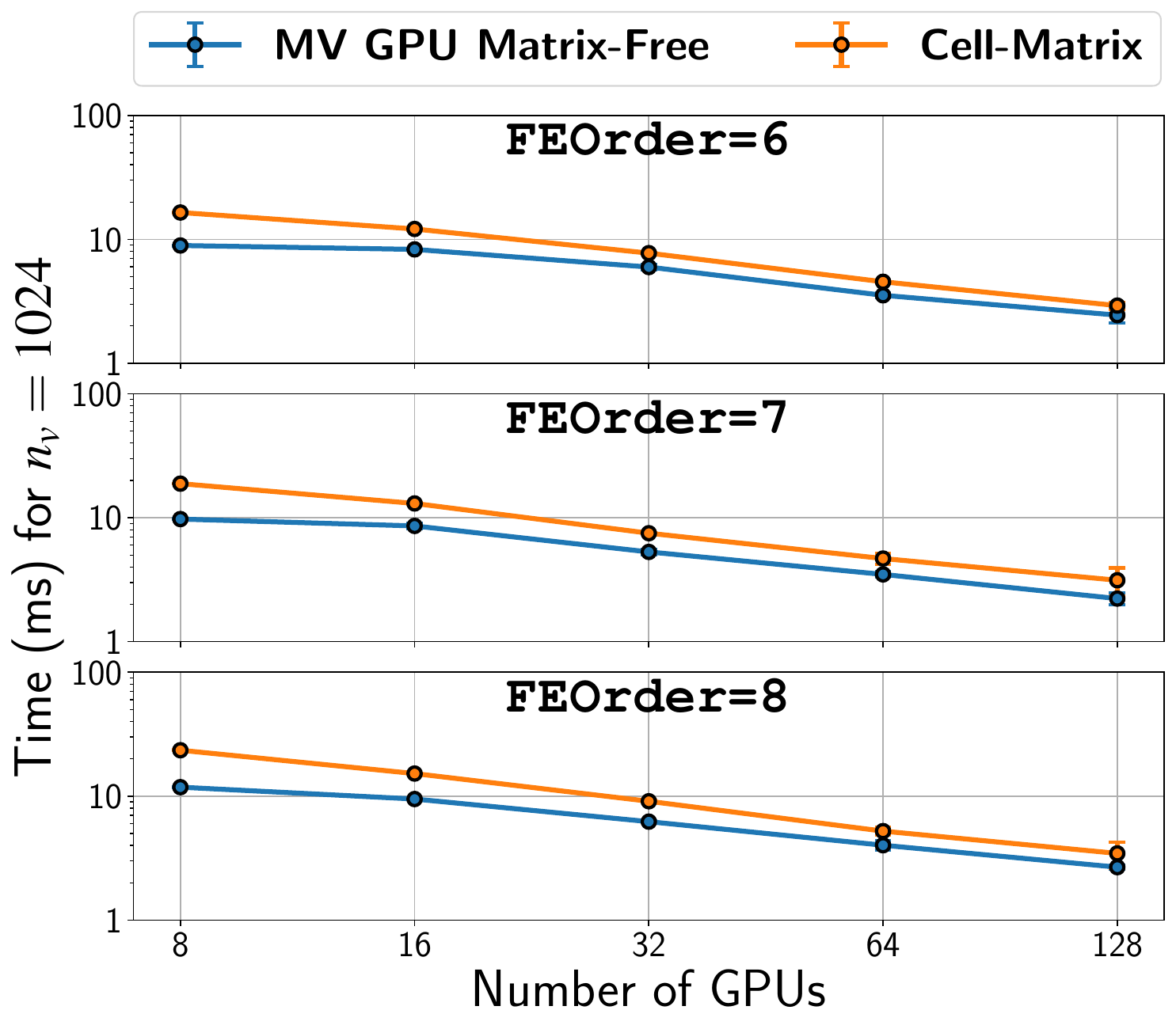}
        \caption{Comparative scaling study of our matrix-free implementation against the cell-matrix method for 1024 vectors.}
        \label{fig:gpuscalingvec1024Selene}
    \end{subfigure}\hfill
    \begin{subfigure}[t]{.32\textwidth}
        \centering
        \includegraphics[width=\textwidth]{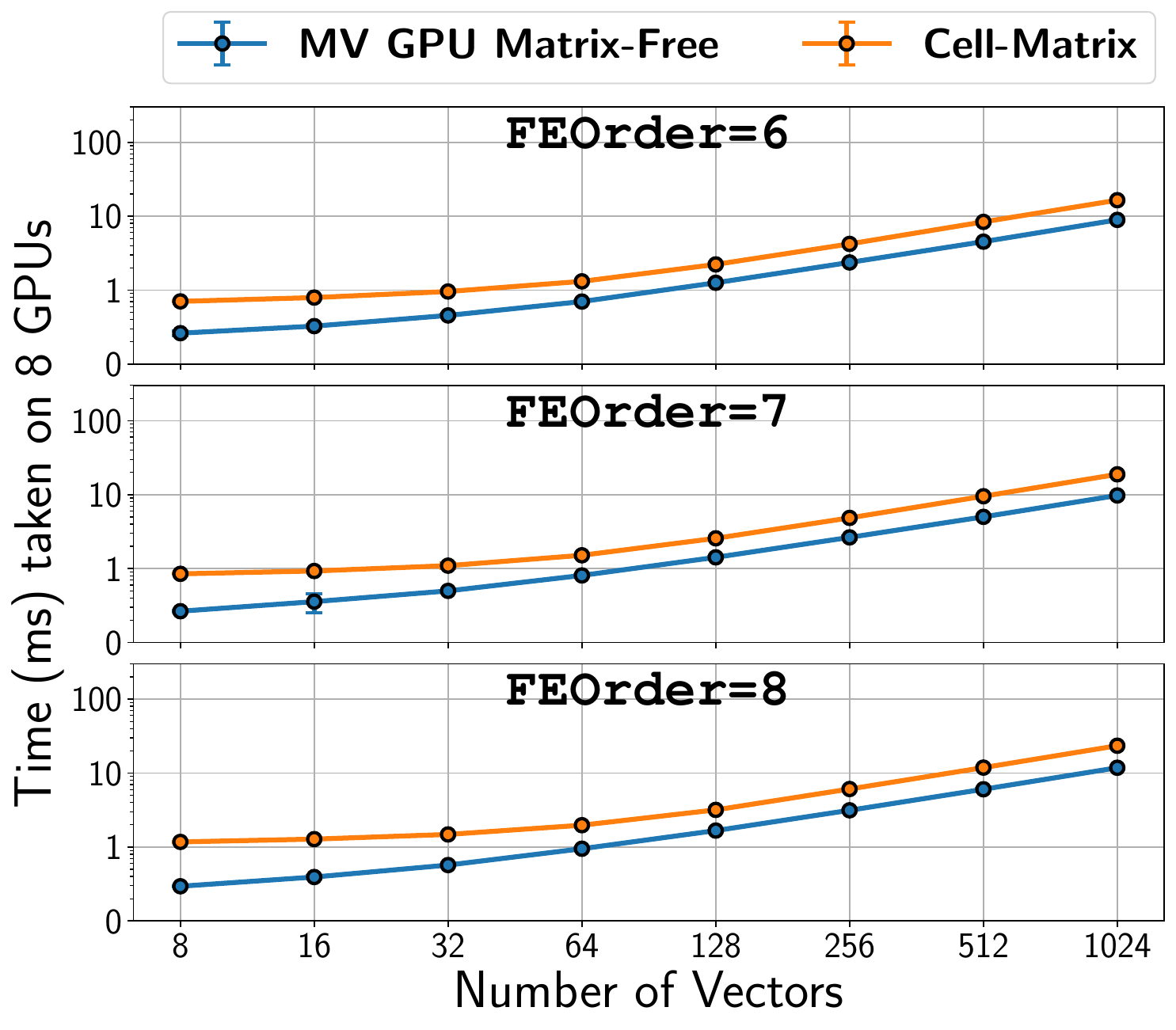}
       \caption{Performance benchmark of our matrix-free implementation against the cell-matrix method on 1 Selene node.}
       \label{fig:gpuscalingN1Selene}
    \end{subfigure}\hfill
    \begin{subfigure}[t]{.32\textwidth}
        \centering
        \includegraphics[width=\textwidth]{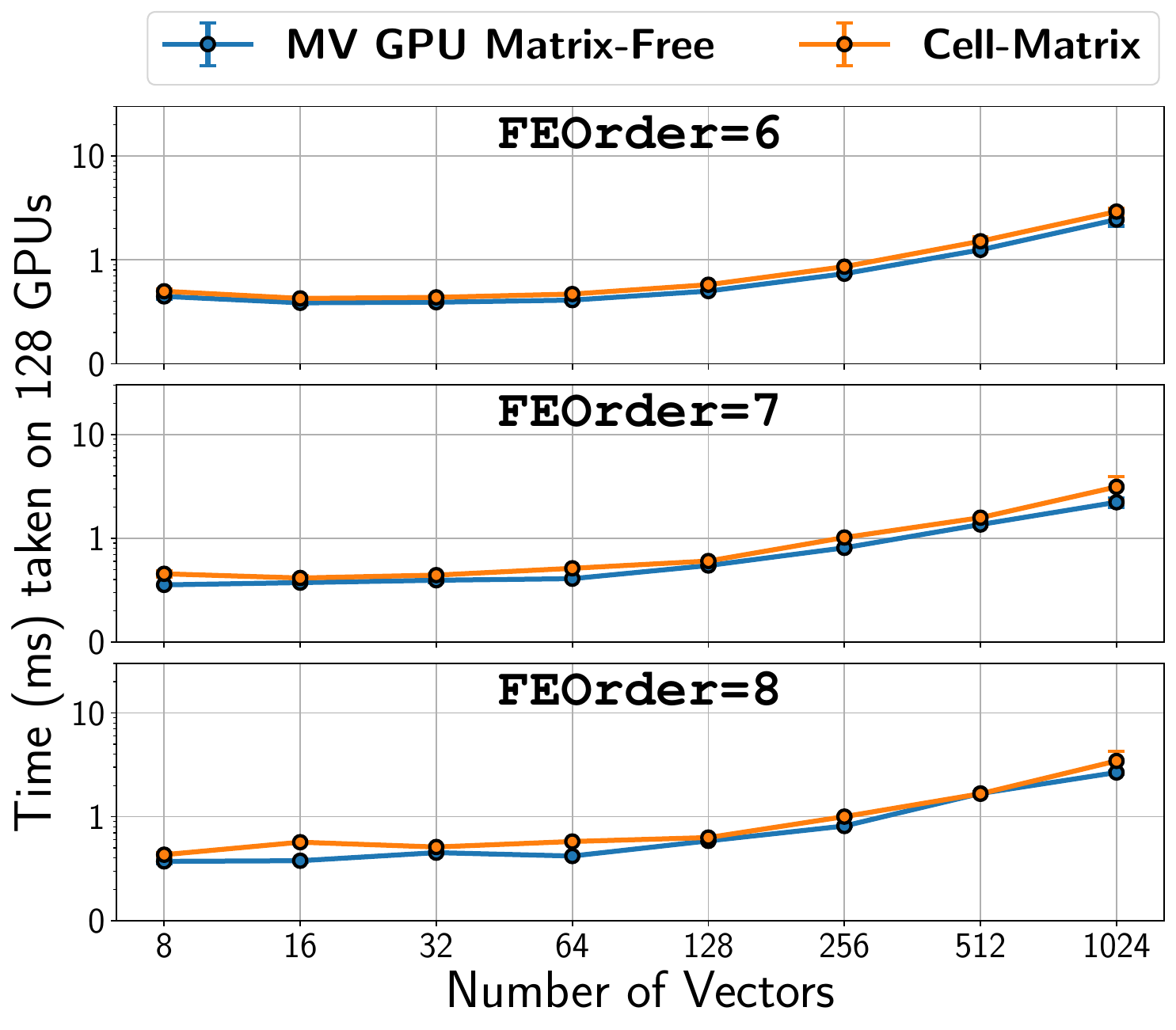}
        \caption{Performance benchmark of our matrix-free implementation against the cell-matrix method on 16 Selene nodes.}
        \label{fig:gpuscalingN16Selene}
    \end{subfigure}%
    
     \caption{Benchmarks of our matrix-free implementation with cell-matrix implementation for the case of $n_q = n_p$ and uniform mesh on NVIDIA\textsuperscript{\tiny\textregistered} Tesla\textsuperscript{\tiny\textregistered} A100 SXM2 80GB. Case studies: 1092727 DoFs (\texttt{FEOrder}=6); 1191016 DoFs (\texttt{FEOrder}=7); 1157625 DoFs (\texttt{FEOrder}=8) for the Helmholtz problem on GPUs.}
\end{figure}

This subsection reports the performance benchmarks obtained using multi-node A100 GPUs on the Selene supercomputer. A single node of the Selene supercomputer has 2 AMD\textsuperscript{\tiny\textregistered} EPYC\textsuperscript{\tiny\texttrademark} 7742 64-Core Processors and 8 NVIDIA\textsuperscript{\tiny\textregistered} A100-SXM4-80GB GPUs with 640 GB HBM2e memory and 156 TFLOP/s performance (A100 FP64). The interconnect is Mellanox\textsuperscript{\tiny\textregistered} ConnectX\textsuperscript{\tiny\textregistered}-6 MT28908, the OS is Ubuntu 20.04.3 LTS and compilers gcc 11.3.0, nvcc 11.8 and Open MPI 4.1.5 with flags \texttt{-O3 -arch=sm\_70 -lcublas}. Employing an uniform mesh with $n_p = n_q$ comprising $\sim$1.2m DoFs, a strong scaling study is conducted to compare the proposed matrix-free multivector implementation with the cell-matrix approach in the case of 1024 vectors as shown in \cref{fig:gpuscalingvec1024Selene}. Our GPU matrix-free implementation has a noticeable performance advantage over the cell-matrix method across all MPI tasks for \texttt{FEOrder} = 6, 7, and 8. In particular, we show the comparisons in more detail (with varying $n_v$) for 1 Selene node (8 GPUs, $\sim$150k DoFs/GPU) and 16 Selene nodes (128 GPUs, $\sim$9k DoFs/GPU) in \cref{fig:gpuscalingN1Selene,fig:gpuscalingN16Selene} respectively. On a single Selene node (8 GPUs, $\sim$150k DoFs/GPU), we observe speedups of up to 1.8x-1.9x for \texttt{FEOrders}=6, 7, 8 over the cell-matrix approach in the case of 1024 vectors. In the case of 8 vectors, we observe a speedup of around 3x-4x for  \texttt{FEOrders}=6, 7, 8 over the cell-matrix approach on the single GPU node. On the other extreme of 16 Selene nodes (128 GPUs, $\sim$9k DoFs/GPU), benchmarks for various number of vectors (\cref{fig:gpuscalingN16Selene}) show performance gains of up to 19\% for \texttt{FEOrder} = 6, a speedup of 41\% for \texttt{FEOrder} = 7, and a speedup of 29\% for \texttt{FEOrder} = 8 against the cell-matrix method for 1024 vectors.



\section{Eigensolver GPU implementations using ChFSI}
On GPUs, the eigensolver employing ChFSI approach has been implemented similar to CPUs following the steps outlined in \cref{sec:cpuchfsi}. 

\twocolumn

\bibliographystyle{elsarticle-num-names} 
\bibliography{references}





\end{document}